\documentclass[journal]{IEEEtran}
\usepackage[depth=subsection]{bookmark}
\usepackage{titlesec}
\titleformat{\paragraph}
{\normalfont\normalsize\bfseries}{\theparagraph}{1em}{}
\titlespacing*{\paragraph}
{0pt}{3.25ex plus 1ex minus .2ex}{1.5ex plus .2ex}
\hypersetup{ colorlinks=true, citecolor=blue}
\usepackage[utf8]{inputenc}
\usepackage{graphicx}
\usepackage[table,xcdraw]{xcolor}
\usepackage{multirow}
\usepackage{hhline}
\usepackage{algorithm,algorithmic,amsbsy,amsmath,amssymb,epsfig,bbm,mathrsfs,multirow,amsthm, verbatim, tikz, cite, makecell, booktabs, color}

\usepackage[switch]{lineno} 
\usepackage{pifont}

\usepackage{array}   
    \newcolumntype{C}[1]{>{\centering\arraybackslash}m{#1}}
    \newcolumntype{P}[1]{>{\centering\arraybackslash}p{#1}}
    \newcolumntype{M}[1]{>{\centering\arraybackslash}m{#1}}

\makeatletter
\renewcommand\paragraph{\@startsection{paragraph}{4}{\z@}%
            {-2.5ex\@plus -1ex \@minus -.25ex}%
            {1.25ex \@plus .25ex}%
            {\normalfont\normalsize\bfseries}}
\makeatother
\ifCLASSINFOpdf
\else
\fi
\begin{document}
%
\title{Deep Reinforcement Learning for Radio Resource Allocation and Management in Next Generation Heterogeneous Wireless Networks: A Survey}
%
%
%

\author{Abdulmalik Alwarafy, Mohamed Abdallah, \IEEEmembership{Senior Member,~IEEE}, Bekir Sait Ciftler, \IEEEmembership{Member,~IEEE}, \\ Ala Al-Fuqaha, \IEEEmembership{Senior Member,~IEEE},  and Mounir Hamdi, \IEEEmembership{Fellow Member,~IEEE} 

\thanks{The authors are with the Division of Information and Computing Technology, College of Science and Engineering, Hamad Bin Khalifa University, Qatar (e-mail: aalwarafy@hbku.edu.qa; moabdallah@hbku.edu.qa; bciftler@hbku.edu.qa; aalfuqaha@hbku.edu.qa; mhamdi@hbku.edu.qa). \textbf{\textit{This work has been submitted to the IEEE for possible publication. Copyright may be transferred without notice, after which this version may no longer be accessible.}}}
}

\maketitle

\begin{abstract}
Next generation wireless networks are expected to be extremely complex due to their massive heterogeneity in terms of the types of network architectures they incorporate, the types and numbers of smart IoT devices they serve, and the types of emerging applications they support. In such large-scale and heterogeneous networks (HetNets), radio resource allocation and management (RRAM) becomes one of the major challenges encountered during system design and deployment. In this context, emerging Deep Reinforcement Learning (DRL) techniques are expected to be one of the main enabling technologies to address the RRAM in future wireless HetNets. In this paper, we conduct a systematic in-depth, and comprehensive survey of the applications of DRL techniques in RRAM for next generation wireless networks. Towards this, we first overview the existing traditional RRAM methods and identify their limitations that motivate the use of DRL techniques in RRAM. Then, we provide a comprehensive review of the most widely used DRL algorithms to address RRAM problems, including the value- and policy-based algorithms. The advantages, limitations, and use-cases for each algorithm are provided. We then conduct a comprehensive and in-depth literature review and classify existing related works based on both the radio resources they are addressing and the type of wireless networks they are investigating. To this end, we carefully identify the types of DRL algorithms utilized in each related work, the elements of these algorithms, and the main findings of each related work. Finally, we highlight important open challenges and provide insights into several future research directions in the context of DRL-based RRAM. This survey is intentionally designed to guide and stimulate more research endeavors towards building efficient and fine-grained DRL-based RRAM schemes for future wireless networks.

\end{abstract}


%
\IEEEpeerreviewmaketitle

\section{Introduction} \label{Sec1:Introduction}
\begin{figure*}[t!]
    \centering
    \includegraphics[height=10.2cm, width=18.1cm]{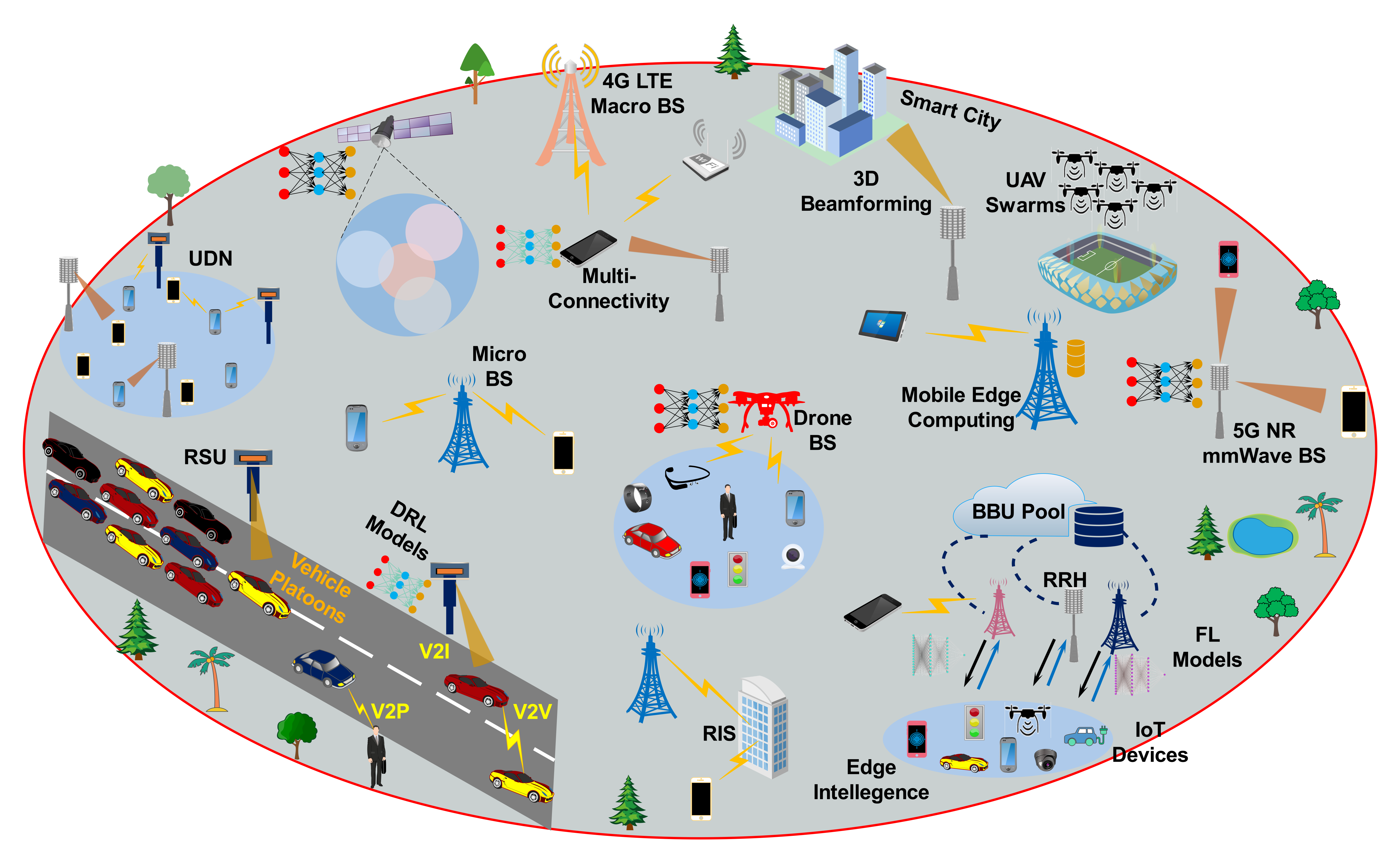}
    \caption{A pictorial illustration of next generation wireless networks characterized by their massive heterogeneity in terms of RANs infrastructures, types and numbers of user devices served, and types of applications and services supported.}
    \label{6G_HetNet_Fig}
\end{figure*}

\IEEEPARstart{R}{adio} resource allocation and management (RRAM) is regarded as one of the essential challenges encountered in modern wireless communication networks \cite{hussain2020machine}. Nowadays, modern wireless networks are becoming more heterogeneous and complex in terms of the types of emerging radio access networks (RANs) they integrate, the explosive number and types of smart devices they serve, and the types of disruptive applications and services they support \cite{RN122, RN234}. It is envisaged that future networks will integrate land, air, space, and deep-sea wireless networks into a single network to meet the stringent requirements of a fully-connected world vision \cite{RN269,6GSummit}, as shown in Fig. \ref{6G_HetNet_Fig}. This will ensure ubiquitous connectivity for user devices with enhanced quality of service (QoS) in terms of coverage, reliability, and throughput. In addition, future user devices will also witness an unprecedented increase in their numbers and types of data-hungry applications they require/support \cite{RN234, forecast2019cisco}. It is expected that by 2023, the number of user networked devices and connections, including smart-phones, tablets, wearable devices, and sensors, will reach 29.3 billion \cite{forecast2019cisco}, and generate a data rate exceeding 50 trillion GB \cite{hussain2020machine}. All these trends will exacerbate the burdens during system design, planning, deployment, operation, and management. In particular, RRAM will become crucial in such complex and large-scale networks in order to guarantee an enhanced communications experience.

RRAM plays a pivotal role during infrastructure planning, implementation, and resource optimization of modern wireless networks. Efficient RRAM solutions will guarantee enhanced network connectivity, increased system efficiency, and reduced energy consumption. The performance of wireless networks heavily relies on two aspects. First, how network radio resources are being utilized, managed, and orchestrated, including transmit power control, spectrum channel allocations, and user access control. Second, how efficiently the system can react to the rapid changes of network dynamics, including wireless channel statistics, users mobility patterns, instantaneous radio resources availability, and variability in traffic loads. Efficient RRAM techniques must efficiently and dynamically account for such design aspects in order to ensure high network QoS and enhanced users' Quality of Experience (QoE). 

\subsection{Motivations of the Paper} \label{paperMotivation}
The main motivations of this work stem from three aspects. First, the paramount importance of allocating radio resources in future wireless networks. Second, the limitations and shortcomings of existing state-of-the-art RRAM techniques. Third, the robustness of Deep reinforcement techniques in alleviating these limitations and providing efficient performance in the context of RRAM. Here we elaborate more on each aspect. 

\subsubsection{Importance of RRAM in Modern Wireless Networks}
The explosive growth in the number and types of modern smart devices, such as smartphones/tablets and wearable devices, has led to the emergence of disruptive wireless communications and networking technologies, such as 5G NR cellular networks, IoT networks, personal (or wireless body area networks), device-to-device (D2D) communications, holographic imaging and haptic communications, and vehicular networks \cite{RN234, RN267, RN273, RN269, RN239, RN264, RN270}. Such networks are envisaged to meet the stringent requirements of the emerging applications and services via supporting high data rates, coverage, and connectivity with significant enhancements in reliability, reduction in latency, and mitigation of energy consumption. 

However, achieving this goal in such large-scale, versatile, and complex wireless networks is quite challenging, as it requires a judicious allocation and management of the networks' limited radio resources \cite{Liang2020deep, nguyen2019non}. In particular, efficient and more advanced RRAM solutions must be developed to balance the tradeoff between enhancing network performance while guaranteeing an efficient utilization of radio resources. Furthermore, efficient RRAM solutions must also strike and intelligent tradeoff between optimizing network radio resources and satisfying users' QoE. For example, RRAM techniques must jointly enhance network spectral efficiency (SE), energy efficiency (EE), and throughput while mitigating interference, reducing latency, and enhancing rate for user devices.  

Efficient and advanced RRAM schemes can considerably enhance the system's SE compared to the traditional techniques by relying on the advanced channel and/or source coding methods. RRAM is essential in broadcast wireless networks covering wide geographical areas as well as in modern cellular communication networks comprised of several adjacent and dense access points (APs) that typically share and reuse the same radio frequencies.

From a cost point of view, the deployment of wireless APs and sites, e.g., base stations (BSs), including the real estate costs, planning, maintenance, and energy, is the most critical aspect alongside with the frequency license fees. Hence, the goal of RRAM is maximizing the network's SE in terms of bits/sec/Hz/area unit or Erlang/MHz/site, under some constraints related to user fairness. For instance, the service grade must meet a minimum acceptable level of QoS, including the coverage of certain geographical areas while mitigating network outages caused by interference, noise, large-scale fading (due to path losses and shadowing), and small-scale fading (due to multi-path). The service grade also depends on blocking caused by admission control, scheduling errors, or inability to meet certain QoS demands of edge devices (EDs).

\subsubsection{Where Do Traditional RRAM Techniques Fail?}
Future wireless communication networks are complex due to their large-scale, versatile, and heterogeneous nature. To optimally allocate and manage radio resources in such networks, we typically formulate RRAM as complex optimization problems. The objective of such problems is to achieve a particular goal, such as maximizing network sum-rate, SE, and EE, given the available radio resources and QoS requirements of user devices. Unfortunately, the massive heterogeneity nature of modern networks poses tremendous challenges during the process of formulating optimization problems as well as applying conventional techniques to solve them, such as optimization, heuristic, and game theory algorithms. 

The large-scale nature of next generation networks makes it quite difficult to formulate RRAM optimization problems that are often intractable non-convex. Also, conventional techniques used to solve the RRAM problems require complete or quasi-complete knowledge of the wireless environment, including accurate channel models and real-time CSI. However, obtaining such information in a real-time fashion in these large-scale networks is quite difficult or even impossible. Furthermore, conventional techniques are often computationally-expensive and incur considerable timing overhead. This renders them inefficient for most emerging time-sensitive applications, such as autonomous vehicles and robotics.

Moreover, game theory-based techniques are unsuitable for future heterogeneous networks (HetNets) as such techniques are devised for homogeneous players. Also, the explosive number of network APs and user devices will create extra burdens on game theory-based techniques. In particular, network players, such as BSs, APs, and user devices, need to exchange a tremendous amount of data and signaling. This will induce unmanageable overhead that largely increases delay, computation, and energy/memory consumption of network elements.

\subsubsection{How Can DRL Overcome these Challenges and Provide Efficient RRAM Solutions?}
Emerging artificial intelligence (AI) techniques, such as deep reinforcement learning (DRL), have shown efficient performance in addressing various issues in modern wireless communication networks, including solving complex RRAM optimization problems \cite{sutton2018reinforcement, RN1, lee2019survey, obite2021overview, gupta2021deep, du2020green, 8917870, xiong2019deep, Feriani2021Single}. In the context of RRAM, DRL methods are mainly used as an alternative to overcome the shortcomings and limitations of the conventional RRAM techniques discussed above. In particular, DRL techniques can solve complex network RRAM optimization problems and take judicious control decisions with only limited information about the network statistics. They achieve this by enabling network entities, such as BSs, RAN's APs, edge servers (ESs), gateways nodes, and user devices, to make intelligent and autonomous control decisions, such as RRAM, user association, and RAN's selection, in order to achieve various network goals such as sum-rate maximization, reliability enhancement, delay reduction, and SE/EE maximization. In addition, DRL techniques are model-free that enable different network entities to learn optimal policies about the network, such as RRAM and user association, based on their continuous interactions with the wireless environment, without knowing the exact channel models or other network statistics \textit{a-priori}. These appealing features make DRL methods one of the main key enabling technologies to address the RRAM issue in modern wireless communication networks \cite{RN234, RN122}.
 
\subsection{Related Work}
There is a limited number of surveys that focus on the role of DRL in RRAM. Existing related surveys are listed in Table 1. The table also summarizes the topics covered in these surveys along with a mapping to the relevant sections of this paper and a categorical discussion of the improvements and value-added in this paper relative to these surveys. In general, as reported in Table 1, these published surveys still have several research gaps that are addressed in this survey. We summarize them as follows.
\begin{itemize}
    \item Some of the existing surveys focus on DRL applications in wireless communications and networking in general, without paying much attention to RRAM \cite{RN1, gupta2021deep}. For example, existing surveys cover topics related to DRL enabling technologies, use-cases, architectures, security, scheduling, clustering and data aggregation, traffic management, etc.
    \item Some of the published surveys focus on RRAM for wireless networks using ML and/or DL techniques without paying much attention to DRL techniques \cite{hussain2020machine, chen2019artificial, Liang2020deep, zappone2019wireless}. For example, they consider ML techniques such as convolutional neural networks (CNN), recurrent neural networks (RNN),  supervised learning, Bayesian learning, K-means clustering, Principal Component Analysis (PCA), etc.
    \item Even the surveys that address DRL for RRAM in wireless networks focus on specific wireless network types or applications \cite{obite2021overview, du2020green, khorasgani2020challenges, lee2019survey, 8917870}, missing some of the recent research, not providing an adequate overview of the most widely used DRL algorithms for RRAM \cite{8917870}, or not covering the RRAM in-depth, but, rather, just covering a limited number of radio resources.
\end{itemize}
  
Hence, the role of this paper to fill these research gaps and overcome these shortcomings. In particular, we provide a comprehensive survey on the application of DRL techniques in RRAM for next generation wireless communication networks. We have carefully cited up-to-date surveys and related research works. We should emphasize here that the scope of this paper is focused only on radio (or communication) resources, and no computation resources are included during the study and analysis. Fig. 2 shows the radio resources or issues addressed in this survey. However, computation resource aspects such as offloading, storage, task scheduling, caching, etc., can be found in other studies such as \cite{rahman2020deep, mohammed2020deep, sheng2021deep, chen2021energy, liu2020edgeslice} and the references therein.

\begin{table*}[t]
\centering
\caption{Relationship Between this Survey and Existing Surveys on DRL-Based RRAM For Wireless Networks}
\label{tab:my-table}
\begin{tabular}{|p{1.4cm}|p{5.4cm}|c|p{7.2cm}|}
\hline
\rowcolor[HTML]{C0C0C0} 
\textbf{Paper} & \textbf{Summary of the survey's contributions} &  \begin{tabular}[c]{@{}l@{}}\textbf{Related contents }\\\textbf{in this paper}\end{tabular}& \textbf{Value added in this paper} \\ \hline\hline
Luong \textit{et al.}\cite{RN1}& Applications of DRL in communications and networking& Section \ref{Sec3:DRL types}/\ref{Sec4:heart}& Particularly focus on DRL usage for RRAM and enhanced list of papers \\ \hline
Hussain \textit{et al.}\cite{hussain2020machine}  & ML- and DL-based resource management mechanisms in cellular wireless and IoT networks & Sections \ref{Sec2:RRAMtechs}/\ref{Sec4:heart} &  In-depth and holistic coverage of DRL algorithms used for RRAM, intensive review of existing papers related to DRL for RRAM, and the coverage of more types of wireless networks \\ \hline
Lin \textit{et al.}\cite{lin2020artificial}& Applications of AI approaches in resource management, such as spectrum, computing, and caching.  & Section \ref{Sec5:future}& Particularly focus on DRL methods, including more radio resources, and intensive literature review  \\ \hline
Liang \textit{et al.}\cite{Liang2020deep} & DL-Based resource allocation with application to vehicular networks & Sections \ref{Sec2:RRAMtechs}/\ref{Sec4:heart} & Focus on DRL techniques for RRAM, in-depth literature review, and include various types of modern wireless networks \\ \hline
Obite \textit{et al.}\cite{obite2021overview}& Spectrum sensing in cognitive radio networks& Section \ref{Sec4:heart}& Detailed investigation of more radio resources, holistic coverage of more wireless networks, and an intensive up-to-date review of existing papers \\ \hline
Chen \textit{et al.}\cite{chen2019artificial}& Applications of ML algorithms in solving wireless networking problems & NA& Focus on applications of DRL in solving RRAM wireless problems, and coverage of more wireless networks \\ \hline
Gupta \textit{et al.}\cite{gupta2021deep}&  General research and simulation tools used for DRL& Section \ref{Sec3:DRL types}& Specifically focus on DRL algorithms along with the related research conducted specifically in the context of RRAM   \\ \hline
Du \textit{et al.}\cite{du2020green}& Investigates how to achieve green DRL for radio resource management via energy allocation based on architecture and algorithm innovations& Section \ref{Sec4:heart}& Further extend to more radio resources and more modern wireless networks  \\ \hline
Pham \textit{et al.}\cite{pham2014resource}& A layered-based classification of resource management techniques in Wireless Access Networks& Section \ref{Sec2:RRAMtechs}& A holistic study of conventional and emerging ML-based techniques for RRAM applied to modern wireless networks and including more radio resources    \\ \hline
Almazrouei \textit{et al.} \cite{almazrouei2020can}& Potential benefits and the challenges when using ML for in radio spectrum& Section \ref{Sec2:RRAMtechs}& Extend the analysis to incorporate more radio resources, focus on DRL methods for RRAM, and provide in-depth literature review\\ \hline
Dhilipkumar \textit{et al.}\cite{dhilipkumar2019comparative}& Discusses various resource allocation scheme for D2D networks in cellular network& Sections \ref{Sec2:RRAMtechs}/\ref{Sec4:heart}& Focus on DRL-based RRAM for various applications in services in wireless networks\\ \hline
Arulkumaran \textit{et al.}\cite{arulkumaran2017deep}, Zhang \textit{et al.}\cite{zhang2019deepOverview}& Overview of DRL approaches in general including, applications and models& Section \ref{Sec3:DRL types}& Focus on DRL approaches utilized in RRAM for wireless networks, and also provide detailed literature review  \\ \hline
Zappone \textit{et al.}\cite{zappone2019wireless}&  Motivations, applications, visions, and case studies for the usage of DL techniques in wireless communication networks & NA& Particularly focusing on DRL techniques for wireless communication networks in the context of RRAM  \\ \hline
Lee \textit{et al.}\cite{lee2019survey}& DRL-based resource management schemes for 5G HetNets in energy harvesting, network slicing, cognitive HetNets, coordinated multi-point transmission, and big data& Section \ref{Sec3:DRL types}& In-depth analysis of DRL methods used for RRAM including; DRL algorithms, types of wireless networks, types of radio resources investigated, and extensive literature review   \\ \hline
Qian \textit{et al.}\cite{8917870} & Applications of RL and DRL in three technologies: mobile edge computing, software defined network, and network virtualization in 5G& NA& Focus on DRL for RRAM for cellular and other emerging wireless networks \\ \hline
Khorasgani \textit{et al.}\cite{khorasgani2020challenges}& Key limitations and challenges in using DRL to address the problem of dynamic dispatching in the mining industry& Section \ref{Sec4:heart}& Extend the investigation to include various wireless networks with an extensive focus on radio resources  \\ \hline
\end{tabular}
\end{table*}

\subsection{Paper Contributions}
\begin{figure}[t!]
    \centering
    \includegraphics[height=5.5cm, width=8.5cm]{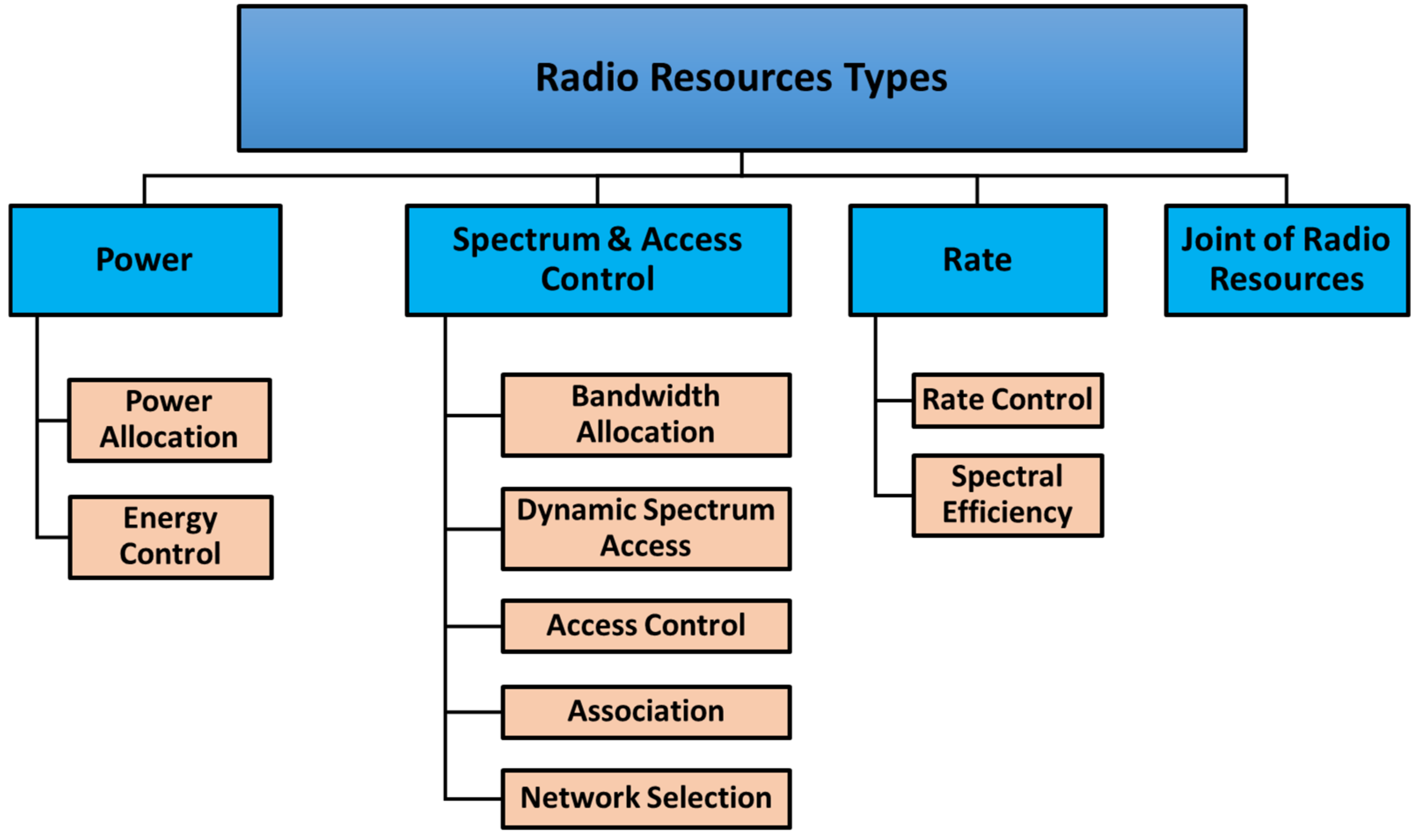}
    \caption{Classification based on radio resources (or issues) addressed in the papers.}
    \label{ClassificationOnResources}
\end{figure}

\begin{figure}[t!]
    \centering
    \includegraphics[height=5cm, width=8.5cm]{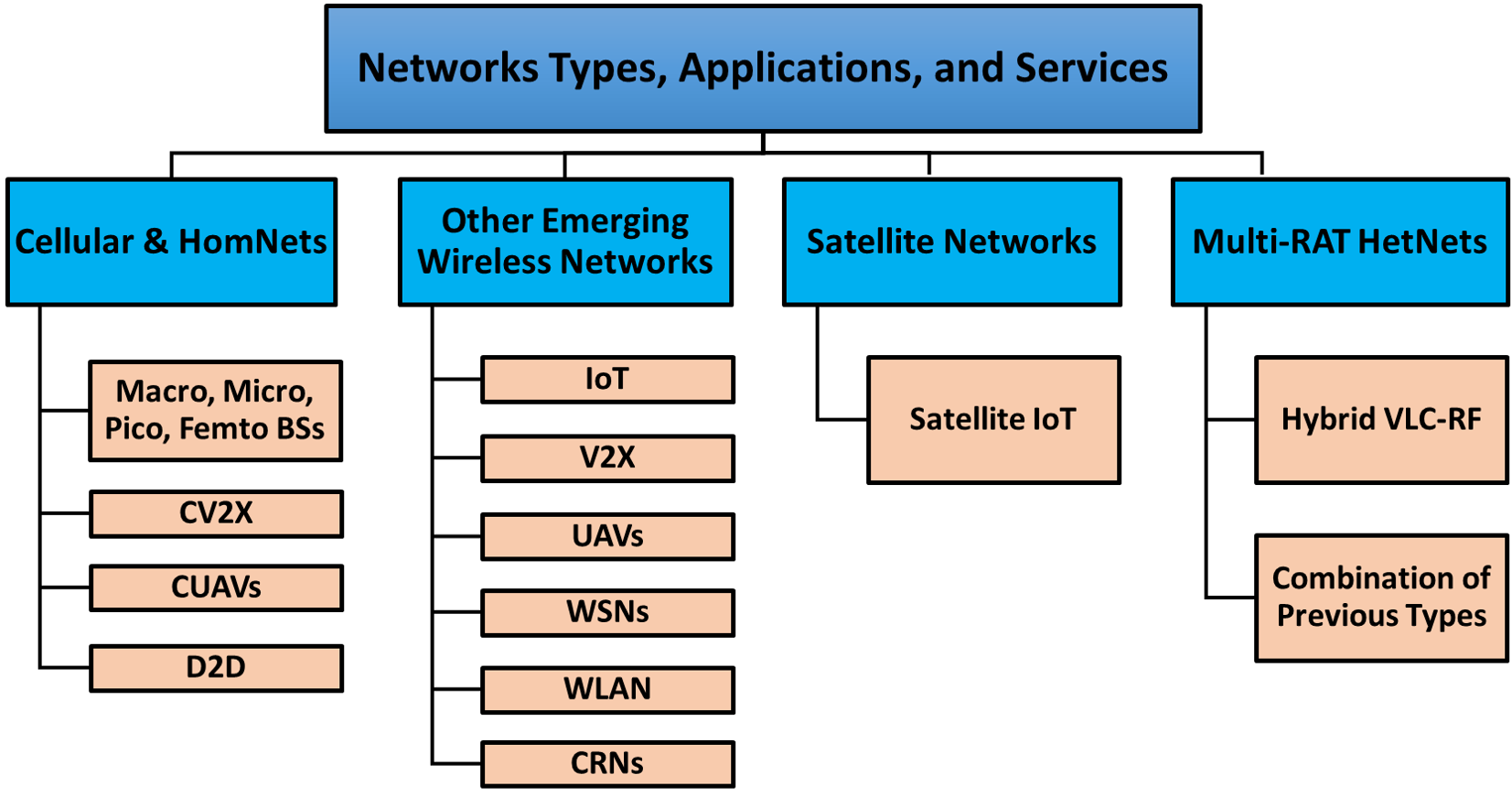} 
    \caption{Classification based on networks types covered in the papers.}
    \label{ClassificationOnNetwoksTypes}
\end{figure}

The main contributions of this paper are summarized as follows.
\begin{enumerate}
    \item We provide a detailed discussion on the state-of-the-art techniques used for RRAM in wireless networks, including their types, shortcomings, and limitations that led to the adoption of DRL solutions.
    \item We identify the most widely used DRL techniques utilized in RRAM of wireless networks and provide a comprehensive overview of them. The advantages, features, and limitations of each technique are discussed. Hence, the reader is provided with an in-depth knowledge of which DRL techniques should be leveraged for each RRAM problem under investigation. 
    \item We conduct an extensive and up-to-date literature review and classify the papers as reported in the literature based on the type of radio resources they address (as shown in Fig. \ref{ClassificationOnResources}) and the types of wireless networks, applications, and services they consider (as shown in Fig. \ref{ClassificationOnNetwoksTypes}). Specifically, for each paper reviewed, we identify the problem it addresses, type of wireless network it investigates, type of DRL model(s) it implements, main elements of the DRL models (i.e., agent, state space, action space, and reward function), and its main findings. This provides the reader with in-depth technical knowledge of how to efficiently engineer DRL models for RRAM problems in wireless communications. 
    \item Based on the papers reviewed in this survey, we outline and identify some of the existing challenges and provide deep insights into some promising future research directions in the context of using DRL for RRAM in wireless networks.
\end{enumerate}

\begin{figure}[htp!]
    \centering
    \includegraphics[height=3.8cm, width=8.8cm]{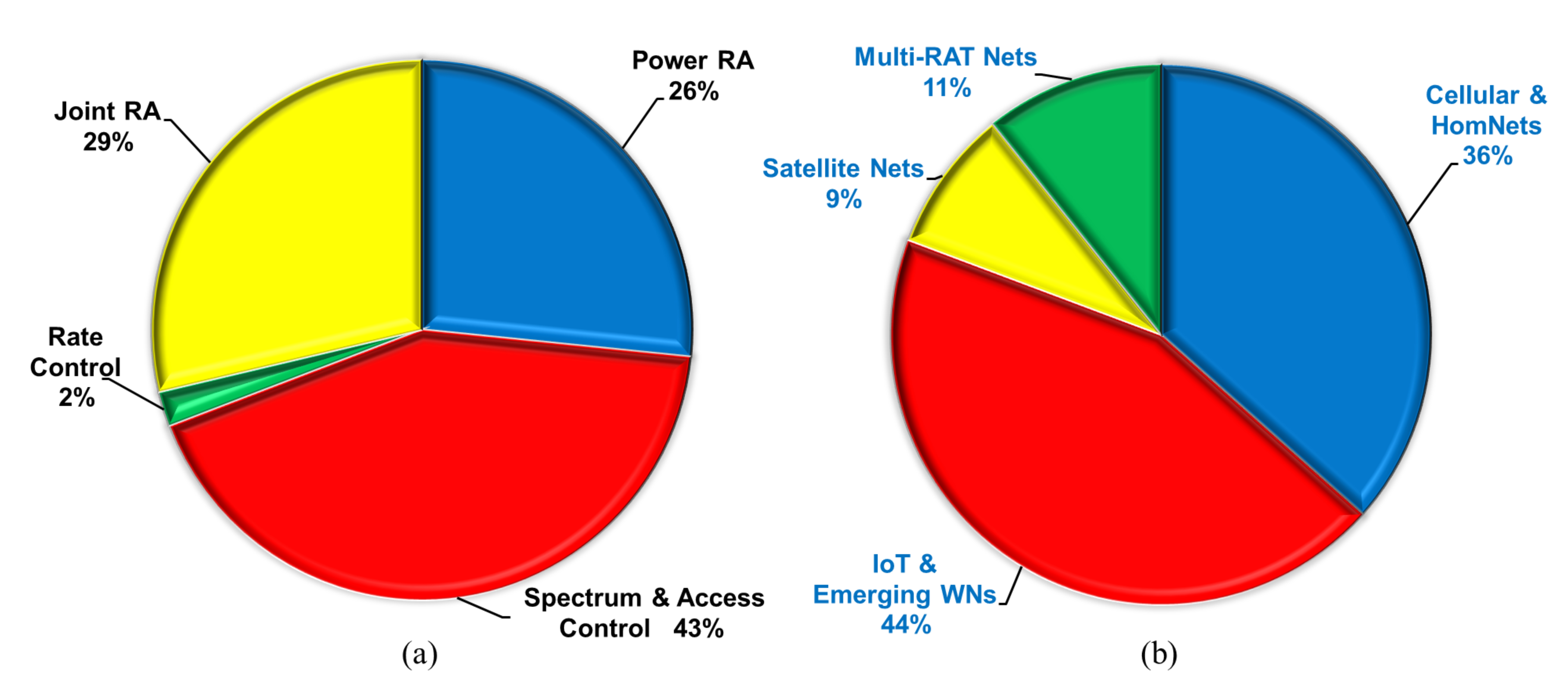}
    \caption{Percentages of related work based on (a) types of radio resources covered and (b) types of networks and application investigated. RA: resource allocation.}
    \label{Percentage_RRAandNetType} 
\end{figure}

\begin{figure}[t!]
    \centering
    \includegraphics[height=8cm, width=8.7cm]{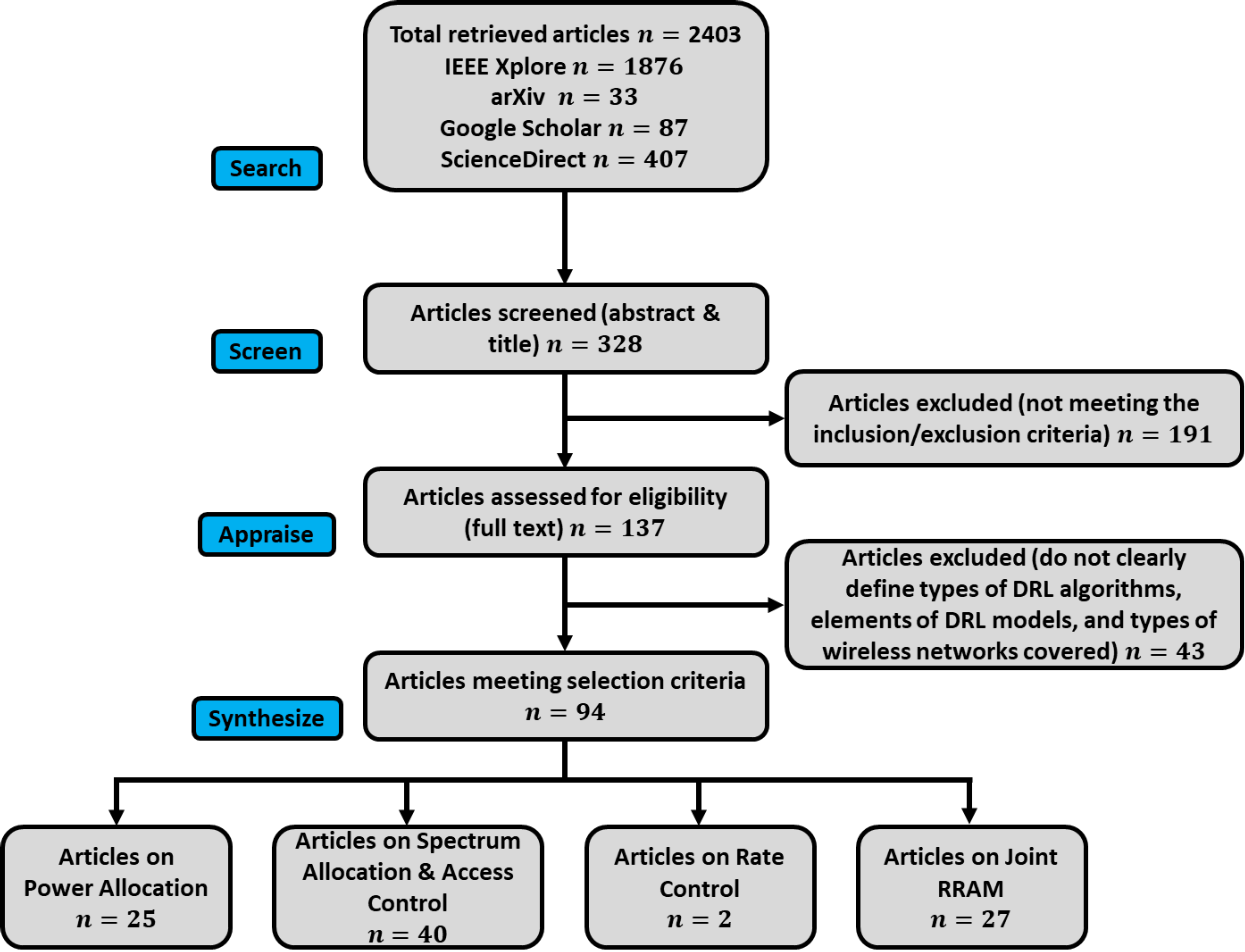}
    \caption{The review protocol followed in this survey.}
    \label{Prisma}
\end{figure}

Fig. \ref{Percentage_RRAandNetType} shows the percentage of the related works, classified based on the types of radio resources discussed in each pape, Fig. \ref{Percentage_RRAandNetType} (a), and based on the types of wireless networks studied in each paper, Fig. \ref{Percentage_RRAandNetType} (b). This survey is designed by carefully following the review protocol illustrated in Fig. \ref{Prisma}. Since this survey mainly focuses on deep reinforcement learning for RRAM in wireless networks, we included the following terms during the search stage along with "AND/OR" combinations of them; "deep reinforcement learning," "DRL," "resource allocation," "resource management," "power," "spectrum," "bandwidth," "access control," "user association," "network selection," "cell selection," "rate control," "joint resources," "wireless networks," "satellite networks," "cellular networks," and "Heterogeneous networks."  The number of papers found and the databases searched are detailed in Fig. \ref{Prisma}. The inclusion criteria are papers that address the use of DRL techniques to manage and allocate the radio resources shown in Fig. \ref{ClassificationOnResources} for the wireless networks shown in Fig. \ref{ClassificationOnNetwoksTypes}. The exclusion criteria are papers that: 1) address computation resources, e.g., task offloading, storage, scheduling, etc., 2) use conventional RRAM approaches, i.e., not using DRL techniques, 3) use ML/DL techniques, or 4) address non-wireless networks, e.g., wired networks, optical networks, etc. In Fig. \ref{Prisma}, the number of papers excluded after a detailed check of the body is 43, which are directly related to our survey but do not clearly identify the types of DRL algorithms used, elements of DRL models (i.e., agents, state space, action space, and reward function), type of wireless networks covered, and/or not well written.

In general, the research questions that this survey aims to address are stated as follows. How can DRL techniques be implemented to address the RRAM problems in modern wireless networks? What are the performance advantages achieved when using DRL tools compared to the state-of-the-art RRAM approaches? What are the challenges and possible research directions that stem from the reviewed papers in the context of using DRL for RRAM in wireless networks? The retrieved papers shown in Fig. \ref{Prisma}, i.e., the 94 papers, are selected carefully to help with answering these questions, as we will elaborate in the next sections.

It is observed from Fig. \ref{Percentage_RRAandNetType} (a) that the majority of related works are on the Spectrum and Access Control radio resources, followed by the Joint radio resources. Also, as shown in Fig. \ref{Percentage_RRAandNetType} (b), the related works on the IoT and Other Emerging Wireless Networks have received more attention than the other wireless network types, followed by the Cellular and Homogeneous Networks (HomNets).

\begin{figure*}[t!]
    \centering
    \includegraphics[height=13cm, width=17cm]{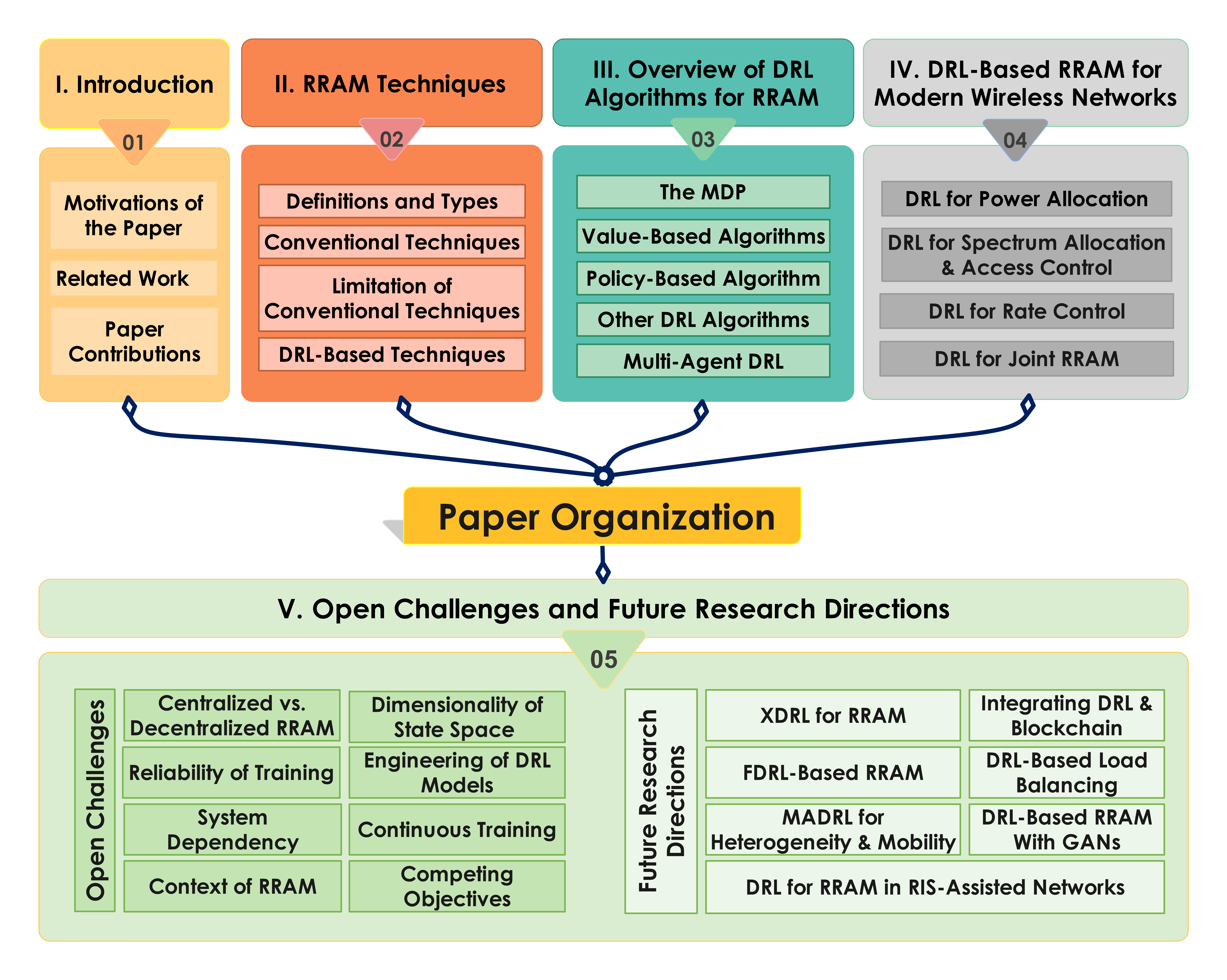}
    \caption{Organization of the paper.}
    \label{Paper_Oraganization}
\end{figure*}

The rest of this paper is organized as follows. Table \ref{Acronyms} lists the acronyms used in this paper and their definitions. Section \ref{Sec2:RRAMtechs} discusses existing RRAM techniques, including conventional methods and DRL-based methods. The definitions, types, and limitations of existing techniques are discussed. Also, the advantages of employing DRL techniques for RRAM are explained. Section \ref{Sec3:DRL types} provides an overview of the DRL techniques widely employed for RRAM, including their types and architectures. In-depth classifications of the existing research works is provided in Section \ref{Sec4:heart}. Existing papers are classified based on the radio resources and the network types they cover. Section \ref{Sec5:future} provides key open challenges along with insights for future research directions. Finally, Section \ref{Conculsion} concludes the paper. The organization of the paper is pictorially illustrated in Fig. \ref{Paper_Oraganization}.

\begin{table*}[htbp]
\caption{List of Acronyms Used and Their Definitions}
\centering
\label{Acronyms}
\begin{tabular}{|c|c||c|c||c|c|}
\hline
\rowcolor[HTML]{C0C0C0} 
\textbf{Acronym} & \textbf{Definition} & \textbf{Acronym} & \textbf{Definition}& \textbf{Acronym} & \textbf{Definition} \\ \hline\hline
RRA & Radio Resource Allocation& RRAM &  Radio Resource Allocation and Management& UE& User End \\ \hline
VLC& Visible Light Communication  &RL  & Reinforcement Learning &EE & Energy Efficiency  \\ \hline
DRL & Deep Reinforcement Learning  &DL  & Deep Learning& TD & Time Difference  \\ \hline
ML & Machine Learning  &DQN  &Deep Q-Network & OU & Ornstein–Uhlenbeck  \\ \hline
MDP &Markov Decision Process  &DDQN  & Double Deep Q-Network & D2D& Device-to-Device  \\ \hline
DNN &Deep Neural Network  & DDPG& Deep Deterministic Policy Gradient & BS& Base Station \\ \hline
QoS &Quality of Service  & RAT & Radio Access Technology & mmWave& Millimeter Wave \\ \hline
A2C &advantage actor-critic  & MADRL &Multi-Agent Deep Reinforcement Learning & RF& Radio Frequency   \\ \hline
NOMA &Non-Orthogonal Multiple Access & FL&  Federated Learning & AP& Access Point  \\ \hline
HetNets & Heterogeneous Networks &  NTNs& Non-Terrestrial Networks & SE& Spectral Efficiency \\ \hline
IAB & Integrated Access and Backhaul& WMMSE & Weighted Minimum Mean Square Error & IoT& Internet of Things \\ \hline
CSI & Channel State Information & WLAN& Wireless Local Area Network  & V2V&Vehicle to Vehicle \\ \hline
M2M & Machine-to-Machine & UAV& Unmanned Aerial Vehicles & V2X& Vehicle to Everything \\ \hline
CRN & Cognitive Radio Network & PPO&  Proximal Policy Optimization& V2I& Vehicle to Infrastructure \\ \hline
DSA & Dynamic Spectrum Access & CV2X & Cellular Vehicular Communication & MeNB & Macro eNodeB \\ \hline
SNR& Signal to Noise Ratio & SINR & Signal to Interference plus Noise Ratio & RSU & Road Side Unit\\ \hline
SIoT& Satellite Internet of Things & A3C & Asynchronous Actor Critic Algorithm & PU& Primary User \\ \hline
IIoT& Industrial Internet of Things & AI&Artificial Intelligence & RB & Resource Block \\ \hline
SU & Secondary User  & OFDM&  Orthogonal Frequency Division Multiplexing& TDD& Time Division Duplex\\ \hline
LTE & Long-Term Evolution & D3QN&  Dueling Double Deep Q-Network& UDN &  Ultra-Dense Network\\ \hline
C-RAN & Cloud Radio Access Network &RRH & Remote Radio Head & BBU& Base-Band Unit \\ \hline
FP& Fractional Programming&ADMM & Alternating Direction Method of Multipliers& QoE & Quality of Experience \\ \hline
OMA & Orthogonal Multiple Access & MCC& Mission-critical communication& LEO& Low Earth Satellite\\ \hline
ANN & Attention Neural Network & RIS& Reconfigurable Intelligent Surface& HSR& High-Speed Railway \\ \hline
RAN& Radio Access Network& VANETs & Vehicular Ad Hoc Networks&PED& Patient Edge Device \\ \hline
RNC&Radio Network Controller& WSN& Wireless Sensor Network & NE& Nash Equilibrium  \\ \hline
DPG & Deterministic Policy Gradient&CUAV&Cognitive Unmanned Aerial Vehicle& XAI&Explainable AI  \\ \hline
GAN& Generative Adversarial Network&KPI& Key Performance Indicator& MCA& Multi-Channel Access \\ \hline
\end{tabular}
\end{table*}

\section{Radio Resource Allocation and Management Techniques} \label{Sec2:RRAMtechs}
In this section, we define the main radio resources of interest and provide a summary of the conventional techniques and tools used for RRAM in wireless networks. Also, the limitations of these conventional techniques that motivate the use of DRL solutions will be highlighted. Then we discuss how DRL techniques can be efficient alternatives to these traditional approaches.

\subsection{Radio Resources: Definitions and Types (or Issues)}
In general, allocation and management of wireless network resources include radio (i.e., communication) and computation resources. This paper focuses only on the RRAM issue. This involves strategies and algorithms used to control and manage wireless network parameters and resources, such as transmit power, spectrum allocation, user association/assignment, rate control, access control, etc. The main goal of wireless networks, in general, is to utilize and manage these available radio resources as efficiently as possible to provide enhanced network QoS, such as enhanced data rate, SE, EE, reliability, connectivity, and coverage while meeting users' QoS demands.

Efficient RRAM schemes can considerably enhance the system’s SE compared to the traditional techniques relying on the advanced channel and/or source coding methods. RRAM is essential in next generation HetNet as it will cover broad geographical areas with ultra-dense network (UDN) deployments. In these UDNs, a massive number of adjacent APs typically require sharing and reusing the same communication resources, such as radio frequencies and channels.

The most crucial radio resources or issues that play a fundamental role in controlling wireless networks' performance are summarized below.

\begin{itemize}
    \item \textbf{Power resource:} which is one of the most critical issues in the RRAM of modern wireless networks. Transmit power allocation in the downlink/uplink from/to network APs, such as BSs, edge servers (ESs), and gateways, is essential to guarantee a satisfactory QoS for communication links. Power control is essential from two perspectives; physical-limitations and communication links perspectives. Practically, the maximum power is limited by the capability of APs' power amplifiers or government regulation. Power control is even more critical in battery-driven user devices, as high transmission power tends to drain battery storage very quickly. Hence, it is a common practice to incorporate the limited power resource as a constraint during the design and implementation of wireless networks. On the other hand, power control is also needed to guarantee enhanced networks' QoS and user devices' QoE. For example, in large-scale and extra-dense modern wireless networks such as the mmWave and THz band systems \cite{RN236, RN234, RN122}, signal attenuation due to path and penetration losses must be accounted for during the power link budget analysis. Also, the coverage of BSs' cells and the inter-and intra-cell interference issues become crucial, which are mainly defined by the transmit power level. Hence, it is essential to develop sophisticated and fine-grained power allocation and interference management strategies to address such challenges. Moreover, many emerging wireless networks, applications, and services are of heterogeneous/homogeneous nature with extremely varying network topology, network traffic, users' QoS demands, and channel characteristics \cite{RN234}. Therefore, it becomes quite challenging to allocate the transmit power adaptively and dynamically in response to the rapid changes of physical channels, network conditions, and users' QoS demands.
    
    \item \textbf{Spectrum resource and access control:} this is also another main issue in RRAM of modern wireless networks. User devices must be allocated frequency channels to start transmitting/receiving data with acceptable SNR. Existing wireless networks, such as the sub 6 GHz, suffers from a severe bandwidth shortage which is even exacerbated with the explosive increase in the user devices \cite{forecast2019cisco}. Fortunately, the mmWave and the emerging THz bands can considerably overcome this shortcoming by providing an extra 3.25 GHz and 10-100 GHz bandwidth, respectively \cite{tataria20206g}. It is also expected that modern wireless user devices will be equipped with advanced capabilities that enable them to aggregate all these three frequency bands, i.e., the sub 6GHz, mmWave, and THz, to support future disruptive technologies and services \cite{alwarafy2021DeepRAT}. However, allocating and managing the radio channels across these three frequency bands across multi-RAN to a massive number of user devices mandate developing advanced signal processing techniques. Unfortunately, such techniques require perfect knowledge of network statistics and CSI, which is quite difficult or even impossible due to the large-scale and massive heterogeneity of modern wireless networks. Hence, it is expected that modern wireless networks will integrate DRL with advanced signal processing techniques to overcome this issue.
    
    \item \textbf{User association:} with the ever-increase in the number of future IoT smart devices and the varying QoS demands of future applications, it becomes necessary to ensure reliable network hyperconnectivity to these devices \cite{RN122, RN236}. In particular, user association defines which BS(s), wireless RAN's AP(s), or wireless edge server that each user device must connect/associate to/with at each time to guarantee its QoS demands. Taken into consideration the multi-RAN and multi-connectivity nature of modern wireless networks \cite{RN234}, it is expected that future smart devices will be equipped with SDR capabilities that enable them to support multi-association/assignment to multiple RANs simultaneously \cite{alwarafy2021DeepRAT}. Based on users' QoS demands, devices can operate in a multi-mode or a multi-homing fashion. In the multi-mode fashion, each device will be associated with a single RAN AP at a time \cite{RN282, alwarafy2021DeepRAT} similar to the traditional fashion. Whereas in the multi-homing fashion, each user device can be simultaneously associated with multiple RANs APs to aggregate their radio resources, such as radio channels and data rate. Performing such a goal, however, is also another challenging issue. Obtaining real-time information on the network statistics such as CSI, traffic load, RANs occupancy, and the number of user devices and their QoS demands require unmanageable and intolerable overhead. This will considerably degrade the performance of the system. Hence, DRL techniques can be adopted efficiently in such a scenario to dynamically learn the channel and perform autonomous user association/assignment decisions. In designing user association schemes, we commonly include constraining parameters for the assignment process, such as the available radio resources of the wireless networks, QoS requirements of users, and the quality of communication links.
    
    \item \textbf{Rate control:} often, the main objective of RRAM is to maximize the QoS of wireless networks in terms of the network's sum-rate or SE. This is typically achieved by formulating complex wireless network optimization problems and deriving their optimal solutions subjected to available network's radio resources, such as power budget and spectrum availability, while respecting the data rate demands of user devices. The optimal solutions obtained represent bound on network performance for practical RRAM algorithms.

	However, accurate solutions for such optimization problems require full knowledge of wireless channel gain, including the large-scale and small-scale fading \cite{lee2020deep}. However, obtaining such parameters in real-time is quite difficult, especially for modern wireless networks, due to their rapid increases in the underlying RANs, the number of user devices, and the type of applications. Moreover, multi-RANs data rate aggregation has also been proposed recently \cite {ciftler2021dqn, alwarafy2021DeepRAT} to support the multi-Gbps data rate requirements of the emerging data-hungry wireless applications. Hence, it becomes imperative to develop efficient schemes that enable rate aggregation while having limited knowledge of the wireless channels.  DRL techniques can be efficiently employed to achieve this goal \cite{ciftler2021dqn, alwarafy2021DeepRAT, RN282}.
\end{itemize}

\subsection{Conventional RRAM Techniques}
In this subsection, we overview the state-of-the-art approaches and tools that are used for RRAM in modern wireless systems. In general, RRAM techniques can be classified into two broad categories based on their adaptivity to the wireless environment; namely, static or dynamic approaches. Each of which can be further classified based on various criteria, such as centralized or distributed, instantaneous or ergodic, optimal or sub-optimal, single-cell or multi-cell, cooperative or noncooperative, in addition to different combinations of these variants. In this paper, we discuss the features of the static and dynamic techniques along with their types. 

RRAM has been one of the major research interests in wireless networks using conventional approaches. It has been extensively surveyed for various wireless networks and systems. Table \ref{Conventional_approaches} lists some of the existing surveys for resource allocation and management using conventional methods along with the types of wireless networks and systems they study.

\begin{table}[]
\centering
\caption{Existing Surveys on Resource Allocation and Management for Wireless Networks and Systems Using Conventional Approaches}
\label{Conventional_approaches}
\begin{tabular}{|p{1.4cm}|p{4.8cm}|}
\hline
\rowcolor[HTML]{C0C0C0} 
\multicolumn{1}{|c|}{\cellcolor[HTML]{C0C0C0}\textbf{Paper}} &
  \multicolumn{1}{c|}{\cellcolor[HTML]{C0C0C0}\textbf{Types of wireless networks and systems studied}}  \\ \hline\hline
\cite{ahmad2015survey, Tanab2017Resource, Naeem2014Resource}     & Cognitive radio networks (CRNs)  \\ \hline
\cite{Manap2020Survey, Peng2015Recent, Teng2019Resource, piamrat2011radio}& Wireless HetNets \\ \hline
\cite{Xia2018Radio}& M2M communication networks      \\ \hline
\cite{Sadr2009Radio, Yaacoub2012Survey, Afolabi2013Multicast, chieochan2009adaptive} & OFDM systems \\ \hline
\cite{Niyato2007Radio} & MIMO-OFDM systems   \\ \hline
\cite{zhao2015resource, song2014game} & D2D communication networks  \\ \hline
\cite {Kawamoto2019Toward} & UAV communications   \\ \hline
\cite{masmoudi2019survey, allouch2021survey}& Vehicular communications (V2X) \\ \hline
\cite{xu2016survey} & Railway communications  \\ \hline\hline
\cellcolor[HTML]{C0C0C0} Value added in this paper &
   Focus on the applications of DRL techniques for RRAM in next generation wireless networks, such as cellular HomNets, IoT networks, satellite networks, multi-RATs networks, HetNet, etc. 
 \\ \hline
\end{tabular}%
\end{table}

\subsubsection{Static Techniques}
Static approaches are designed based on a priori statistical information and cannot adapt to wireless network parameters, such as traffic load, users' mobility pattern, channel conditions/quality, network spectrum occupancy, and users' QoS demands. These techniques are simple; however, they suffer from several degradations, such as severe underutilization of radio resources, increased network outage, reduced network throughput, and poor network QoS.

Static RRAM techniques are employed in several traditional wireless communication networks, such as cellular communication networks and WLANs. Examples of static RRAM techniques include circuit-mode communication using frequency division multiple access (FDMA) and time division multiple access (TDMA) schemes and fixed radio resource allocation, such as fixed power and channel allocation.

\subsubsection{Dynamic Techniques}
On the contrary, dynamic or adaptive RRAM approaches are more efficient as they can dynamically adjust the network radio resources to accurately track variations in propagation conditions and user QoS requirements. 

Dynamic RRAM schemes are widely utilized in designing modern wireless systems. They have shown efficient results in reducing the expensive manual network planning and achieving tighter radio resource utilization, which will lead eventually to enhanced network SE. Some RRAMs schemes are centralized, where several BSs, ESs, wireless APs, and network gateways are controlled by a central Radio Network Controller (RNC). Others are distributed, either autonomous algorithms implemented in smart user devices, BSs, ESs, or APs, or coordinated by exchanging information among these network entities. Examples of dynamic RRAM schemes include power control algorithms, dynamic spectrum/channel allocation algorithms, multi-access control schemes, link adaptation algorithms, precoding schemes, traffic adaption algorithms, channel-dependent scheduling schemes, and cognitive radio approaches.

In dynamic RRAM, we typically formulate the wireless radio resource allocation problem as a complex optimization problem. The main objective of such a problem is maximizing/minimizing some utility/cost functions, e.g., network sum-rate, EE, and SE, while constraining the available network's radio resources such as the available power and bandwidth. The state-of-the-art approaches to solve these wireless RRAM optimization problems are heuristic-based, optimization-based, and game theory-based approaches. Such approaches employ advanced algorithms to solve the RRAM problem either optimally or sub-optimally. 

\paragraph{Heuristic-Based Techniques}
These techniques allocate radio resources in a sub-optimal fashion and without any performance guarantee. They are typically used to provide approximate and sub-optimal solutions in cases the solution of the formulated optimization problem is quite complex or intractable. Modern wireless communication systems such as 4G LTE implements some types of greedy heuristics \cite{dhilipkumar2019comparative}. Examples of heuristic algorithms include the recursive branch-and-bound state-space search algorithm \cite{clausen1999branch}, alpha-beta search algorithm \cite{nau_relation}, and particle swarm optimization (PSO). 

\paragraph{Optimization-Based Techniques}
Typically, most of the RRAM optimization problems in modern wireless networks are non-convex (e.g., continuous power allocation) \cite{nasir2020deepjoint}, combinatorial (e.g., user association and channel access) \cite{Liang2020deep}, or mixed-integer nonlinear programming (MINP) (e.g., combined of continuous- and discrete-type problems) \cite{alwarafy2021DeepRAT}. Many algorithms have been developed to systematically solve such problems and find either the global optimum \cite{Liang2020deep} solution or sub-optimal solution. Such algorithms include, fractional programming (FP) \cite{shen2018fractional, nasir2020deepjoint}, genetic \cite{kim2015efficient}, Weighted Minimum Mean Square Error (WMMSE) \cite{shen2018fractional, nasir2020deepjoint}, among others.

These algorithms are extremely computationally-extensive and typically executed in a central RNC with full and real-time information about network statistics and CSI.

\paragraph{Game Theory-Based Methods}
Game theory techniques are typically used for distributed RRAM of modern wireless networks when network entities (i.e., players) cooperate or compete on radio resources. Such techniques have shown efficient results, and they are widely used as tools to model complex wireless optimization problems in a decentralized fashion \cite{hussain2020machine}. In particular, the RRAM problem is formulated as a cooperative or non-cooperative game/optimization problem between network entities, such as BSs, RANs' APs, and user devices. In cooperative game techniques, players collaboratively solve the underlying RRAM game using heuristic- or optimization-based techniques to achieve a specific network goal such as sum-rate or SE maximization. However, in non-cooperative game techniques, players try to solve the RRAM game in a greedy and non-collaborative fashion in order to achieve their own goal (e.g., to satisfy their own QoS demands). The main goal of most game theory algorithms is to find the Nash Equilibrium (NE) solution for the underlying RRAM problem.

\subsection{Limitation of Conventional RRAM Techniques}
Unfortunately, all these state-of-the-art approaches will encounter severe limitations in future wireless networks, which mainly motivate the usage of DRL in RRAM. Here we summarize the main limitations, and the interested reader can also refer to \cite{hussain2020machine}.

\begin{itemize}
    \item Most of these approaches require complete or quasi-complete knowledge of the wireless environment, including accurate channel models and real-time channel state information (CSI). However, obtaining such accurate information in next generation wireless networks is quite difficult or even impossible due to the large-scale, ultra-dense, and massive heterogeneity nature of the system.
    \item These approaches are generally not scalable, and they encounter several challenges when the number of user devices becomes very large or when used in ultra-dense wireless networks (UDNs) with multi-cell multi-objective homogeneous/heterogeneous scenarios. The main reason is that the optimization space becomes prohibitively large to cater to the whole wireless network. This will lead to a significant increase in computational complexity when finding optimal solutions. With this large-scale and massive heterogeneity nature of future wireless networks, it becomes essential to engineer and devise more efficient and practical implementations from computation and performance perspectives. Also, it becomes even quite challenging in many scenarios to mathematically formulate RRAM optimization problems, or we may end up with non-well-defined or even intractable optimization problems. Such cases are typically encountered for many reasons, such as the uncertain nature of wireless channels, network traffic load, users' mobility patterns, etc. Hence, new innovative RRAM solutions must be developed to address such challenges. In this context, the data-driven artificial intelligence (AI)-based RRAM techniques are feasible solutions in such scenarios, and they have shown efficient adaptivity when applied on the dynamic wireless networks.
    \item Such approaches are heavily system-dependent and will not be accurate for rapidly varying systems or wireless environments. They need, however, reconfiguration to reflect the new system settings. Unfortunately, modern wireless networks need to support highly dynamic systems characterized by massive rapidities, such as vehicular and railway networks. This renders conventional approaches impractical for such scenarios.
    \item Most of these approaches are computationally expensive and incur considerable timing overhead. This renders them inefficient for most emerging time-sensitive applications, such as autonomous vehicles/drones applications. Also, the computational complexity of these approaches proportionally increases with the increase in the network size, making them unscalable and unsuitable for modern large-scale wireless networks. Furthermore, since most of the conventional algorithms are computationally expensive, they can be implemented only in sophisticated infrastructures with high computational capabilities, such as supercomputers and servers. Hence, tiny and self-powered user devices will not be able to support them.  
    \item RRAM optimization problems in wireless networks are generally complex and non-convex \cite{alwarafy2021DeepRAT}. Hence, leveraging conventional optimization algorithms to solve such problems will likely result in local optimal solutions rather than global ones. This case is regularly encountered in wireless optimization problems, which have too many local optima.
    \item Game theory-based techniques are unsuitable for networks characterized by massive heterogeneity in system architecture and user devices. In particular, NE solutions are obtained based on the assumption that all players are homogeneous, have statistically equal capabilities, and have complete network information. Unfortunately, this is not the case in modern wireless networks, in which network entities are massively heterogeneous in terms of physical, communication, and computational capabilities.
    \item Finally, the complexity of game theory-based techniques and the amount of information exchanged between competing players are proportional to the number of playing nodes. Unfortunately, future wireless networks are expected to be prohibitively large-scale in terms of the number of network APs and user devices \cite{RN122, forecast2019cisco}. Hence, such techniques will fail. In particular, exchanging and updating the tremendous amount of data and signaling among the massive number of players will create extra and unmanageable overhead as well as a drastic increase in delay, computation, and energy/memory consumption of network players.
\end{itemize}

\subsection{Advantages of Using DRL-Based Techniques for RRAM}
Emerging AI tools, such as ML, DL, and DRL methods, have been recently used to effectively address various problems and challenges in different areas of wireless communications and networking, including RRAM \cite{hussain2020machine, zappone2019wireless, chen2019artificial,naeem2020gentle, sun2018learning, xiong2019deep, Liang2020deep, RN1, lee2019survey}. Next generation wireless networks will generate a tremendous amount of data related to network statistics, such as user traffic, channel occupancy, channel quality, etc. AI algorithms can leverage this data to develop automated and fine-grained schemes to optimize network radio resources. This paper is solely dedicated to providing a comprehensive survey on DRL applications for RRAM in modern wireless networks. However, the applications of ML and DL techniques in various wireless fields can be found in \cite{hussain2020machine, Liang2020deep, chen2019artificial, zappone2019wireless} and the references therein.  

DRL is an advanced data-driven AI technique that combines neural networks (NNs) with traditional reinforcement learning (RL). It is mainly utilized to enhance the learning rate of RL algorithms and address wireless communication and networking problems having high dimensionality \cite{lee2019survey, arulkumaran2017deep, obite2021overview, zhang2019deepOverview}. DRL techniques have gained considerable fame lately to their superiority in making judicious control decisions in uncertain environments like the wireless channels. They enable various network components such as BSs, RAT APs, edge servers (ESs), gateways nodes, and user devices to make autonomous and local decisions, such as RRAM, RATs selection, caching, and offloading, that achieve the objectives of various wireless networks, including sum-rate maximization and SE/EE maximization. Since traditional approaches will not be able to address the RRAM issue of future wireless networks, DRL methods have been proposed lately to be alternative solutions. In particular, DRL techniques are appealing for next generation wireless communication networks due to the following distinct features. 

First, they enable network controllers to solve complex network optimization problems, including RRAM and other wireless control problems, with only limited information about the wireless networks. Second, DRL methods enable network entities (e.g., BSs, RAT APs, ESs, gateways nodes, and user devices) to act as agents (i.e., decision-makers) to learn and build knowledge about the wireless environment. This is achieved by learning optimal policies, such as radio resource allocation, RATs selection, and scheduling decisions, based on continuous interaction between agents and the wireless environment, without knowing the accurate channel models or statistics of the underlying systems \textit{a-priori}. DRL algorithms employ the data collected during the continuous interaction with the environment as a training data-set to train their models. Once DRL agents learned the optimal policies, they can be deployed in an online fashion to make intelligent and autonomous decisions based on local observations made on the wireless environment.

DRL techniques provide efficient solutions from both the network and user devices' points of view to overcome the problems of the conventional RRAM approaches. By employing DRL techniques, various network entities are enabled to learn wireless environments in order to optimize system configuration. Networks entities will be able to optimally and autonomously allocate the optimal transmitting power to mitigate signals interference and reduce energy consumption. For this purpose, advanced DRL techniques such as the deep deterministic policy gradient (DDPG) method and its variants can be utilized. On the other hand, DRL can also enable smart devices to autonomously access the radio channels. For this purpose, deep Q-network (DQN) and its variants can be leveraged. The wireless channels are extremely stochastic due to, e.g., the rapid mobility of user devices and channel objects. Hence, accurate and real-time knowledge of channel state information (CSI) becomes quite difficult, and DRL techniques can be efficiently used to learn wireless channel statistics. 

Finally, spectrum prediction and forecasting is also another promising field enabled by DRL techniques. Emerging DL models, such as recurrent neural networks (RNNs) and convolutional neural networks (CNNs), can be integrated with DRL to add the "prediction" capability to the DRL algorithms. Also, conventional optimization techniques do not incorporate the context, and hence they cannot adapt and react according to the sudden variations and changes in the wireless environments. Therefore, such conventional approaches will result in unreliable and poor resource management and utilization. DRL techniques can, however, dynamically adapt and learn the context of wireless environments, which makes their RRAM solutions more accurate and reliable. 

To sum up, DRL techniques are required in RRAM problems in four main scenarios; when there is insufficient knowledge about the statistics of the wireless networks, accurate mathematical models do not exist, inference information is required to be incorporated into the decision process, or a mathematical model exists, but applying conventional algorithms is not possible. In general, most of the RRAM problems in modern wireless networks fall under the above scenarios. The main reason is the large-scale and massive heterogeneity nature of networks in terms of types and numbers of underlying infrastructures, user devices, and QoS demands of applications. 

All the aforementioned unique features of DRL techniques make them one of the leading AI-based enabling technologies that can be leveraged to address the RRAM in future wireless communication networks \cite{RN122, RN234}.

\section{Overview of DRL Techniques Used for RRAM} \label{Sec3:DRL types}
\begin{figure}[htp]
    \centering
    \includegraphics[height=7.4cm, width=8.7cm]{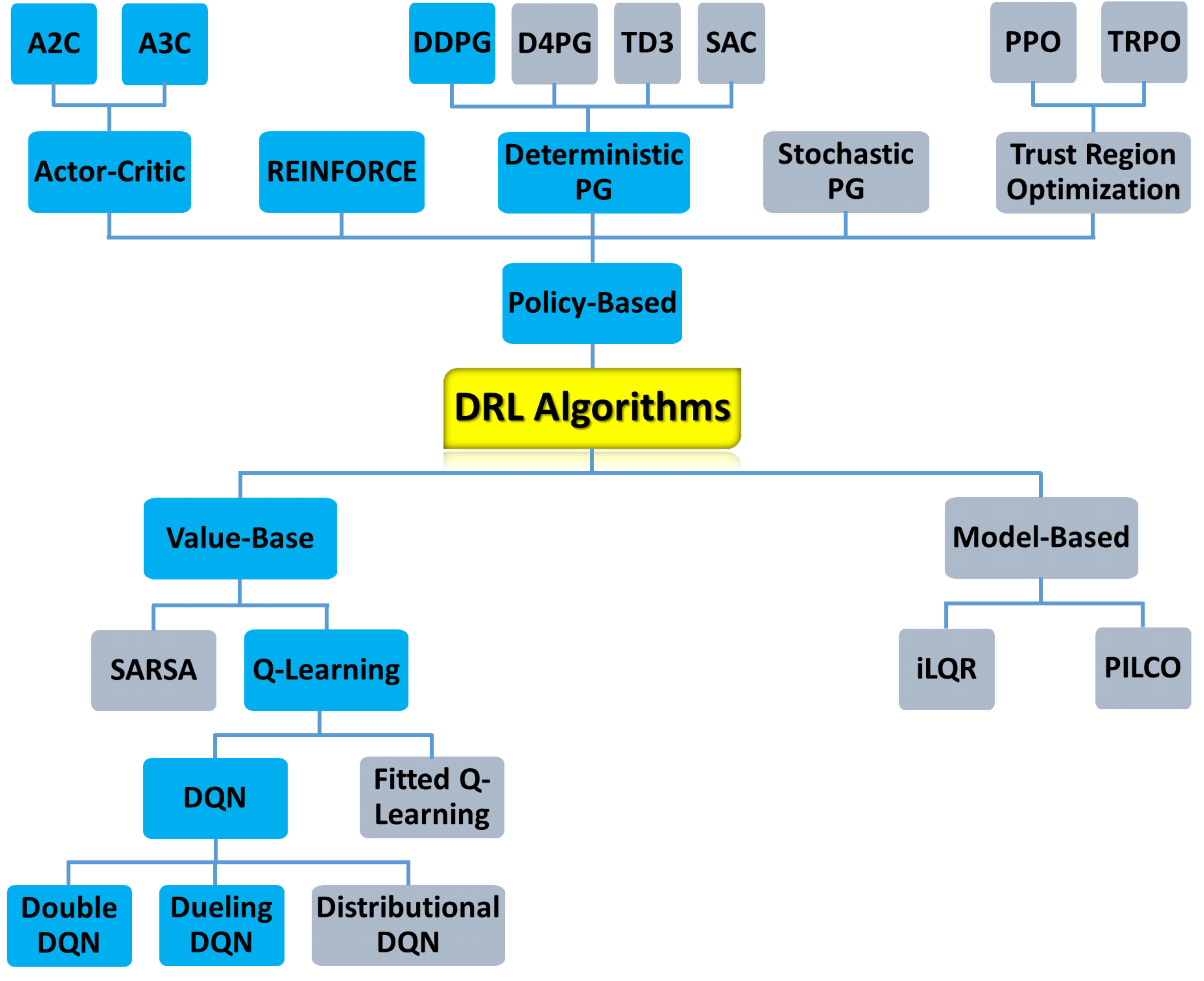}
    \caption{Taxonomy of all DRL algorithms. Algorithms colored in blue are covered in Section \ref{Sec3:DRL types}.}
    \label{DRL_Algorithms}
\end{figure}

In this section, we briefly review the foundations of DRL, such as the Markov Decision Process (MDP), and show how RRAM problems can be modeled as MDPs. Fig. \ref{DRL_Algorithms} shows a detailed taxonomy of existing DRL techniques/algorithms. Reviewing all these techniques is beyond the scope of this paper, and we rather focus on the most widely used ones in the literature to address RRAM problems. Interested readers, however, can refer to \cite{sutton2018reinforcement, RN1} for a thorough review of the remaining algorithms. Furthermore, we briefly review other emerging technologies used for RRAM problems, such as federated/distributed learning (FL) and multi-agent DRL (MARDL) models. Hence, this section is deliberately designed to provide the reader with adequate knowledge of the basics, advantages, limitations, and use-cases of the most widely used DRL techniques employed in the RRAM field.

Table \ref{Used_Algorithms} lists the most widely used DRL techniques/algorithms in RRAM of modern wireless networks. Note that all of them are model-free learning algorithms, which means that the agent does not build a model of the wireless environment or reward; instead, it directly maps states to the corresponding actions. 

Depending on the dimensionality of the RRAM problem, we can select the most appropriate DRL algorithm that fits the problem settings. For example, RRAM problems could have discrete action space, such as channel access, user association, RAN assignment, etc., or could have continuous action space, such as power allocation and continuous spectrum allocation.

\begin{table*}[htbp]
\centering
\caption{List of the Model-Free DRL Algorithms That Are Widely Used in RRAM for Modern Wireless Networks}
\label{Used_Algorithms}
\begin{tabular}{|l|c|c|c|}
\hline
\rowcolor[HTML]{C0C0C0} 
Family                         & \multicolumn{1}{c|}{\cellcolor[HTML]{C0C0C0}Algorithm} & Action Space           & Policy Type           \\ \hline
                               & $Q$-Learning     & Discrete (also discrete state space) &     \\ \cline{2-3}
                               & DQN & Discrete  &  \\ \cline{2-3}
                               & Double DQN & Discrete  &  \\ \cline{2-3}
\multirow{-3}{*}{Value-Based} & Dueling DQN                                            & Discrete & \multirow{-3}{*}{Off} \\ \hline
                               & REINFORCE & Discrete \& Continuous  & On  \\ \cline{2-4} 
                               & A2C-A3C        & Discrete \& Continuous               & On  \\ \cline{2-4} 
\multirow{-3}{*}{Policy-Based} & DDPG           & Continuous                           & Off \\ \hline
\end{tabular}%
\end{table*}

\subsection{The Markov Decision Process (MDP)}
Under the uncertain and stochastic environments of modern wireless networks, the problem of RRAM, or any decision-making problem including control problems, are typically modeled by the so-called Markov Decision Process (MDP). It provides a mathematical framework for modeling decision-making problems whose outcome is random and controlled by a decision-maker, aka agent. The MDP also has another variant, called partially observable MDP (POMDP), which models decision-making problems in partially observable wireless environments. 

The general practice in RRAM is to formulate the radio resource allocation as an optimization problem whose objective is to maximize/minimize some network utility/cost function while constraining on the available network radio resources and optional QoS demands of user devices. However, as we discussed in Section \ref{Sec2:RRAMtechs}, tremendous challenges are encountered during formulating such problems or/and even during solving them, which renders conventional approaches inapplicable. Hence, RL/DRL techniques are utilized instead. 

\begin{figure}[tb!]
    \centering
    \includegraphics[height=7cm, width=8.8cm]{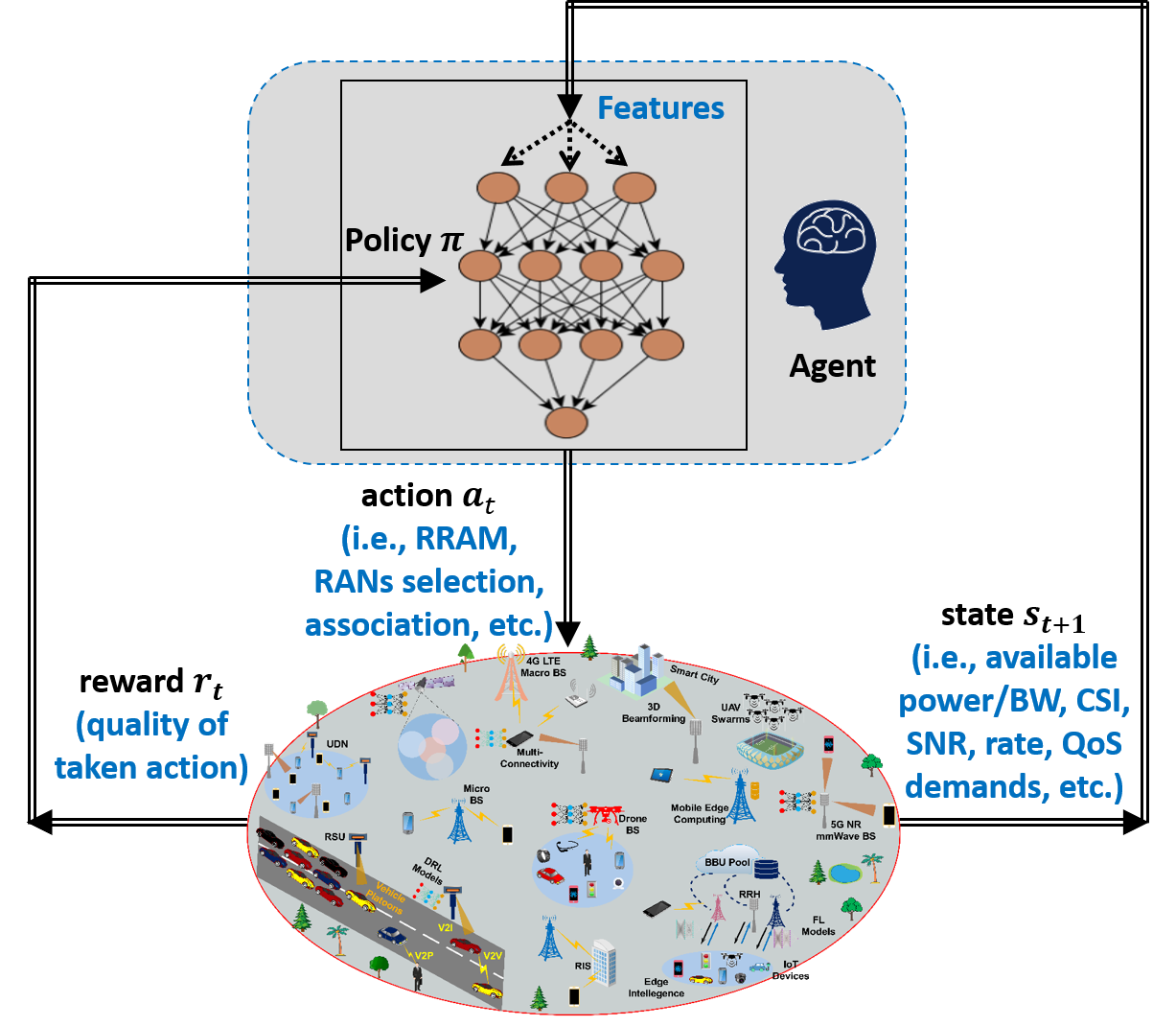}
    \caption{Framework of DRL models.}
    \label{DRL_Framework}
\end{figure}

In order to apply DRL to solve radio resource allocation (RRA) problems, we need first to convert the formulated optimization problem into the MDP framework. The resultant MDP-based model must contain seven elements: the agent(s), environment, action space $\mathcal {A}$, state space $\mathcal {S}$, instantaneous reward function $r$, a transition probability $p$, and policy $\pi$, as shown in Fig. \ref{DRL_Framework}. The MDP is typically represented mathematically by the tuple ($\mathcal {S}$, $\mathcal {A}$, $p$, $r$). 

In RRAM problems, the dynamicity of the agent's learning process according to the MDP framework is shown in Fig. \ref{DRL_Framework}, and described as follows. At time $t$, the agent observes a state $s_t$ from the state space $ \mathcal {S}$. The state space should contain useful and effective information about the wireless network environment, such as available radio resources, SNR, the number of user devices, required QoS, rate, etc. Then, the agent takes action $a_t$ from the action space $\mathcal {A}$ such as the RRA, RAN assignment, etc. The taken action must achieve network utility goal, such as sum-rate maximization, SE/EE maximization, etc. Then the state moves to a new state $s_{t+1}$ with a transition probability $p$, and the agent receives a feedback numerical instantaneous reward $r_t$ which quantifies the quality of the taken action. This interaction, i.e., $(s_t, a_t, r_t, s_{t+1})$, between the agent and the wireless environment repeatedly continues, and the agent will utilize the received instantaneous reward signal to adjust its strategy until it learns the optimal policy $\pi^*$. The agent's policy $\pi$ defines the mapping from states to the corresponding actions $\mathcal {S}$ $\leftarrow$ $\mathcal {A}$ , i.e., $a_t = \pi(s_t)$. Typically, we define the long-term reward as the expected accumulated discounted instantaneous reward over the time horizon $T$, which is given by ${\cal R} = \mathbb{E} \left[ \sum_{t=1}^{T} {\gamma r_t (s_t, \pi(s_t))} \right] $. The parameter $0\leq \gamma \leq1$ is the discounted factor, which trades-off between instantaneous and future rewards. The main goal of the agent in MDP is to obtain the optimal decision policy $\pi^*$ (i.e., selecting optimal radio resources) that maximizes the long-term reward, i.e., $\pi^{*} = \underset{\pi}{\text{max}} ~ {\cal R}$.

Next, we discuss the most widely used DRL algorithms to handle MDP problems, i.e., RRAM problems. As shown in Fig. \ref{DRL_Algorithms}, these algorithms belong to two main families of methods; namely, the value-based and the policy-based methods. 

\subsection{Value-Based Algorithms}
This family of methods is used to estimate the value function of the agent. This value function is then utilized to implicitly and greedily obtain the optimal policy. Two value functions exists; the value function $V^\pi (s)$ and the state-action function $Q(s_t, a_t)$. Both represent the expected accumulated discounted rewards received when taking action $a_t$ (in state $s_t$ for the value function) (or at pair $(s_t, a_t)$ for the state-action function) and then following the policy $\pi$ thereafter. These functions are quite important as they represent the link between the MDP mathematical formulation and the DRL formulation, and they are given by \cite{sutton2018reinforcement}:

\begin{equation*}
    V^\pi (s) = \mathbb{E} \left[ \sum_{t=0}^{\infty} \gamma^t r_t (s_t, a_t, s_{t+1}) | a_t \sim \pi(.|s_t), s_0 = s \right],
\end{equation*}

\begin{gather*}
    \begin{split}
    Q^\pi (s, a)  = \mathbb{E} \Big[ \sum_{t=0}^{\infty} \gamma^t r_t (s_t, a_t, s_{t+1}) | \\
    a_t \sim \pi(.|s_t), s_0 = s, a_0 = a \Big].
    \end{split}
\end{gather*}

The optimal value function $V^*(s)$ and state-action function $Q^*(s, a)$ are obtained by solving the following Bellman optimality equations \cite{sutton2018reinforcement, Feriani2021Single, RN1}:
\begin{equation*}
    V^* (s) = \underset{a_{t}}{\text{max}} \Big[ r_t (s_t, a_t) + \gamma \mathbb{E}_{\pi} V^*(s_{t+1})    \Big],
\end{equation*}

\begin{equation*}
    Q^* (s, a) = r_t (s_t, a_t) + \gamma \mathbb{E}_{\pi} \left[ \underset{a_{t+1}}{\text{max}} Q^*(s_{t+1}, a_{t+1})   \right].
\end{equation*}

Recall that the main goal of MDP is to obtain the optimal policy $\pi^*$ (i.e., mapping states to optimum actions), which is given by $\pi^{*} = \underset{\pi}{\text{argmax}} ~ {\cal R} = \underset{\pi}{\text{argmax}} ~ = \mathbb{E} \left[ \sum_{t=1}^{T} {\gamma r_t (s_t, \pi(s_t))} \right] $. Hence, the optimal actions can be obtained to be the ones that maximize the above value functions, and the optimal policy will be the one that maximizes these values functions \cite{sutton2018reinforcement}. In particular, the $Q$-function $Q^\pi (s, a) $ is commonly used, and the problem of obtaining the optimal policy becomes $\pi^*(s) = \underset{a}{\text{argmax}} ~ Q^{\pi^*} (s_t, a_t) $. The ultimate goal of all the value-based DRL algorithm is to approximate this function as discussed in the following subsections.
 
\subsubsection{$Q$-Learning Technique}
In RL, $Q$-learning is one of the most widely used algorithms to address MDPs. It obtains the optimal values of the $Q$-function iteratively using the following Bellman equation updating rule:  

\begin{equation*}
    \begin{split}
    Q(s_{t}, a_{t}) & = Q(s_t, a_t) + \\ 
    &  \alpha_t \left[r_t (s_t, a_t) + \gamma  \underset{a_{t+1}}{\text{max}} Q(s_{t+1}, a_{t+1}) - Q(s_t, a_t)    \right]    
    \end{split}
\end{equation*}
where $\alpha_t$ is the learning rate that defines how much the new information contributes to the existing $Q$-value. The main idea of this Bellman rule relies on finding the Temporal Difference (TD) between the current $Q$-value ($Q(s_t, a_t)$) and the predicted $Q$-value ($r_t (s_t, a_t) + \gamma  \underset{a_{t+1}}{\text{max}} Q(s_{t+1}, a_{t+1}) - Q(s_t, a_t)$). The $Q$-learning algorithm uses this Bellman equation to construct a table of all possible $Q$ values for each stat-action pair. The algorithm terminates when we reach a certain number of iterations or when all Q-values have converged. In such a case, the optimal policy will determine the optimal action to take at each state such that $Q^{\pi^*}(s_t, a_t)$ is maximized for all states in the state space, i.e., $\pi^* = \underset{a_{t+1}}{\text{argmax}} ~ Q^{\pi^*} (s_t, a_t) $. This optimal policy says that at any state we take the action that will eventually obtain the highest cumulative reward.

However, the $Q$-learning algorithm has many limitations when applied for RRAM in modern wireless networks. First, it is applicable only to problems with low dimensionality of both state and action spaces, making it unscalable. Second, it is applicable only on RRAM with discrete state space and action space, such as channel access and RANs assignment. If, however, they are applied to problems with continuous action space nature, such as power allocation, the action space must be digitized. This renders them inaccurate due to the quantization error.

\subsubsection{Deep Q Network (DQN) Technique}
\begin{figure}[tb!]
    \centering
    \includegraphics[height=10cm, width=8.8cm]{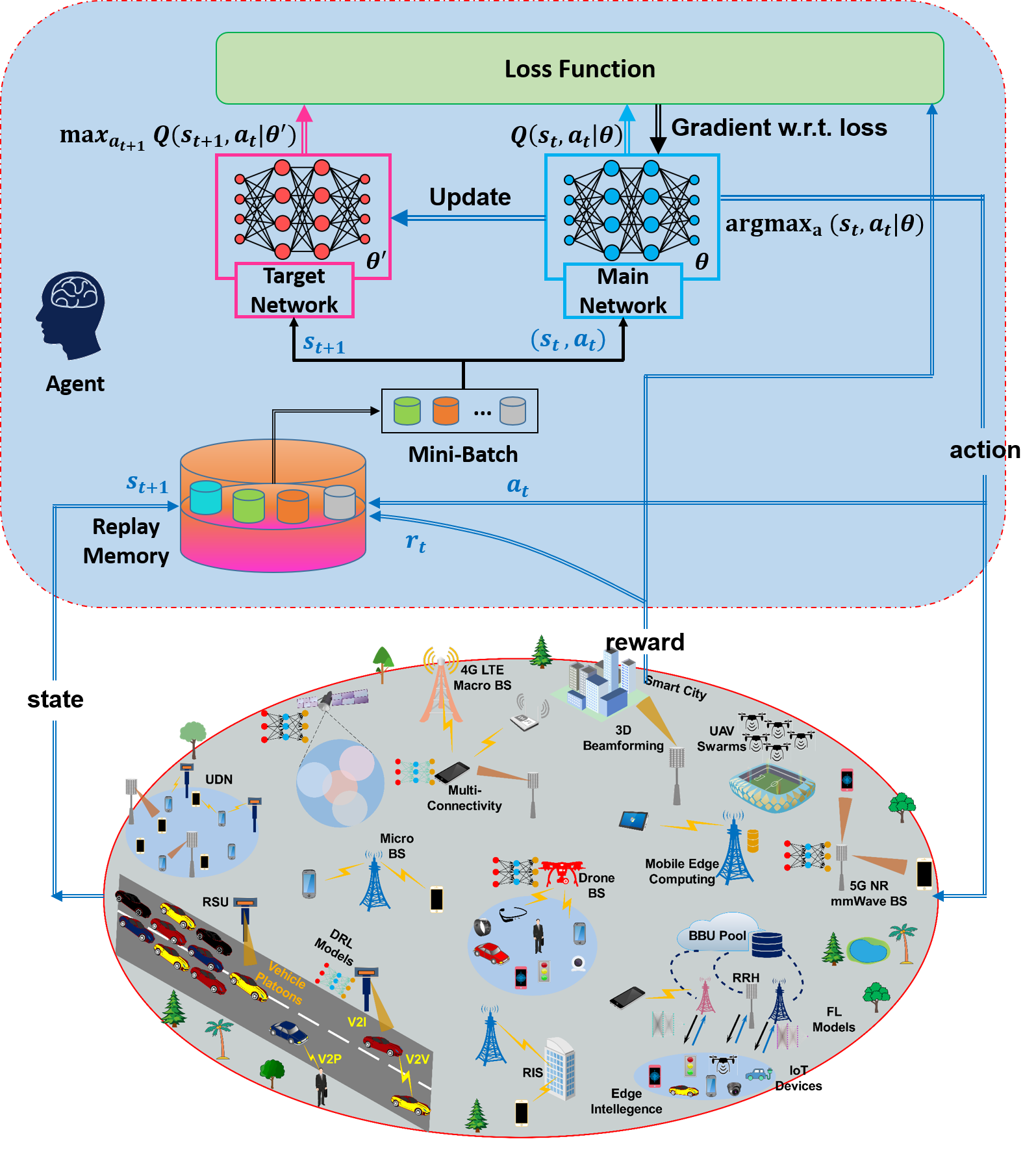}
    \caption{Illustration of the DQN architecture.}
    \label{DQN_Architcture}
\end{figure}

Since the $Q$-learning algorithm relies on building a table for the $Q$ values, it will fail to obtain the optimal policy when the state space and action space become prohibitively large. This case is commonly encountered in the RRAM problems of modern wireless systems. To overcome this issue, the DQN algorithm has been developed , which inherits the advantages of $Q$-learning and DL techniques. The main idea is to replace the table in the $Q$-learning algorithms with a DNN that tries to approximate the $Q$ values. Hence, the DNN is also called the function approximator and denoted as $Q(s_t,a_t| \theta)$, where $\theta$ represents the training parameters (i.e., weights) of the DNN. Fig. \ref{DQN_Architcture} shows the DQN architecture. The replay memory is denoted by $\mathcal{D}$, and it is mainly used to break the correlation between the training samples or transitions, i.e., ($s_t, a_t, r_t, s_{t+1}$), by making them independently and identically distributed i.i.d. During the learning process of the policy, we store the training transitions that are generated during the interaction with the wireless environment in $\mathcal{D}$. The DQN's agent will then randomly select minibatch samples of transitions from $\mathcal{D}$ to train its DNN. To enhance the stability of the DQN model, the target $Q$ network is used, whose weights will be periodically updated to track those of the main $Q$ network. 

Since the DQN algorithm is mainly used to learn the optimal policy, i.e., $\pi^* = \underset{a}{\text{argmax}} ~ Q^{\pi^*} (s_t, a_t) $, the optimal $Q$-function is derived from the following iterative Bellman equation: 

\begin{equation*}
    Q(s_{t}, a_{t}) = r_t (s_t, a_t) + \gamma  \underset{a_{t+1}}{\text{max}} Q(s_{t+1}, a_{t}),
\end{equation*}
and the DQN algorithm is then optimized by iteratively updating the weights $\theta$ of its DNN to minimize the following Bellman loss function;

\begin{equation*}
    \begin{split}
    L(\theta_t) & = \mathbb{E}_{{s_t, a_t, r_t, s_{t+1}} \in \mathcal{D}} \Big[ r_t (s_t, a_t) + \\
    & \gamma  \underset{a_{t+1}}{\text{max}} Q(s_{t+1}, a_{t}| \theta^{'}) - Q(s_{t}, a_{t} | \theta) \Big]^2  .
    \end{split}
\end{equation*}
where $\theta^{'}$ is the weights of the target $Q$ network.

The DQN algorithm is applicable to a wide variety of RRAM problems, specifically for problems characterized by their discrete action space. As we will elaborate in-depth in Section \ref{Sec4:heart}, the DQN technique can be used efficiently for channel allocation, access control, spectrum access, user association, and RANs assignment. The DQN algorithm can also be used for RRAM problems with continuous action space, such as power control, by discretizing the action space. However, such a methodology makes DQN vulnerable to serious quantization error that may considerably deteriorate its accuracy. There are also other limitations in the basic DQN, and various $Q$-learning algorithms have been proposed to overcome them, as we discuss in the following sections.

\subsubsection{Double DQN Algorithm}
The Double DQN technique has been proposed in \cite{van2016deep} to enhance the basic DQN algorithm.  The DQN algorithm tends to overestimate the $Q$ values, which can degrade the training process and lead to suboptimal policies. The overestimation results from the positive bias caused by the max operation employed in the Bellman equation. Specifically, the root cause is that the same training transitions are utilized in selecting and evaluating an action. As a solution to this problem, the authors in \cite{van2016deep} propose to use two $Q$ value functions, one for selecting the best action and the other to evaluate the best action. The action selection is still based on the online weights parameters $\theta$, while the second weights parameters $\theta^{'}$  are used to evaluate the value of this policy. So, as in the conventional $Q$ learning, the value of the policy is still estimated based on the current $Q$ values. The weights parameters $\theta^{'}$ are updated via switching between $\theta$ and $\theta^{'}$. 

Hence, the target $Q$ values are derived from the following modified Bellman equation \cite{van2016deep}: 
\begin{equation*}
    Q(s_{t}, a_{t}) = r_t (s_t, a_t) + \gamma Q(s_{t+1}, \underset{a_{t+1}}{\text{argmax}} Q(s_{t+1}, a_{t} | \theta_t), \theta_{t}^{'}) ,
\end{equation*}
and the Double DQN algorithm uses the following modified Bellman loss function to update its weights;

\begin{equation*}
    \begin{split}
    L(\theta_t) & = \mathbb{E}_{{s_t, a_t, r_t, s_{t+1}} \in \mathcal{D}} [ r_t (s_t, a_t) + \\
    & \gamma Q(s_{t+1}, \underset{a_{t+1}}{\text{argmax}} Q(s_{t+1}, a_{t} | \theta_t), \theta_{t}^{'}) - Q(s_{t}, a_{t} | \theta_t)]^2 .
    \end{split}
\end{equation*}
 
The Double DQN algorithm is also widely used in RRAM problems, as we will discuss in the next section. Although this algorithm has advantages over the basic DQN algorithm, they both share the same shortcomings.

\subsubsection{Dueling DQN Algorithm}
This algorithm is another enhancement to the basic DQN proposed in \cite{wang2016dueling}. Recall that the goal of the network is to estimate the $Q$ values, i.e., $Q(s_t, a_t)$. This function can be divided into two terms; the state-value function $V(s)$, which tells the importance of being in a particular state, and the action-value function (or the advantage function) $A(s, a)$, which tells the importance of selecting a particular action among all available actions. Hence, the $Q$ value function can be written as $Q(s, a) = V(s) + A(s, a)$. The authors in \cite{wang2016dueling} utilized this concept and suggested having two main independent paths of fully-connected layers instead of having only a single path as the case in the basic DQN. One path will estimate $V(s)$, and the other path will estimate $A(s, a)$. The two paths will eventually be combined to produce one single output, which is $Q(s, a)$ as \cite{wang2016dueling}:
\begin{equation*}
    \begin{split}
    Q(s_t,a_t; \theta, \alpha, \beta) & = V (s_t; \theta, \beta) + \\
    & \left( A(s_t, a_t; \theta, \alpha) - \frac{1}{\mathcal{A}} \sum_{a_{t+1}} {A(s_t, a_{t+1}; \theta, \alpha)}     \right)
    \end{split}
\end{equation*}
where $\mathcal{A}$ is the number of actions in the action space, and  $\beta$ and $\alpha$ are the weights of the fully-connected layers of the two paths $V (s_t; \theta, \beta)$ and $A(s_t, a_t; \theta, \alpha)$, respectively. Here, the loss function is obtained similar to the DQN and Double DQN algorithms.

\subsection{Policy-Based Algorithm}
The policy-based techniques are part of the policy gradient family of methods. They provide an alternative way to solve the MDP problems having high dimensionality and continuous action spaces. Recall that the main idea of the value-based methods discussed in the previous subsection is the state-action value function $Q(s, a)$. This function is defined as the expected total discounted reward received by taking a particular action from the state. If these $Q$ values are known, the optimal policy is defined by selecting actions that maximize the $Q$ values in each state. However, in environments with continuous action spaces, such as power control in wireless systems, the $Q$ function cannot be obtained as it is impossible to conduct a full search in a continuous action space to obtain the optimal action. Hence value-based approaches inapplicable to problems characterized by their continuous action space, and the policy-based methods are applied instead.

In policy-based approaches \cite{sutton2000policy, sutton2018reinforcement}, we avoid calculating $Q$ values and directly obtain the optimal policy $\pi_{\theta}(a|s)$ that maximizes the agent's expected accumulated reward $J$, i.e.,  $J(\theta) = \mathbb{E}_{\pi_\theta} \left[ \sum_{t=0}^{\infty} \gamma^t r_t (s_t, a_t) \right]$. The policy gradient approaches learn the optimal network weights $\theta^*$ via performing gradient ascent on the function $J$. In particular, the policy gradients are derived from trajectories obtained via the current policy, and they are given by \cite{sutton2000policy}:  
\begin{equation*}
    \begin{split}
    \triangledown_{\theta} J(\theta) = \mathbb{E}_{\pi_\theta} \left[ \sum_{t=0}^{T} {\triangledown \log \pi_\theta (a_t|s_t) Q^{\pi_\theta} (s_t, a_t)} \right].
    \end{split}
\end{equation*}
In each gradient update, the agent interacts with the environment to collect new and fresh trajectories, and this is why policy-gradient methods are called on-policy algorithms. In this formula, the function $Q^{\pi_\theta} (s_t, a_t)$ is unknown, and some of the algorithms used to estimate it are as follows:

\subsubsection{REINFORCE Algorithm}
The main idea of this algorithm is to increase the probabilities of good actions and reduce the probabilities of bad actions. In general, the REINFORCE algorithm differs from the $Q$ learning methods in three core aspects. First, REINFORCE algorithm does not need a replay buffer $\mathcal{D}$ during training as it belongs to the on-policy family, which requires only fresh training transitions. Although this enhances its convergence speed, it needs much more interaction with the environment. Second, the REINFORCE algorithm implicitly performs the exploration process, as it depends on the probabilities returned by the network, which incorporate uniform random agent behavior. Third, no target network is required in the REINFORCE algorithm as the $Q$ values are obtained from the experiences in the environment. 

The weights $\theta$ of the network in the REINFORCE algorithm are updated to minimize the following loss function:
\begin{equation*}
    \begin{split}
    L = - \sum_{t=0}^{T} {Q^{\pi_\theta} (s_t, a_t) \log \pi_\theta (a_t|s_t)}.
    \end{split}
\end{equation*}

The disadvantage of the REINFORCE algorithm is that it suffers from high variance, meaning that any small shift in the return leads to a different policy. This limitation motivated the actor-critic algorithms.

\subsubsection{Actor-Critic Algorithm}
The actor-critic methods are mainly developed to enhance the convergence speed and stability (i.e., reducing the variance) of the policy-gradient method. Like the policy-based methods, it utilizes the accumulated discounted reward $J$ to obtain the gradient of the policy $\triangledown J$, which provides the direction that enhances the policy. This algorithm learns a critic to reduce the variance of gradient estimates since it utilizes various samples, whereas the REINFORCE algorithm utilizes only a single sample trajectory.

To select the best action in any state, the total discount reward of the action is used, i.e., $Q(s, a)$. The total reward can be decomposed into state-value function $V(s)$ and advantage function $A(s, a)$, i.e., as $Q(s, a) = V(s) + A(s, a)$. So, another DNN is utilized to estimate $V(s)$, which is trained based on the Bellman equation. The estimated $V(s)$ is then leveraged to obtain the policy gradient and update the policy network such that the probabilities of actions with good advantage values are increased. Hence, the \textit{actor} is the policy network $\pi(a|s)$ that takes actions by returning the probability distribution of actions, while the \textit{critic} network evaluates the quality of the taken actions, $V(s)$. This algorithm is also called the advantage actor-critic method (A2C).

In the A2C algorithm, the weights of the actor network $\theta_\pi$ and critic network $\theta_{\text{v}}$ are updated using the accumulated policy gradients $\partial \theta_\pi$ and value gradients $\partial \theta_{\text{v}}$, respectively, according to the following formulas \cite{mnih2016asynchronous}: 
\begin{equation*}
    \begin{split}
\partial \theta_\pi \leftarrow \partial \theta_\pi + \triangledown_\theta \log \pi_\theta(a_t|s_t) (R - V_\theta (s_t)), \\
\partial \theta_{\text{v}} \leftarrow \partial \theta_{\text{v}} +
\frac{\partial \left(R - V_\theta (s_t) \right)^2 }{\partial \theta_{\text{v}}}.
    \end{split}
\end{equation*}
where $ R \leftarrow r_t(s_t, a_t) + R$. The weights are updated to move in the direction of the policy gradients and in the opposite direction of the value gradients. 

\subsubsection{A3C Algorithm}
The asynchronous advantage actor-critic (A3C) algorithm is an extension of the basic A2C \cite{mnih2016asynchronous}. This algorithm is used to solve the high variance issue in gradients that results in non-optimal policies. A3C algorithm conducts a parallel implementation of the actor-critic algorithm, where the actor and critic share the network layers. A global NN is trained to output action probabilities and an estimate of the advantage function $A(s_t, a_t|\theta_\pi, \theta_{\text{v}})$ given by $\sum _{i=0}^{k-1} \gamma^i r_{t+1} + \gamma^{k} V(s_{t+k}|\theta_{\text{v}}) - V(s_t|\theta_{\text{v}})$, where $k$ depends on the state and upper-bounded by the maximum number of time steps. The update rule in A3C is given by \cite{mnih2016asynchronous}:
\begin{equation*}
    \begin{split}
\triangledown_{\theta_{\pi}^{'}} \log \pi (a_t|s_t;\theta^{'}) A(s_t, a_t|\theta_\pi, \theta_{\text{v}}).
    \end{split}
\end{equation*}
where $\theta_{\pi}^{'}$ is thread-specific weights. 

Several parallel actor learners are instantiated with copies of both the environment and global NN weights. Each learner independently interacts with its environment and gathers training transitions to derive the gradients with respect to its NN weights. Learners will then propagate their gradients to the global NN to update its weights. This mechanism ensures a periodic update of the global model with diverse transitions from each learner. 

\subsubsection{Deep Deterministic Policy Gradient (DDPG) Algorithm}
DDPG is one of the most widely used DRL techniques in addressing RRAM problems for wireless networks characterized by their high dimensionality and continuous action space \cite{lillicrap2015continuous}. DDPG algorithm belongs to the actor-critic family, and it combines both $Q$-learning and policy gradients algorithms. It consists of actor and critic networks. The actor network takes the state as its input, and it outputs the exact "deterministic" action, not probability distribution over actions as in the actor-critic algorithm. Whereas the critic is a $Q$-value network that takes both the state and action as inputs, and it outputs the $Q$-value as a single output. 

The deterministic policy gradient (DPG) algorithm is proposed in \cite{pmlr-v32-silver14} to overcome  the limitation caused by the $max$ operator in the $Q$-learning algorithm, i.e., $\underset{a_{t+1}}{\text{max}} Q(s_{t+1}, a_{t})$. It simultaneously learns both the $Q$-function and the policy. In particular, the DPG algorithm has a parameterized actor function $\mu(s| \theta^\mu)$ with weights $\theta$, which learns the deterministic policy that gives the optimal action corresponding to $\underset{a_{t+1}}{\text{max}} Q(s_{t+1}, a_{t})$. The critic $Q(s,a)$ is leaned via minimizing the Bellman loss function as in the $Q$-learning algorithm. 

The learning process of the actor policy is updated using gradient ascent with respect to $\theta^\mu$ in order to solve the objective given by the following chain rule \cite{pmlr-v32-silver14}: 
\begin{gather*}
J(\theta) = \mathbb{E}_{s \in \mathcal{D}} \Big[Q(s, \mu(s| \theta^\mu))    \Big], \\
\triangledown_{\theta^{\mu}} J = \mathbb{E}_{s \in \mathcal{D}} \Big[ \triangledown_a Q (s, a|\theta^Q)|_{s=s_t, a=\mu(s_t)} \triangledown_{\theta^\mu} \mu(s|\theta^\mu)|_{s=s_t}    \Big].
\end{gather*}

\begin{figure}[tb!]
    \centering
    \includegraphics[height=10cm, width=8.8cm]{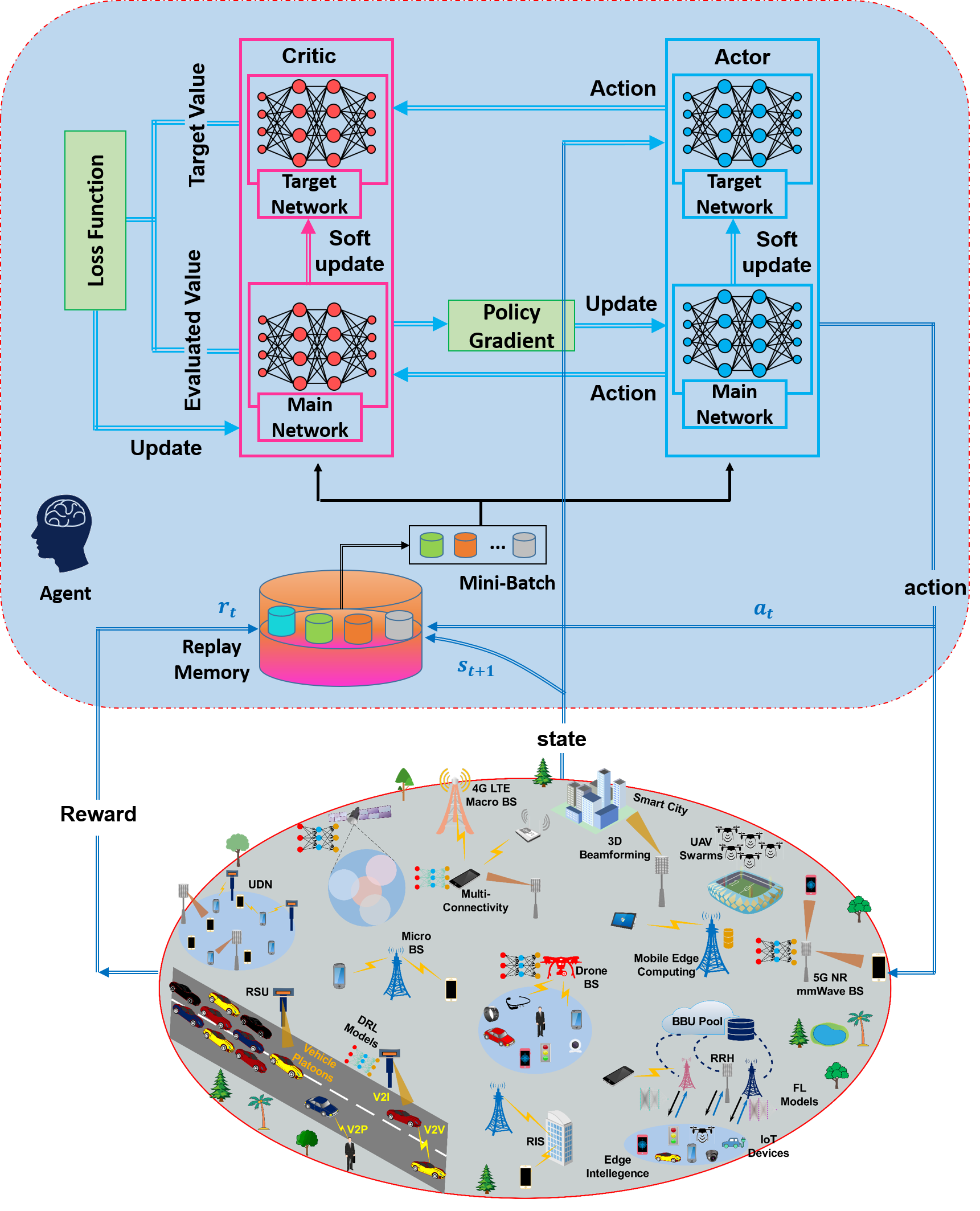}
    \caption{Illustration of the DDPG actor-critic architecture.}
    \label{DDPG_Architecture}
\end{figure}

The DDPG algorithm proposed in \cite {lillicrap2015continuous} is built based on the DPG algorithm, where both the policy and critic are DNNs, as shown in Fig. \ref{DDPG_Architecture}. The DDPG algorithm creates a copy of both the actor and critic networks, $Q^{'}(s, a|\theta^{Q^{'}})$ and $\mu^{'}(s|\theta^{\mu^{'}})$, respectively, to compute the target values. The weights of these two target networks, $\theta^{Q^{'}}$ and $\theta^{\mu^{'}}$, are then updated to slowly track the weight of the learned network to provide more stable training using $\theta^{'} \leftarrow \tau \theta + (1-\tau) \theta^{'} $ with $\tau \ll 1 $. The critic network is updated to minimize the following Bellman loss function \cite {lillicrap2015continuous};
\begin{equation*}
    \begin{split}
L(\theta^{Q}) & = \mathbb{E}_{s_t, a_t, r_t, s_{t+1} \in \mathcal{D}}  
\Bigg[  \bigg(   r_t(s_t,a_t) + \\ 
& \gamma \underset{a_{t+1}}{\text{max}}    Q(s_{t+1}, \mu(s_{t+1}|\theta^{\pi^{'}})|\theta^{Q^{'}}) - 
Q(s_t, a_{t}|\theta^{Q})  \bigg)^{2}   \Bigg]  .
    \end{split}
\end{equation*}
Note that the DDPG algorithm is off-policy, which means that we use a replay buffer $\mathcal{D}$ to store training transitions.

The exploration-exploitation issue is addressed by adding the Ornstein–Uhlenbeck (OU) process or some Gaussian noise $\mathcal{N}$ to the action selected by the policy, i.e., $\mu(s_t|\theta_{t}^{\mu}) + \varepsilon \mathcal{N}$ \cite{lillicrap2015continuous}.

\subsection{Other DRL Algorithms}
The DRL algorithms discussed above are the commonly used approaches to address the problem of RRAM in wireless networks, as we will discuss in the next section. Although there are several other algorithms, they are rarely utilized for such types of problems. Therefore, they are not included in this article. However, generally speaking, all the other variants are mainly developed to enhance the performance of the basic algorithms discussed above. For completeness, this section highlights some of these variants for the interested reader.

Other variants of the value-based algorithms are developed to enhance the performance of vanilla DQN algorithm in terms of stability, convergence speed, implementation complexity, sample/learning efficiency, etc. Such variants include prioritized experience replay DQN \cite{schaul2015prioritized},  distributed prioritized experience replay DQN \cite{horgan2018distributed}, distributional DQN \cite{bellemare2017distributional},  Rainbow DQN \cite{hessel2018rainbow}, and recurrent DQN \cite{hausknecht2015deep}.

For the policy-based algorithms, several algorithms are envisioned to enhance the overestimation issue, such as the Twin Delayed DDPG (TD3) \cite{fujimoto2018addressing}, enhance stability and robustness, such as the Soft Actor-Critic (SAC) \cite{haarnoja2018soft}, and to enhance stability, convergence, and sample efficiency, such as the distributed distributional DDPG (D4PG) \cite{barth2018distributed}. 

\subsection{Multi-Agent DRL Algorithms}
Multi-agent DRL (MADRL) is a natural generalization of the single-agent DRL that allows multiple agents to concurrently learn optimal policies based on their interactions with the environment and with each other. These agents can either be deployed cooperatively, in which all agents interact with each other to learn the same global policy, or non-cooperatively, in which each agent learns its own policy. MADRL provides several performance advantages over the single-agent case regarding the quality of the learned policies, convergence speed, etc. However, it encounters several challenges such as scalability, partial observability, and agents' non-stationarity. Nguyen \textit{et al.} \cite{nguyen2020deep} provide a survey on MADRL systems and their applications. Different methods are reviewed along with their advantages and disadvantages. In \cite{zhang2019multi}, the authors provide a selective overview of the theories and algorithms for MARL. 

MADRL is widely employed in addressing various problems in wireless networks. The authors in \cite {Feriani2021Single} provide an overview of the MADRL algorithms and highlight their applications in future wireless networks. The learning frameworks in MADRL are also investigated. The application of MARL in solving problems for vehicular networks is studied in \cite{althamary2019survey}. In \cite{lee2020optimization}, an overview of the evolution of cooperative MARL algorithms is presented with an emphasis on distributed optimization. 

Most of the RRAM problems in modern wireless networks are of a multi-agent nature \cite{Feriani2021Single}. Network entities such as user devices, BSs, and wireless APs can act as cooperative/non-cooperative multi-agents to learn optimal radio resource allocation policies and solve complex network optimization problems. For example, channel access control may be formulated as a MADRL problem in which each user device represents a learning agent that senses the radio channels and coordinates with other agents to avoid collisions.

In the following sections, we discuss how RRAM problems in wireless communication networks are formulated and solved using these algorithms.

\section{DRL-Based Radio Resource Allocation and Management for Next Generation Wireless Networks} \label{Sec4:heart}
This section provides an extensive and in-depth review of the related works for RRAM using DRL techniques. We classify them based on the radio resources (or issues) they investigate as well as based on the wireless network types they cover, as shown in Figs. \ref{ClassificationOnResources} and \ref{ClassificationOnNetwoksTypes}, respectively. It must be noted that this survey is dedicated to only the application of DRL algorithms for radio resources, i.e., no computation resources are covered, which can be found in \cite{RN1}.

DRL algorithms enable various network entities to efficiently learn the wireless networks, which allows them to make optimal control decisions that achieve some network utility function. For example, DRL methods can be deployed to maximize network sum-rate, minimize network energy consumption, or enhance spectral efficiency. In this section, we review the applications of DRL methods in the following RRAM issues: power allocation, spectrum allocation and access control, rate control, and the joint use of these radio resources.

\subsection{DRL for Power Allocation} 
\begin{figure}[tb!]
    \centering
    \includegraphics[height=9cm, width=8.8cm]{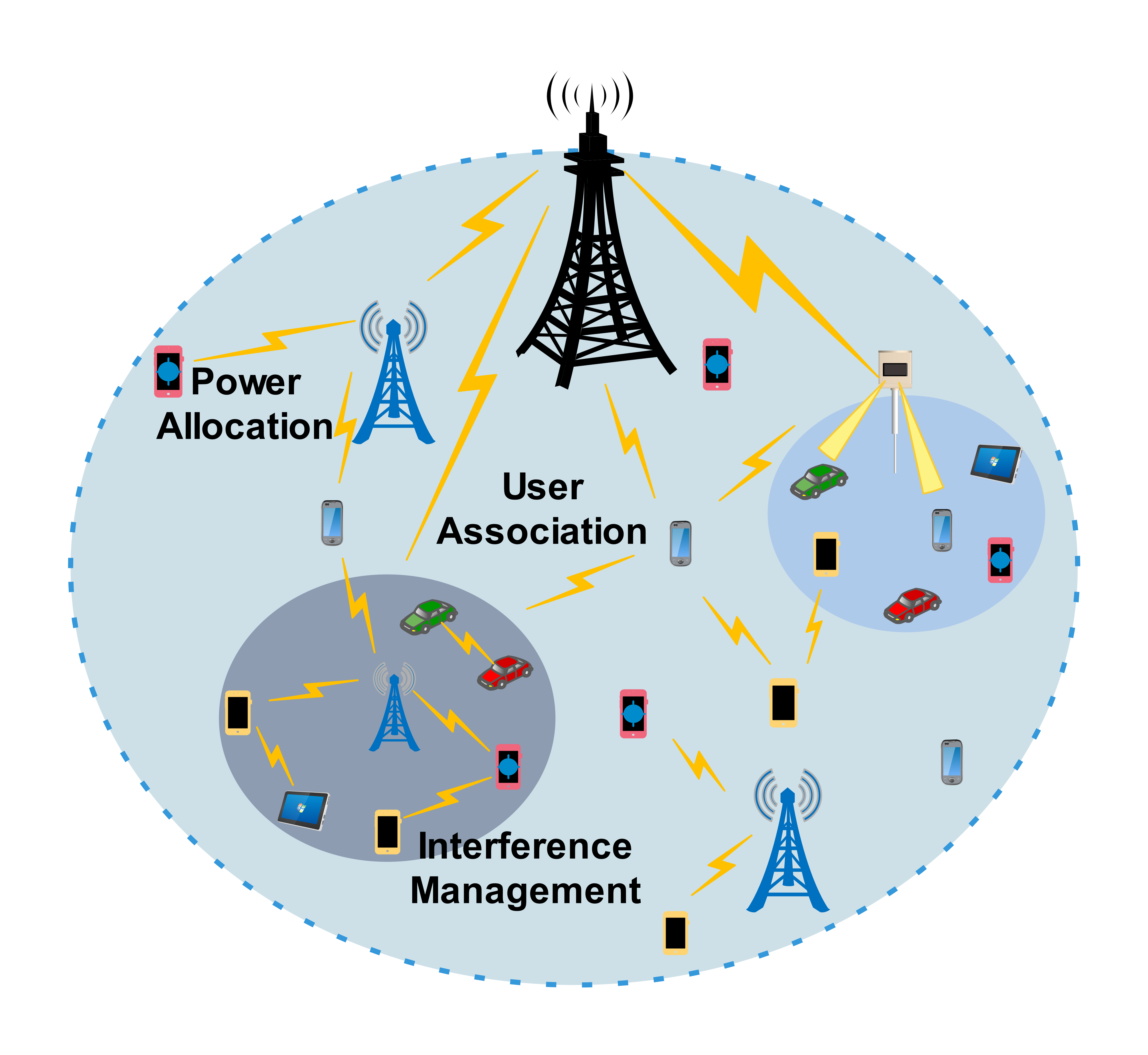}
    \caption{Importance of power allocation in modern wireless communication networks.}
    \label{Power_Allocation}
\end{figure}

\
Energy-efficient communication is one of the main objectives of modern wireless networks. It is achieved via efficient power allocation to ensure high QoS, better coverage, and enhanced data rate, as shown in Fig. \ref{Power_Allocation}. Power allocation is mainly involved in vital network operations such as modulation and coding schemes, path loss compensation, interference management, etc. On the other hand, almost all modern user devices and IoT sensors are battery-powered with very limited battery capacity and charging capabilities. Hence, designing energy-efficient resource allocation schemes, protocols, and algorithms becomes fundamental in dynamic wireless network environments.

Several conventional approaches have been applied for power allocation and management. Most of them rely on solving power-constrained optimization problems, such as FP algorithm \cite{shen2018fractional} and WMMSE algorithm \cite{shi2011iteratively}. These approaches are iterative and model-driven, which means that they need a mathematically tractable and accurate model. They are typically executed in a centralized fashion in which a network controller has full CSI. In such a mechanism, BSs, wireless APs, and/or user devices require to wait until the centralized controller's iterations converge and send the outcome back over backhaul links. However, as discussed in Section \ref{Sec2:RRAMtechs}, such approaches become impractical due to the large-scale nature of modern wireless networks and the difficulty in obtaining accurate and instantaneous CSI. Hence, DRL techniques are used instead due to their superiority in obtaining optimal power allocation policies based on limited CSI.

\subsubsection{In Cellular and HomNets} \
In the following paragraphs, we review works that employ DRL algorithms to address the power allocation problem in cellular, cellular IoT, and wireless homogeneous networks (HomNets) depicted in Fig. \ref{ClassificationOnResources}. 

In \cite{ghadimi2017reinforcement}, the authors propose a multi-agent $Q$-learning-based RL model to address the problem of downlink power control and rate adaptation in cellular systems. The agent is a network entity that resides in the cell, such as an eNB. The action space is discrete, corresponding to allocating downlink transmit power of the cell. The state space comprises four elements; cell power, average reference signal received power, average interference, and cell reward. The reward function is continuous, defined based on the $\alpha$-fair resource allocation utility function. It is defined by the radio measurement or performance indicator associated with all users in the radio cell. System-level simulations show that their proposed method quickly learns the power control policy to provide significant energy savings and fairness across the network users.

The authors in \cite {hu2020deep} propose a single-agent DQN method to address the problem of power resource use in fuel cell powered data center power grids. In the model, the agent is the data center architecture whose state space is discrete and comprised of the following elements; 1) the number of data centers in the previous section at the current and next time slots, 2) the action at the current time slot, 3) and the reward function at the current time slot. The action space is choosing the data center, and the reward is a function of the sum of all the data centers’ power demands in each time slot. Their approach is compared to the state-of-the-art approaches and evaluated based on real-world traces, which shows that it reduces the energy gap and saves more energy at around 5\%.

In \cite{khan2020centralized}, the authors propose single- and multi-agent actor-critic DRL methods to tackle the problem of downlink sum-rate maximization through power allocation in multi-cell, multi-user cellular networks. In their model, the agents are the base stations (BSs), whose state space is continuous and comprises network CSI and the transmit power allocation by previous BSs. The action space is continuous, which represents the power allocation, while the reward function is the cellular network sum SE. Experimental results demonstrate that their proposed DRL-based method can both achieve higher SE than conventional optimization algorithms, such as fractional programming (FP) and weighted minimum mean-squared error (WMMSE) while performing two times faster than these conventional methods. 

The work in \cite{nasir2020deep} proposes a distributive multi-agent DDPG-based DRL algorithm to address the problem of sum-rate maximization via continuous power control in wireless mobile networks. The agents are each transmitter (e.g., mobile devices, links, etc.) whose state is a combination of three feature groups; the local information, interfering neighbors, and interfered neighbors feature groups. Each agent's action is choosing the transmit power level, while the reward is a function of the sum-rate maximization problem. Simulation results show that their proposed method gives better performance results than the conventional FP methods and comparable results with the WMMSE methods.

D2D communication has emerged as one of the main enabling technologies for modern wireless networks. In \cite{bi2020deep}, the authors present a centralized multi-agent DQN-based DRL algorithm to address the problem of power allocation of D2D cellular communication in a time-varying environment. The agents are the D2D transmitters, whose state space is continuous, comprised of the SINR and channel gain of users. The action space is discrete, representing the transmit power of each D2D user, while the reward is a function of system throughput. Simulation results show that their proposed algorithm outperforms the traditional RL methods in terms of network capacity and user's achieved QoS.

5G UDNs are characterized by their high vulnerability to inter-cell interference, which can be greatly reduced via judicious power management. Towards this, Saeidian \textit{et al.} \cite{saeidian2020downlink} propose a data-driven approach based on a multi-agent DQN algorithm to tackle the downlink power control in dense 5G cellular networks. The agents are the BSs, whose state space is continuous, comprised of path-gain, SINR, downlink rate, and downlink power. The action space is discrete, representing the downlink power, while the reward is a function of the network-wide harmonic-mean of throughput. Simulation results indicate that their approach can improve data rates at the cell edge while ensuring a reduced transmitted power compared to the baseline fixed power allocation approaches.

Non-orthogonal multiple access (NOMA) technology has recently emerged as an efficient tool to enhance the QoS and EE of millimeter-wave (mmWave) communication systems via enhancing the power level of received signals. In this context, the authors in \cite{zhang2020deep} propose a multi-agent DQN-based DRL framework to optimize the EE in downlink full-duplex cooperative NOMA of mmWave UDNs. The agents are the relay near users, whose state space is continuous, consisting of information related to wireless environment and channel, the user's battery capacity, energy power transfer coefficient, self-interference cancellation residue coefficient, and the buffer size of nearby relay users. The action space is to specify the required user pairing between the near relay user group and edge user group, along with the pre-processing of EE power allocation. The reward is a function of the EE of the mmWave network. Experimental results are compared with a conventional centralized iteration algorithm, which demonstrate both the superiority of their proposed algorithm in terms of the convergence speed and the efficiency to provide near-optimal results.

Due to the importance of the power allocation in multi-user cellular networks, the authors in \cite{meng2020power} address this issue by building on their initial investigation in \cite{meng2019power}. A multi-agent DQN-based DRL algorithm is proposed in which each BS-user link is considered as an agent. The state space is continuous, comprised of a logarithmic normalized interferer, the link's corresponding downlink rate, and the transmitting power. The action space is discrete, corresponding to the downlink power allocation, while the reward is continuous, which is a function of the downlink data rate of the communication link. Experimental results indicate that their proposed DQN outperforms benchmark algorithms such as FP, WMMSE, random power allocation, and maximum power allocation in terms of achievable averaged sum-rate and the convergence time when considering different user densities.

In another interesting work in \cite{zhang2020deeppower}, the authors design a multi-agent DQN and DDPG-based DRL framework to address the problem of power control in HetNets. A centralized-training-distributed-execution algorithm is designed in which the APs are the agents, each of which implements a local DNN. The state space of each local DNN is continuous, representing the local state information, while the local action space is continuous, representing the transmit power. Then, multiple-actor-shared-critic method (MASC) is proposed to separately train each of these local DNN in an online fashion. The main idea is that the MASC training method is composed of multiple actor DNNs and a shared critic DNN. An actor DNN is first established in the core network for each local DNN, and the structure of each actor DNN is the same as the corresponding local DNN. Then, a shared critic DNN is established in the core network for these actor DNNs. Historical global information is provided into the critic DNN, and the output of the critic DNN will evaluate whether the output power of each actor DNN is optimal or not from a global view. The reward function is continuous, representing the data rate between each AP and its associated user. Simulation results show that their proposed algorithm outperforms the WMMSE and FP algorithms in terms of both convergence rate and computational complexity.

Managing radio resources in virtualized networks is attracting more attention lately. In this context, the authors in \cite{mohsenivatani2020throughput} address the problem of throughput maximization in C-RANs via power allocation in virtualized 5G networks. The authors propose a multi-agent DQN-based DRL algorithm to solve the problem in which the agents are each link between RRH and user. The action space is discrete, corresponding to the transmit power. The state space is continuous, representing the current partial CSI and the respective power set. The reward of each slice is discrete, defined as a function of the sum of its tenants' rates. Via simulation, the authors show that their proposed scheme achieves a higher sum-rate compared to greedy search-based power allocation approaches.

\subsubsection{In IoT and Other Emerging Wireless Networks} \ 
In the following paragraphs, we review works that utilize DRL algorithms to address the power allocation issue in IoT and other emerging wireless networks shown in Fig. \ref{ClassificationOnResources}.

Developing efficient spectrum sharing schemes is regarded as one of the main persistent objectives and challenges in CRNs. In \cite{li2018intelligent}, the authors propose a non-cooperative single-agent DQN-based DRL scheme to address the problem of spectrum sharing via power control in CRNs. In their model, the agent is the SU, whose action space is discrete, corresponding to selecting the transmit power from a pre-defined power set. The state space is discrete, defined by four parts; the transmit power of PU and SU, the path loss between PU and a sensor that measures the RSS, the path loss between the SU and a sensor that measures the RSS, and some Gaussian random variable. The reward is a discrete function defined by the achieved SINR level and the minimum SINR requirements of both PU and SU. Simulation results show that their proposed algorithm is robust against random variation in state observations, and the SU interacts with PU efficiently until they reach a state in which both users successfully transmit their own data.

In other work in \cite{nguyen2019non}, the authors present a non-cooperative multi-agent algorithm to address the problem of power allocation in D2D communication networks based on three DQNs, namely, DQN, Double DQN, and Dueling DQN. The agents are the D2D transmitters in each D2D pair, whose state space is discrete, comprised of the level of the interference indicator function. The action space is discrete, representing the set of transmitting power levels, while the reward is a function of the system EE. Simulation results show the ability of such DQN-based models to provide energy-efficient power allocation for the underlying D2D network.

In \cite{zhang2018power}, a DRL algorithm based on multi-agent DQN is proposed to tackle the problem of downlink power allocation in multi-cell cellular networks. The agents are the base stations (BSs), whose stat space is hybrid, comprised of the number of users connected to the BSs and the corresponding power allocation. The action space corresponds to selecting the adjustment value of the power on the subcarriers of the BSs, while the reward is a function of the overall network capacity. Experimental results demonstrate that their proposed algorithm is better than both the water-filling power allocation and $Q$-learning methods in terms of model stability and convergence speed.

UAV networks are attracting considerable attention recently due to their ability to provide enhanced QoS communication in harsh and vital environments. However,  power management is one of the key challenges in such networks. In this context, the authors in \cite{li2020downlink} address the problem of downlink power control in ultra-dense UAV networks with the aim of maximizing the network's EE. A multi-agent DQN-based DRL model is proposed in which the agents are the UAVs in the network. The state space is continuous, representing the remaining energy of the UAV and the interference caused by neighboring UAVs. The action space is discrete, representing the set of possible discrete transmit power values, while the reward function is the EE of the UAV network. Simulation results are compared with $Q$-learning and random algorithms, which show the superiority of their proposed scheme in terms of both the convergence speed and EE.

Other inserting work \cite{nasir2019multi} proposes a multi-agent DQN-based DRL method to study the problem of transmit power control in wireless networks. The agents are the transmitters whose state space is continuous, consisting of three main feature groups; local information, interfering neighbors, and interfered neighbors. The action space is discrete corresponds to discrete power levels, while the reward is a function of the weighted sum-rate of the whole network. Experimental results demonstrate that the proposed distributed algorithm provides comparable and even better performance results to the state-of-the-art optimization-based algorithms available in the literature.

Another interesting work is proposed in \cite{wangsmoothing}. The author proposes a single-agent smoothing DQN-network to control power allocation in cognitive radio-based wireless networks. The agent is the sensor, whose state space is discrete, comprising of a set of received signal strength (RSS) information. The action space is discrete, corresponding to selecting the transmit power, while the reward is a function of the SINR. Using simulation results, the author shows that the proposed smoothing DQN outperforms the conventional DQN in cognitive-based radio networks.

In \cite{zhao2020multi}, the authors propose a multi-agent DDPG-based DRL to address the problem of joint trajectory design and power allocation in multi-UAV wireless networks. In their scheme, the agents are the UAVs, whose state space is a discrete binary indicator function representing whether the QoS of the user ends (UEs) are satisfied or not. The action space is also discrete, corresponding to selecting UAVs' trajectory and transmission power. The reward is a continuous function defined by the joint trajectory design and power allocation as well as the number of UEs covered by the UAVs. Simulation results show that the proposed algorithm achieves higher network utility and capacity than the other optimization methods in wireless UAV networks with reduced computational complexity. 

High-speed railway (HSR) systems are one of the emerging IoT applications for next-generation wireless networks. Such systems are characterized by their rapid variations in the wireless environment, which mandate the development of light-weighted RRAM solutions. As a response to this, Xu \textit{et al.} \cite{xu2021experience} propose a multi-agent DDPG-based DRL model to address the problem of sum-rate maximization via power allocation in hybrid beamforming-based mmWave HSR systems. In their approach, each mobile relay (MR) acts as an agent. The action space is continuous, corresponding to the transmit power level of each MR agent. Also, the state space is continuous, defined by; each MR own signal channel, local observation information of each MR, i.e., beamforming design, each MR achievable rate, and each MR transmit power in the last time step. The reward function is continuous, defined by the achievable sum-rate of the network. Simulation results demonstrate that the SE of their proposed algorithm is comparable to the full digital beamforming scheme, and it outperforms conventional approaches such as maximum power allocation, random power allocation, DQN, and FP.

Federated deep reinforcement learning (FDRL) is an emerging technique that integrates FD and DRL methods. FDRL can be utilized as an efficient tool to enhance the RRAM solutions in large-scale distributed systems). As an example, an interesting approach is proposed in \cite{yan2020federated}, in which the authors propose a cooperative multi-agent actor-critic-based FDRL framework for distributed wireless networks. The authors particularly address the problem of energy/spectrum efficiency maximization via distributed power allocation for network edge users. In their proposed model, the agents are the edge users, whose action space is continuous, defined as the power allocation strategies. The state space is continuous, defined by the allocated transmit power, the SINR on the assigned RBs, and the reward of the previous training time step. The system is defined in terms of a local continuous cost function expressed in terms of SINR, power, path loss, and environmental noise. Using simulation results, the authors demonstrate that their proposed framework achieves better performance accuracy in terms of power allocation than other approaches such as greedy, non-cooperation power allocation, and traditional FL.

\subsubsection{In Satellite Networks}\ 
In the following paragraphs, we review works that employ DRL techniques to address the power allocation issue in satellite networks as well as emerging satellite IoT systems. 

Managing downlink transmit power in satellite networks is also one of the major persistent challenges. To this end, the authors in \cite{luis2019deepJournal} extended their work in \cite{luis2019deep} and present a single-agent Proximal Policy Optimization (PPO)-based DRL model to solve the problem of power allocation in multi-beam satellite systems. In their model, the agent is the processing engine that allocates power within the satellite, whose state space is continuous, comprises the set of demand requirements per beam, and the optimal power allocations for the two previous time steps. The action space is continuous, representing the allocation of the power for each beam, while the reward is a function of both the link data rate achieved by the beam and the power set of the agent. Experimental results demonstrate the robustness of their proposed DRL algorithm in dynamic power resource allocation for multi-beam satellite systems.

NOMA technique has shown efficient results in improving the performance of terrestrial mmWave cellular systems \cite{Maraqa2020Survey}. This has motivated the use of NOMA for satellite communication systems. However, managing the radio resources in such a system becomes an imperative issue. In this context, Yan \textit{et al.} \cite{yan2020delay} study the problem of power allocation for NOMA-enabled SIoT using a single-agent DQN-based DRL scheme. In their system, the agent is the satellite, whose action space is discrete, corresponding to selecting the power allocation coefficient for each NOMA user. The state space is continuous, consisting of the average SNR, link budget, and delay-QoS requirements of NOMA users, while the reward is discrete, which is a function of the effective capacity of each NOMA user. Experimental results demonstrate that their proposed DRL-based power allocation scheme can produce optimal/near-optimal actions, and it provides superior performance to both the fixed power allocation strategies and OMA scheme.

\subsubsection{In Multi-RAT HetNets} \ 
Multi-RAT wireless HetNets is one of the main enabling technologies for modern wireless systems, including 6G networks \cite{RN234}. In HetNets, several RATs with different operating characteristics coexist to enhance network coverage and reliability while providing enhanced QoE to users. The underlying RATs have non-overlapping radio resources; therefore, there would not be typically interference in the network. 

Since a stand-alone network with a single RAT would not be able to support the stringent requirements of emerging disruptive applications, modern smart user devices are equipped with advanced capabilities that enable them to aggregate various radio resources to boost their QoE. Modern user devices can operate in a multi-mode scenario, in which each user device can be connected to a single RAT at any time. Alternatively, user devices can operate in a multi-homing scenario such that they can be connected simultaneously to various RATs to aggregate their radio resources, such as bandwidth and data rate. Multi-RAT networks include the coexistence of RATs, such as the licensed band networks, unlicensed bands networks, hybrid systems, and any combination of the wireless networks that are shown in Fig. \ref{ClassificationOnNetwoksTypes}. 

Visible Light Communication (VLC) is a promising RAT that can support multi-Gbps of data rates over wireless links \cite{Alwarafy2018Performance}. It is mainly developed for indoor applications; however, it is gaining considerable attention lately for outdoor applications as well \cite{Memedi2021Vehicular}. This has motivated researchers to propose solutions that integrate VLC with conventional radio systems to boost aggregate data rates. Managing radio resources in these integrated systems, however, becomes a challenge. In this context, in \cite{RN282}, the authors propose a multi-agent $Q$-learning-based two-time scale scheme to address the power allocation issue for multi-Homing hybrid RF/VLC networks. In their technique, the agents are the RF and VLC APs, whose action space is discrete, corresponding to selecting the downlink power level that ensures the QoS's satisfaction of the multi-homing users. The state space is discrete, which is a function of users' achievable and target rates from the RF and VLC APs. The reward is also discrete, which is a function of the achieved and target rates from all RF and VLC APs. Experimental results demonstrate that not only the users' target rates are satisfied, but also the ability of their algorithm to adapt to the network's dynamics.

For the same network settings in \cite{RN282}, Ciftler \textit{et al.} \cite{ciftler2021dqn} propose a DRL-based scheme to enhance the results and overcome the shortcomings. In particular, the authors in \cite{ciftler2021dqn} propose a non-cooperative multi-agent DQN-based algorithm to address the problem of power allocation in hybrid RF/VLC networks. The agents are the RF and VLC APs whose action space is discrete, representing the transmit power. The state space is continuous, comprised of the actual and target rates, while the reward function is continuous and is a function of target rate band, target rate, and actual rate. Using simulation results, the authors demonstrate that the DQN-based algorithm converges with the rate of 96.1\% compared with the $Q$-learning-based algorithm’s convergence rate of 72.3\%.

\paragraph*{Synthesis and Reflections}
In this section, we review the applications of DRL techniques for power allocation and management in modern wireless networks. The reviewed papers are summarized in Table \ref{PowerClass}. We observe that various DRL techniques can efficiently solve the power allocation optimization problems in diversified wireless network scenarios, and their performance outperforms the state-of-the-art heuristic approaches. Besides, as we discussed in the previous paragraphs, DRL methods can provide comparable results to the conventional centralized optimization-based approaches that have full knowledge of the wireless environments as reported in \cite{nasir2020deep}, or even better results as reported in \cite{zhang2020deeppower}.

We also observe that most of the papers implement multi-agent DRL interactions, and the value-based DRL algorithms, such as DQN and $Q$-learning, are utilized more than the policy-based counterparts. Since the power allocation problem falls in the continuous action space, the use of value-based algorithms to address these types of problems makes the learned policies vulnerable to discretization errors that degrade the accuracy and reliability of the learned models. Hence, the policy-based algorithms, such as DDPG and actor-critic, have received more attention lately, and they have shown more accurate and reliable results compared to the value-based counterparts with additional complexity, as discussed in \cite{nasir2020deep, khan2020centralized, zhang2020deeppower, xu2021experience}. In addition, most of the papers consider the rate maximization, SE, and EE as key performance metrics (e.g., \cite{xu2021experience, khan2020centralized, zhang2020deep}). However, other KPI metrics must be considered as well during the design of DRL frameworks, such as latency, reliability, and coverage, especially for emerging real-time and time-sensitive IoT applications.

Moreover, we observe from Table \ref{PowerClass} that both the cellular HomeNets and emerging IoT wireless networks gain more attention than satellite and multi-RAT networks that still in their early stages and require more in-depth investigation. 

\begin{table*}[h]
\centering
\caption{A Summary List of Papers Related to DRL for Power Allocation.}
\label{PowerClass}
\resizebox{\textwidth}{!}{%
\begin{tabular}{|l|l|c|c|c|c|}
\hline
\rowcolor[HTML]{C0C0C0} 
\multicolumn{2}{|l|}{\cellcolor[HTML]{C0C0C0}} & \cellcolor[HTML]{C0C0C0} & \cellcolor[HTML]{C0C0C0} & \multicolumn{2}{c|}{\cellcolor[HTML]{C0C0C0} \textbf{Learning Algorithm }} \\ \cline{5-6} 
\rowcolor[HTML]{C0C0C0} 
\multicolumn{2}{|l|}{\multirow{-2}{*}{\cellcolor[HTML]{C0C0C0} \textbf{Network Type}}} & \multirow{-2}{*}{\cellcolor[HTML]{C0C0C0} \textbf{Ref.}} & \multirow{-2}{*}{\cellcolor[HTML]{C0C0C0} \textbf{Radio Resource (or Issues Addressed)}} & \textbf{Mode} & \textbf{Algorithm} \\ \hline\hline
 \parbox[t]{2mm}{\multirow{10}{*}{\rotatebox[origin=c]{90}{Cellular \& HomNets}}} &  Cellular systems& Ghadimi \textit{et al.}\cite{ghadimi2017reinforcement}&  power control \& rate adaptation&  Multi-agent& $Q$-learning \\ \cline{2-6}
 & Data center power grids& Hu \textit{et al.}\cite{hu2020deep}& Power resource use& Single-agent& DQN \\ \cline{2-6} 
 & Multi-cell cellular& Khan \textit{et al.}\cite{khan2020centralized}& Downlink power allocation& Single- \& multi-agent & Actor-critic \\ \cline{2-6}
 & Wireless mobile networks& Nasir \textit{et al.}\cite{nasir2020deep}& Continuous power control & Multi-agent & DDPG \\ \cline{2-6} 
 & D2D cellular & Bi \textit{et al.}\cite{bi2020deep}& Power allocation& multi-agent & DQN \\ \cline{2-6} 
 & Dense 5G cellular& Saeidian \textit{et al.}\cite{saeidian2020downlink}& Downlink power control& Multi-agent & DQN \\ \cline{2-6} 
 & NOMA mmWave UDNs& Zhang \textit{et al.}\cite{zhang2020deep}& EE power allocation& Multi-agent & DQN \\ \cline{2-6} 
 & Cellular networks& Meng \textit{et al.}\cite{meng2019power, meng2020power}& Downlink power allocation& multi-agent & DQN \\ \cline{2-6} 
 & HetNets& Zhang \textit{et al.}\cite{zhang2020deeppower}& Power control& multi-agent & DQN \& DDPG \\ \cline{2-6} 
 & Virtualized 5G networks& Mohsenivatani \textit{et al.}\cite{mohsenivatani2020throughput}& Power allocation& multi-agent & DQN \\ \Xhline{4\arrayrulewidth} 
 
 \parbox[t]{2mm}{\multirow{9}{*}{\rotatebox[origin=c]{90}{IoT \& Emerging Nets}}}& CRNs& Li \textit{et al.}\cite{li2018intelligent}& Power control & Single-agent & DQN   \\ \cline{2-6} 
 & D2D networks& Nguyen \textit{et al.}\cite{nguyen2019non}& Power allocation& multi-agent & DQN, DDQN, \& Dueling DQN  \\ \cline{2-6} 
 & Multi-cell cellular& Zhang \textit{et al.}\cite{zhang2018power}& Downlink power allocation & Multi-agent & DQN \\ \cline{2-6} 
 & Ultra-dense UAV& Li \textit{et al.}\cite{li2020downlink}& Downlink power control& Multi-agent & DQN \\ \cline{2-6} 
 & Wireless Networks& Nasir \textit{et al.}\cite{nasir2019multi}& Transmit power control& Multi-agent & DQN \\ \cline{2-6} 
 & CRNs& Wang \textit{et al.}\cite{wangsmoothing}& Power allocation& Single-agent & DQN \\ \cline{2-6} 
 & Multi-UAV& Zhao \textit{et al.}\cite{zhao2020multi}& Power allocation & Multi-agent & DDPG \\ \cline{2-6} 
 & mmWave HSR systems& Xu \textit{et al.}\cite{xu2021experience}& Power allocation& Multi-agent & DDPG \\ \cline{2-6} 
 & Distributed networks & Yan \textit{et al.}\cite{yan2020federated}& Distributed power allocation& Multi-agent & Actor-critic  \\ \Xhline{4\arrayrulewidth}  
  
 \parbox[c][6mm]{2mm}{\multirow{5}{*}{\rotatebox[origin=c]{90}{\quad \quad Satellite}}}& Multi-beam satellites& Luis \textit{et al.}\cite{luis2019deep, luis2019deepJournal}& Power allocation& Single-agent & PPO  \\ \cline{2-6} 
  \parbox[c][6mm]{2mm} {}& NOMA-enabled SIoT& Yan \textit{et al.}\cite{yan2020delay}& Power allocation& Single-agent & DQN \\ \Xhline{4\arrayrulewidth}
 
 \parbox[c][6.7mm]{2mm}{\multirow{5}{*}{\rotatebox[origin=c]{90}{\quad \quad \quad Multi-RAT}}} & Hybrid RF/VLC networks& Kong \textit{et al.}\cite{RN282}& Power allocation&  Multi-agent & $Q$-learning   \\  \cline{2-6}
\parbox[c][6.7mm]{2mm} {} & Hybrid RF/VLC networks& Ciftler \textit{et al.}\cite{ciftler2021dqn}& Power allocation& multi-agent & DQN \\  \hline
\end{tabular}%
}
\end{table*}  

\subsection{DRL for Spectrum Allocation and Access Control}\
One of the significant challenges in modern wireless communication networks that still needs more investigation is spectrum management and access control. In this context, DRL techniques have attracted considerable research interest recently due to their robustness in making optimal decisions in dynamic and stochastic environments. This section presents the related works to the applications of DRL algorithms for radio spectrum allocation in modern wireless networks. This includes issues, such as dynamic network access, user association or cell selection, spectrum access or channels selection/assignment, and the joint of any of these issues, as shown in Fig. \ref{ClassificationOnNetwoksTypes}.

In modern wireless networks, a massive number of user devices may request to access the wireless channel simultaneously. This may drastically overload and congest the channel, causing communication failure and unreliable QoS. Hence, efficient communication schemes and protocols must be developed to address this issue in channel access via employing various access scheduling and prioritization techniques. RRAM for modern wireless networks requires considering dynamic load balancing and access management methods to support the massive capacity and connectivity requirements of the future wireless networks while utilizing their radio resources efficiently. DRL methods have been used recently to address these issues, and they have demonstrated efficient results in the context of massive channel access. 

On the other hand, user devices in cellular networks are required to associate or be assigned to BS(s) or network AP(s) to get a service. The association process could be symmetric, i.e., both uplink and downlink are from the same BS/AP, or it may be asymmetric in which the uplink and downlink may associate to different BSs/APs. This association or cell selection process must be carefully addressed as it strongly affects the allocation of network radio resources. Unfortunately, such types of problems are typically non-convex and combinatorial \cite{alwarafy2021DeepRAT} and need accurate network information to obtain the optimal solution. In this context, DRL techniques have also shown efficient results in addressing user association and cell selection issues for modern wireless networks.

\subsubsection{In Cellular and HomNets} \
In the following paragraphs, we review works that employ DRL algorithms to address the spectrum and access control problem in cellular and HomNets depicted in Fig. \ref{ClassificationOnResources}.

RRAM in UAV-assisted cellular networks is also one of the main emerging challenges. Towards this end, in \cite{RN18}, a multi-agent DQN model addresses the joint user association, spectrum allocation, and content caching in an LTE network consisting of UAVs serving ground users. In their model, the agents are the UAVs, which have storage units and have the ability to cached contents in LTE-BSs. These UAV agents can access the licensed as well as the unlicensed spectrum bands, and a remote cloud-based server is used to control them. The licensed cellular spectrum band is used in the transmissions from the cloud to the UAVs. Each UAV agent has to obtain 1) its user association, 2) bandwidth assignment indicators in the licensed spectrum band, 3) time slot indicators in the unlicensed spectrum band, and 4) content that the users request. The input of the DQL is the other agents’ actions (the UAV-user association schemes), and the output is the set of users that the UAV can handle. Simulation results demonstrate that their proposed DQL strategy enhances the number of users up to 50\% compared to the standard $Q$-learning strategy.

Based on their initial work in \cite{zhao2018deep}, the authors in \cite{RN19} propose a multi-agent Dueling Double deep Q-network (D3QN)-based DRL model to handle the joint BS and channel selections in macro and femto BS networks sharing a set of radio channels. In their scheme, the agents are the UEs, whose state space is a discrete binary vector that shows whether UEs' SINR higher than the minimum QoS requirement or not. The action space is discrete, corresponding to the BS and channel association. The reward function is discrete in which the UE agent will receive a utility as a reward if the QoS is met; otherwise, it will receive a negative value for the reward. Simulation results demonstrate that their proposed strategy outperforms the standard $Q$-learning strategy in terms of generalization, system capacity, and convergence speed.

The problem of user association problem in cellular IoT networks is studied in \cite{zhang2020intelligent}. The goal is to assign IoT devices to particular cellular users to maximize the sum-rate of the IoT network. Two single-agent DQN DRL algorithms are proposed; the first one utilizes global information to makes decisions for all IoT devices at one time, while the other algorithm uses local information to make a distributed decision for only a single IoT device at one time. In their model, the BS acts as the agent whose state space is continuous, consisting of both historical CSI and interference information. The action space is discrete, representing both all possible association schemes of the network and the individual association for only a single IoT device. The reward function of the first DQN algorithm is the sum-rate of all IoT devices, while for the second DQN includes both the current transmission rate of IoT devices and the interference with other IoT devices. Experimental results demonstrate that their proposed DRL framework both scalable and achieves performance comparable to the optimal user association policy.

In another interesting work in \cite{lei2020deep}, the authors study the problem of spectrum allocation in the emerging integrated access and backhaul (IAB) cellular networks characterized by their dynamic environment and large-scale deployment. The problem is formulated as a non-convex mix-integer and non-linear programming, and a DRL framework based on single-agent Double DQN and actor-critic algorithms is proposed to solve it. In their model, the agent is a center-located controller or distributed UE. The state space is discrete, indicating the status of UEs' QoS, and the action space is discrete, corresponding to the allocation matrix for the donor BS and IAB nodes. The reward function is modeled to optimize the proportional fairness allocation of the network. Experimental results demonstrate that their framework has promising results compared to other conventional spectrum allocation policies.

The problem of load balancing in large-scale and dynamic wireless networks is also another important issue. In this context, the authors in \cite{li2017user} present a multi-agent $Q$-learning-based algorithm to address the problem of user association for load balancing in vehicular networks. In their scheme, the agents are the BSs, whose action space is discrete, representing the associations with the network's vehicles. The state space is a hybrid (continuous and discrete), consisting of the service resources and its service demands, SINR matrix, and association matrix. The reward is a continuous function defined through the association and SINR matrices. Using experiments on real-field taxi movements, the authors evaluate the performance of their proposed algorithm, and they show that it provides higher quality load balancing compared to conventional association methods.

In other work in \cite{perez2017machine}, the authors propose a multi-agent $Q$-learning-based algorithm for the cognitive RAT selection issue in 5G HetNets. In their model, the agents are the terminal devices whose action is discrete, corresponding to identifying the available RATs. The state space is also discrete that corresponds to the currently active RAT association, while the reward function is continuous, which is defined by the average measured throughput and handovers call drops. Using experimental results, the authors show that their proposed framework outperforms the random and max-SINR decision approaches.

Zheng \textit{et al.} \cite{zheng2021channel} propose a single agent actor-critic-based DRL algorithm to address the problem of channel assignment for hybrid NOMA-based 5G cellular networks. The agent is the BS, whose action space is discrete, corresponding to assigning channels for users. The state space is a hybrid (continuous and discrete) comprised of three elements; the CSI matrix, achieved users’ data rate in the previous time slot, and the assigned channels in the previous time slot. The reward is a discrete function defined in terms of users' SE, the number of channels that use NOMA for transmission, and the number of users whose data rate is zero. Simulation results demonstrate that their proposed method outperforms some conventional approaches, such as greedy, random, match theory-based, and Genetic Algorithms, in terms of both network SE and sum-rate.

The problem of spectrum management in wireless DSA is addressed in \cite{song2021deep} based on distributed multi-agent DQN. In their approach, the agents are each DSA user, whose action space is discrete, corresponding to the transmit power change for each channel. The state space is discrete, defined as the transmit power on wireless channels. The reward is a continuous function defined by the SE and the penalty caused by the interference to PUs. Experimental results show that their proposed model with echo state network-based DQN achieves a higher reward with both the achievable data rate and PU protections.

\subsubsection{In IoT and Other Emerging Wireless Networks} \ 
In the following paragraphs, we review works that employ DRL algorithms to address the spectrum and access control problem in IoT and emerging wireless networks illustrated in Fig. \ref{ClassificationOnResources}.

In \cite{RN11, wang2018deep, wang2017deep}, the authors propose a single-agent DQN-based DRL scheme to tackle the problem of dynamic channel access for IoT sensor networks. In their scheme, the agent is the sensor, and its action is discrete, corresponding to selecting one channel to transmit its packets at each time slot. The state space is discrete, which is a combination of rewards and actions in the previous time slots. The reward function is also discrete, which is "+1" if the selected channel is in low interference in such case a successful transmission occurs; otherwise, the reward is "-1" in such case the selected channel is in high interference, and a transmission failure occurs. Simulation results show that their proposed scheme achieves an average reward of 4.4 compared to 4.5 obtained using the conventional myopic policy \cite{RN243}, which needs a compact knowledge of the transition matrix of the system.

Other work in \cite{RN13} develops a single-agent DQN-based DRL model to address the channel selection in energy harvesting-based IoT sensors. In that work, the agent is one BS, which controls the channel assignments for energy harvesting-enabled sensors. The problem of the agent is to predict the battery level of the sensors and to assign channels to sensors such that the total rate is maximized. The DQL model used to solve this problem has two long-short-term-memory (LSTM) neural network (NN) layers, one for predicting the sensor’s battery state and one for obtaining channel access policy based on the predicted states obtained from the first layer. The agent’s action is all the available sensors that require to access the channels. The state contains the history of channel access scheduling, true and predicted battery information history and the current sensor’s CSI. Simulation results show that the total rates obtained using the DQL scheme are 6.8 kbps compared to 7 kbps obtained from the optimal scheme rate.

A multi-agent DQN-based DRL model is proposed in \cite{RN21} in order to address the cooperative spectrum sensing issue in CRNs. In their scheme, the agents are the SUs whose action is discrete, corresponding to sensing the spectrum for possible transmission without interfering with the PUs. The state space is discrete, and it is comprised of four elements representing cases when the spectrum is sensed as occupied, the spectrum is not sensed in a particular time slot, the spectrum is sensed as idle, and one of the other SUs broadcast the sensing result first. The reward function is the binary indicator, which is "+1" if the spectrum is sensed as idle and "0" otherwise. Simulation results show that their proposed algorithm has a faster convergence speed and better reward performance than the conventional $Q$-learning algorithm.

Xu \textit{et al.} \cite{RN290} propose a single-agent DQN and DDQN-based DRL approaches to address the problem of dynamic spectrum access in wireless networks. In their model, the agent is a wireless node (e.g., a user) whose action is discrete, corresponding to sensing the discrete frequency channel for possible data transmission. The state space is discrete, defining if the channel is occupied or idle at time slot $t$. The reward function is discrete, which is ranging from 0 to 100 for successful transmission; otherwise, the reward is "-10" if the channel state is occupied and the user transmission fails. It is shown using simulation results that both DQN and DDQN can learn different nodes’ communication patterns and achieve near-optimal performance without prior knowledge of system dynamics.

Based on their initial work in \cite{liang2019multi}, the authors in \cite{liang2019spectrum} propose a distributed single- and multi-agent DQN-based DRL schemes to address the spectrum sharing problem in V2X networks. In their proposed system, multiple V2V links reuse the frequency spectrum preoccupied with V2I links. The agents are the V2V links whose action space is discrete, corresponding to spectrum sub-band and power selection. Each agent's local observation space includes local channel information (such as its own signal channel gain, interference channels from other V2V transmitters, interference channel from its own transmitter to the BS, and the interference channel from all V2I transmitters), the remaining V2V payload, and the remaining time budget. The reward is continuous, which is a function of both the instantaneous sum capacity of all V2I links and the effective V2V transmission rate until the payload is delivered. Experimental results show that the agents cooperatively learn a policy that enables them to simultaneously improve the sum capacity of V2I links and payload delivery rate of V2V links. The authors also show that their proposed models for the single-agent and multi-agent settings provide very close performance to the conventional exhaustive search.

The authors in \cite{RN12} propose a single-agent DQN model to address the joint channel access and packet forwarding in a multi-sensor scenario. In the proposed scheme, one sensor is the agent, which acts as a relay to forward packets arriving from its surrounding sensors to the sink. The agent has a buffer to store arriving packets. The agent’s action is to choose channels for the packet forwarding, the packets transmitted on these channels, and a modulation scheme at each time slot to maximize its utility (defined as the ratio of the transmitted packets number to the transmit power). The state is the combination of the buffer and channel states. The input of the DQL model is the state, while the output is the corresponding action selection. Simulation results demonstrate that the proposed DQL scheme enhances system utility (i.e., 0.63) compared to the conventional random action selection scheme (i.e., 0.37).

Khan \textit{et al.} \cite{khan2019reinforcement} propose a multi-agent A3C-based DRL to address the problem of vehicle-cell association in mmWave V2X networks. The agents are the RSUs whose action is discrete, corresponding to determining the optimal vehicle-RSU association for RSU. The state space is a hybrid (discrete and continuous) defined in terms of the last channel observations, rate threshold violation indicator, and experienced data rate of vehicles. The reward function is continuous, which is defined in terms of the average rate per vehicle and threshold rate. Using experimental results, it is shown that their proposed algorithm achieves around 15\% sum-rate gains and a 20\% reduction in vehicular user outages compared to baseline approaches.

The problem of dynamic spectrum access in CRNs is investigated in \cite{yang2018dynamic} through combining DRL and evolutionary game theory. In particular, uncooperative multi-agent DQN is considered in which the agents are the SUs whose action is discrete, corresponding to selecting the access channel. The state space is discrete, which includes two main parts; the channel selected by the agent and the utility obtained after transmission on the selected channel. The reward function is defined in terms of evolutionary game theory. Simulation results indicate the performance enhancement of their proposed algorithm over the case without learning in terms of average system capacity.

Another interesting work is presented in \cite{cao2020deep} in which the authors propose a multi-agent DQN-based DRL algorithm to address the problem of optimum multi-user access control in Non-Terrestrial Networks (NTNs). In their model, UEs are the independent agents that report their experiences and local observations to a centralized trainer controller located at the backhaul network. The latter will then utilize the collected experiences to update the global DQN parameter. The agent's state space is continuous, comprised of the connected NT-BS of UEs at the previous time slots, the RSS of UEs, the number of connected UEs of each NT-BS, and the transmission rate of UEs. The action space is discrete, representing the binary indicator functions of UEs, while the reward is a function of the transmission rate of UEs. Experimental performance results show that their proposed scheme is efficient in addressing the fundamental issues in the deployment of NTNs infrastructure, and it outperforms the traditional algorithms in terms of both the data rate and the number of handovers.

In \cite{yang2020partially}, the authors propose a multi-agent DQN-based DRL scheme to address the problem of dynamic joint spectrum access and mode selection (SAMS) in cognitive radio networks (CRNs). The agents are the secondary users (SUs) whose action space is discrete, corresponding to selecting the access channel and access mode. The state space of each SU agent is discrete, comprised of the action taken by the $m$th SU agent, the ACKs of all SUs agents, and the ACK of the $m$th SU agent. The reward function is discrete, which is "1" if the action selection process is successful; otherwise, there is a collision, and the agent will receive a "0" reward.

Tomovic \textit{et al.} \cite{tomovic2020novel} propose a single-agent DRL model based on the integration of Double deep $Q$-learning architecture and RNN to address the problem of DSA in multi-channel wireless networks. In particular, the agent is the SU node, whose action space is discrete, representing the selection of a channel for sensing. The state space is also discrete, comprised of a history of the binary observations and history of taken actions. The reward function is a discrete binary function, which is "1" if the observation is "1" and "0" otherwise. Simulation results show that their proposed method is able to find a near-optimal policy in a smaller number of iterations, and it can support a wide range of communication environment conditions.

In other work in \cite{RN292}, the authors propose both a single-agent and multi-agent deep actor-critic DRL-based framework for dynamic multi-channel access in wireless networks. In their system, the agents are the users whose action space is discrete, corresponding to selecting a channel. The observation space is also discrete, which is defined based on the status of the channel and collision status. The reward function is discrete, which is "+1" if the selected channel is good; otherwise, it is "-1". Using simulation results, the authors show that their proposed actor-critic framework outperforms the DQN-based algorithm, random access, and the optimal policy when there is full knowledge of the channel dynamics.

The problem of DSA for the CRN is studied in \cite{tondwalkar2019deep} based on an uncoordinated and distributed multi-agent DQN model. The agents are CRs, whose action is discrete, representing the possible transmit powers. The state space is discrete, reflecting whether the limits for DSA are being met or not, depending on the relative throughput change at all the primary network links. The reward is also discrete, which is a function of the throughput of the links and the environment's state. Experimental results reveal that their proposed scheme finds policies that yield performance within 3\% of an exhaustive search solution, and it finds the optimal policy in nearly 70\% of cases.

Industrial IoT (IIoT) has emerged recently as an innovative networking ecosystem that facilitates data collection and exchange in order to improve network efficiency, productivity, and other economic benefits \cite{Gumaei2018Survey}. RRAM in such a sophisticated ecosystem is also a challenge that needs more investigation. In this context, recently in \cite{shi2020deep}, the authors propose a solution for spectrum resource management for IIoT networks, with the goal of enabling spectrum sharing between different kinds of UEs. In particular, a single-agent DQN algorithm is proposed in which the agent is the BS. The action space is discrete, which corresponds to the allocation of spectrum resources for all UEs. The observation space is a hybrid (continuous and discrete) consisting of four elements; the current action (i.e., the allocation of spectrum resources), the data priority of type I UEs, the buffer length of type II UE, and the communication status of the first type of UEs. The reward function is continuous, defined to address their optimization problem. It is divided into four objectives; 1) maximizing the spectrum resource utilization; 2) quickly transmitting the high-priority data; 3) meeting the threshold rate requirement of the first type of UEs; 4) ensuring that the second type of UEs completes the transmission in time. Using simulation results, it is demonstrated that their proposed algorithm achieves better network performance with a faster convergence rate compared with other algorithms.

In \cite{janiar2021deep}, the authors propose a multi-agent Double DQN-based DRL model to address the problem of DSA in distributed wireless networks. In particular, they design a channel access scheme to maximize channel throughput regarding fair channel access. The agents in their scheme are the users. The action space is discrete, which is "0" if the user does not attempt to transmit packets during the current time slot, and it is "1" if it has attempted to transmit. The state space is discrete, consisting of four main elements; each user action taken on the current time slot, channel capacity (which could be negative, positive, or zero), a binary acknowledgment signal showing if the user transmits successfully or not, and a parameter that enables each user to estimate other users’ situations. The reward is a discrete binary function that takes the value of "1" if the user transmits successfully; otherwise, it is "0" meaning that the user transmitted with collision. It is shown using simulation results that their scheme can maximize the total throughput while trying to make fair resource allocation among users. Also, it is shown that their proposed scheme outperforms the conventional Slotted-Aloha scheme in terms of sum-throughput.

Wang \textit{et al.} \cite{wang2021intelligent} address the problem of DSA in VANETs, by proposing an interesting scheme that combines multi-hop forwarding via vehicles and DSA. The optimal DSA policy is defined to be the joint maximization of channel utilization and minimization of the packet loss rate. A multi-agent DRL network structure is presented that combines RNN and DQN for learning the time-varying process of the system. In their scheme, each user acts as an agent whose action space is discrete, corresponding to choosing a channel for transmission at time slot $t$. The state space is discrete, composed of three components; a binary transmission condition $\eta$, which is "1" if the transmission is successful and "0" otherwise, the channel selection action, and the channel status indicator after each dynamic access process. The reward is a discrete binary function, which takes a positive value if $\eta = 1$, otherwise it takes the value of "0". Simulation results show that their proposed scheme: 1) reduces the packet loss rate from 0.8 to around 0.1, 2) outperforms Slotted-Aloha and DQN in terms of reducing collision probability and channel idle probability by about 60\%, and 3) enhances the transmission success rate by around 20\%.

Most recently in \cite{jiang2021multi}, the authors propose multi-agent DQN-based DRL schemes to address the problem of joint cooperative spectrum sensing and channel access in clustered cognitive UAV (CUAV) communication Networks. Three algorithms are proposed: 1) a time slot multi-round revisit exhaustive search based on virtual controller (VC-EXH), 2) a  $Q$-learning based on independent learner (IL-Q) and 3) a DQN based on independent learner (IL-DQN). The agents are the CUAVs in the network. The action space of any CUAV agent is a discrete function defined by the steps that this agent moves clockwise in time slot $t$ relative to the channel selected in time slot $t-1$ on the PU channel ring. The state space is a discrete set consisting of two main elements: 1) the number of CUAVs agents that have selected a particular channel to sense and access in the previous time slot and 2) a binary indicator function that shows the occupancy status of a particular channel in the previous time slot. The reward is a discrete function defined in terms of spectrum sensing, channel access, utility, and cost. Experimental results show that all the three algorithms proposed show efficient results in terms of convergence speed and the enhancement of utilization of idle spectrum resources.

The authors in \cite{xu2020application} propose a multi-agent deep recurrent Q-network-based DRL model to address the problem of DSA in dynamic heterogeneous environments with partial observations. In their work, the authors consider a case-study with multiple heterogeneous PUs sharing multiple independent radio channels. The agents are the SUs, whose action space is discrete, corresponding to deciding whether to transmit in a particular band or wait during the next time slot. The state space is discrete, representing whether the channels are occupied, idle, or unknown. The reward function is discrete, represented by two values; 100 per channel for successful transmission and -50 per channel for collision. Using simulation results, the authors show that their proposed algorithm handles various dynamic communication environments, and its performance outperforms the myopic conventional methods and very close to the optimization-based approaches.

\subsubsection{In Satellite Networks}\ 
In the following paragraphs, we review works that employ DRL algorithms to address the spectrum and access control problem in satellite networks and emerging satellite IoT systems. 

The work in \cite{RN17} proposes a single-agent DQN-based DRL algorithm that considers the problem of channel assignment in multi-beam satellite systems. In their scheme, the agent is the satellite, whose action is discrete, including an index that indicates the channel that the newly arrived user has occupied. The agent’s reward is discrete, which contains a positive value if the service is satisfied, and a negative value if the service is not satisfied or blocked. The state space is also discrete, which comprises three elements; the current users, the current channel assignment matrix, and a list of the new user arrivals. Experimental results demonstrate that their proposed scheme decreases the blocking probability and improves the carried traffic up to 24.4\% as well as enhances the spectrum efficiency compared to the conventional fixed channel assignment approach.

The authors in \cite{hu2018deep} propose a single-agent DQN-based DRL algorithm to address the problem of dynamic channel allocation in multi-beam satellite systems. In particular, an image-like tensor formulation on the system environments is considered in order to extract traffic spatial and temporal features. The agent in their model is the satellite, whose action space is discrete, corresponding to determining the resource allocation schemes. The state space is continuous, consisting of two elements; the system resource allocation state and the users' service request state. The reward function is discrete, which is defined in terms of the optimization objective function. 

Recently, the authors in \cite{zhao2020deep} propose a single-agent DQN-based DRL approach for energy-efficient channel allocation in SIoT. The agent in their model is the LEO satellite, whose action space is discrete, corresponding to mapping from newly coming node tasks to channels to be allocated. The state space is discrete, including information about user tasks, such as the size and location of tasks. The reward is continuous, which is divided into two normalized reward function components; the power efficiency reward and the normalized value of the service blocking rate. Both of these reward components are functions of power set up by the agent, the optimal power decided by the location of the beam, and the number of served nodes. Experimental results demonstrate that their proposed algorithm saves energy consumption of around 67.86\%  compared to some conventional approaches.

In \cite{liu2020dynamic}, the authors propose a centralized single-agent DQN-based scheme to address the problem of dynamic channel allocation in SIoT. The agent in their model is the satellite, whose action is discrete, corresponding to selecting which sensors to allocate channels to. The state space is discrete, comprised of three parts; the number of tasks in each time step, the bandwidth that a sensor node requires, and the duration of a new task. The reward is continuous, which is a function of the duration of data transmission for the sensor. Using simulation results, it is shown that their proposed algorithm both provides higher transmission success rates and reduces data transmission latency by at least 87.4\% compared to the conventional channel allocation algorithms.

Zheng \textit{et al.} \cite{zheng2020leo} propose a single-agent $Q$-learning-based RL model to address the problem of combination allocation of fixed channel pre-allocation and dynamic channel scheduling in a network architecture of LEO satellites that utilizes a centralized resource pool. In their model, the satellite serves as an agent whose action is discrete, corresponding to assigning channels to users. The state space is discrete, defined by the channel assignment of users in each beam. The reward is continuous, which is a function of the user's supply-demand ratio. Experimental results demonstrate that their proposed approach enhances system supply-demand ratio by 14\% and 18\% compared to the static channel allocation the Lagrange algorithm channel allocation method.

\subsubsection{In Multi-RAT HetNets} \ 
In the following paragraphs, we review works that employ DRL algorithms to address the spectrum and access control problem in multi-RAT HetNets. This includes the coexistence of various variants of the wireless networks shown in Fig. \ref{ClassificationOnResources}. 

In \cite{RN15}, a multi-agent DQN-based model that jointly tackles the dynamic channel selection and interference management in Small Base Stations (SBSs) cellular networks that share a set of unlicensed channels in Long Term Evolution (LTE) networks. In the proposed scheme, the SBSs are the agents who choose one of the available channels for transmitting packets in each time slot. The agent’s action is channel access and channel selection probability. The DQL input includes the channels’ traffic history of both the SBSs and Wireless Local Area Networks (WLAN), while the output is the agent’s predicted action vectors. Simulation results reveal that their proposed DQL strategy enhances the average data rate by up to 28\% when compared to the conventional $Q$-learning scheme.

In \cite{RN20}, a single-agent DQN-based model is proposed to tackle the dynamic spectrum allocation for multiple users that share a set of $K$ channels in the same network settings of \cite{RN15}. In their scheme, the agent is the user whose action is either choosing a channel with a particular attempt probability or selecting not to transmit. The agent’s state includes the history of the actions of the agent and its current observations. The DQL model input is the previous actions along with their observations, while the output is the Q-values corresponding to the actions. Simulation results demonstrate that their proposed DQL strategy achieves a double data rate compared to the state-of-the-art Slotted-Aloha scheme. 

Wang \textit{et al.} \cite{wang2019deep} propose a single-agent prediction-DDPG-based DRL algorithm to study the problem of the dynamic multichannel access (MCA) for the hybrid long-term evolution and wireless local area network (LTE-WLAN) aggregation in dynamic HetNets. The agent is the central BS controller, whose state space is continuous, consisting of both the channels’ service rates and the users’ requirement rates. The action space, on the other hand, is discrete, representing the users’ index. Two reward functions are provided; online traffic real reward and online traffic prediction reward, each of which are functions of users’ requirements, channels’ supplies, degree of system fluctuation, the relative resource utilization, and the quality of user experience. Using simulation results, the authors demonstrate the efficiency of the proposed prediction-DDPG model in solving the dynamic MCA problem compared to conventional methods.

Other interesting work in \cite{peng2020deep}, the authors investigate the joint allocation of the spectrum, computing, and storing resources in multi-access edge computing (MEC)-based vehicular networks. In particular, the authors propose multi-agent DDPG-based DRL algorithms to address the problem in a hierarchical fashion considering a network comprised of macro eNodeB (MeNB) and Wi-Fi APs. The agents are the controller installed at MEC servers. The agents' action space is discrete including the spectrum slicing ratio set, spectrum allocation fraction sets for the MeNB and for each Wi-Fi AP, computing resource allocation fraction, and storing resource allocation fraction. The state space is discrete representing information of the vehicles within the coverage area of the MEC server, including vehicles' number, x-y coordinates, moving state, position, and task information. The reward function is discrete, defined in terms of the delay requirement, and requested storing resources required to guarantee the QoS demands of an offloaded task. Provided experimental results reveal that their proposed schemes achieve high QoS satisfaction ratios compared with the random assignment techniques.

Recently in \cite{challita2021deep}, the authors propose a single-agent DQN algorithm based on Monte Carlo Tree Search (MCTS) to address the problem of dynamic spectrum sharing between 4G LTE and 5G NR systems. In particular, the authors used the MuZero algorithm to enable a proactive BW split between 4G LTE and 5G NR. The agent is a controller located at the network core, whose action space is discrete, corresponding to a horizontal line splitting the BW to both 4G LTE and 5G NR. The state space is discrete, defined by five elements: 1) an indicator if the user is an NR user or not, 2) the number of bits in the user's buffer, 3) an indicator of whether the user is configured with multimedia broadcast single frequency network (MBSFN) or not, 4) the number of bits that can be transmitted for the user in a given subframe, and 5) the number of bits that will arrive for each user in the future subframes. The reward function is a continuous function defined as a summation of the exponential of the delayed packet per user. Experimental results show that their proposed scheme provides comparable performance to the state-of-the-art optimal solutions.

\paragraph*{Synthesis and Reflections}
This section reviews the applications of DRL for dynamic spectrum allocation and access control in modern wireless networks. These types of radio resources are inherently coupled with user association, network/RAT selection, dynamic multi-channel access, and DSA. Table \ref{SpectrumClass} summarizes the reviewed papers in this section. In general, the application of DRL for spectrum allocation and access control problems has received considerable attention lately. We observe that most DRL algorithms, when deployed for non-IoT networks, are implemented in centralized fashions at network controllers, such as BSs, RSUs, and satellites \cite{zhang2020intelligent, khan2019reinforcement, RN17}. This is done to utilize the controllers' powerful and advanced hardware capabilities in collecting network information and designing cross-layer policies. Hence, we observe that DRL models are deployed as a single-agent at the network controllers \cite{shi2020deep}. On the contrary, DRL provides a flexible tool in diversified IoT networks and systems, conventionally involving dynamic system modeling and multi-agent interactions, such as CRNs and distributed systems.

In addition, the management of such types of radio resources falls in the discrete action space. Therefore, the value-based algorithms are utilized more than the policy-based ones. We also observe from Table \ref{SpectrumClass} that the use of DRL techniques for IoT and emerging wireless networks receives more attention than other networks, especially for the cognitive radio-based systems as in \cite{jiang2021multi}.

The exponential increase in smart IoT devices mandates to take autonomous decisions locally, especially for delay-sensitive IoT applications and services. In this context, we anticipate that the research on spectrum allocation and access control using distributed multi-agent DRL algorithms for future IoT networks will attract more attention as in \cite{liang2019spectrum, khan2019reinforcement, wang2021intelligent, jiang2021multi}


\begin{table*}[h]
\centering
\caption{A Summary List of Papers Related to DRL for Spectrum Allocation and Access Control.}
\label{SpectrumClass}
\resizebox{\textwidth}{!}{%
\begin{tabular}{|l|l|c|c|c|c|}
\hline
\rowcolor[HTML]{C0C0C0} 
\multicolumn{2}{|l|}{\cellcolor[HTML]{C0C0C0}} & \cellcolor[HTML]{C0C0C0} & \cellcolor[HTML]{C0C0C0} & \multicolumn{2}{c|}{\cellcolor[HTML]{C0C0C0} \textbf{Learning Algorithm }} \\ \cline{5-6} 
\rowcolor[HTML]{C0C0C0} 
\multicolumn{2}{|l|}{\multirow{-2}{*}{\cellcolor[HTML]{C0C0C0} \textbf{Network Type}}} & \multirow{-2}{*}{\cellcolor[HTML]{C0C0C0} \textbf{Ref.}} & \multirow{-2}{*}{\cellcolor[HTML]{C0C0C0} \textbf{Radio Resource (or Issues Addressed)}} & \textbf{Mode} & \textbf{Algorithm} \\ \hline\hline
 \parbox[t]{2mm}{\multirow{8}{*}{\rotatebox[origin=c]{90}{Cellular \& HomNets}}} &  UAV-assisted LTE&  Chen \textit{et al.}\cite{RN18}&Joint user association, spectrum allocation, \& content caching  & Multi-agent & DQN \\ \cline{2-6} 
 &  Macro \& femto BS&  Zhao \textit{et al.} \cite{RN19, zhao2018deep}&  Joint BS \& channel selections&  Multi-agent& Dueling DDQN \\ \cline{2-6} 
 &  Cellular IoT&  Zhang \textit{et al.}\cite{zhang2020intelligent}&  User association&  Single-agent& DQN \\ \cline{2-6} 
 &  IAB cellular& Lei \textit{et al.}\cite{lei2020deep}&  Dynamic spectrum allocation&  Single-agent& DDQN \& actor-critic \\ \cline{2-6} 
 &  CV2X& Li \textit{et al.}\cite{li2017user}&  User association&  Multi-agent&$Q$-learning  \\ \cline{2-6} 
 &  5G HetNets&  Perez \textit{et al.}\cite{perez2017machine}&  User association&  Multi-agent&$Q$-learning  \\ \cline{2-6} 
 &  Hybrid NOMA-based 5G&  Zheng \textit{et al.}\cite{zheng2021channel}&  Dynamic spectrum allocation&  Single-agent& actor-critic \\ \cline{2-6} 
 &  Wireless DSA& Song \textit{et al.}\cite{song2021deep}&  Dynamic spectrum allocation&  Multi-agent&DQN  \\ \Xhline{4\arrayrulewidth}
 
 \parbox[t]{2mm}{\multirow{18}{*}{\rotatebox[origin=c]{90}{IoT \& Other Emerging Wireless Networks}}}& IoT sensor networks&  Wang \textit{et al.}\cite{RN11, wang2018deep, wang2017deep}&  Dynamic multi-channel access&  Single-agent& DQN  \\ \cline{2-6} 
 & Energy harvesting-based IoT sensors & Chu \textit{et al.}\cite{RN13}&  Dynamic spectrum allocation&  Single-agent& DQN \\ \cline{2-6}
 & CRNs & Zhang \textit{et al.}\cite{RN21} &  Dynamic multi-channel access &  Multi-agent& DQN \\ \cline{2-6}
 &  Wireless networks& Xu \textit{et al.}\cite{RN290}&  DSA&  Single-agent& DQN \& DDQN \\ \cline{2-6}
 &  V2X& Liang \textit{et al.}\cite{liang2019spectrum, liang2019multi}&  Dynamic spectrum sharing&  Single-\& multi-agent& DQN \\ \cline{2-6}
  & Multi-sensor scenario & Zhu \textit{et al.}\cite{RN12}&  Joint channel access \& packet forwarding&  Single-agent& DQN \\ \cline{2-6}
 &  mmWave V2X& Khan \textit{et al.}\cite{khan2019reinforcement}&  User association& Multi-agent &A3C \\ \cline{2-6}
 &  CRNs& Yang \textit{et al.}\cite{yang2018dynamic}&  DSA&  Multi-agent& DQN \\ \cline{2-6}
 &  NTNs& Cao \textit{et al.}\cite{cao2020deep}&  Dynamic multi-channel access&  Multi-agent& DQN \\ \cline{2-6}
 & CRNs & Yang \textit{et al.}\cite{yang2020partially}&  Joint dynamic spectrum access \& mode selection&  Multi-agent& DQN \\ \cline{2-6}
 &  Multi-channel wireless networks& Tomovic \textit{et al.}\cite{tomovic2020novel}&  DSA&  Single-agent& RNN-base DDQN \\ \cline{2-6}
 &  Wireless networks& Zhong \textit{et al.}\cite{RN292} &  Dynamic multi-channel access&  Single-\& multi-agent& actor-critic  \\ \cline{2-6}
 &  CRNs& Tondwalkar \textit{et al.}\cite{tondwalkar2019deep}& DSA&  Multi-agent& DQN \\ \cline{2-6}
 & IIoT networks & Shi \textit{et al.}\cite{shi2020deep}&  Dynamic spectrum sharing&  Single-agent& DQN \\ \cline{2-6}
 &  Distributed wireless networks& Janiar \textit{et al.}\cite{janiar2021deep}&  DSA&  Multi-agent& DDQN \\ \cline{2-6}
 & VANETs & Wang \textit{et al.}\cite{wang2021intelligent}& DSA &  Multi-agent& RNN-based DQN \\ \cline{2-6}
 &  CUAV& Jiang \textit{et al.}\cite{jiang2021multi} &  Joint spectrum sensing \& channel access& Multi-agent & DQN \\ \cline{2-6}
 & Heterogeneous environments & Xu \textit{et al.}\cite{xu2020application}&  DSA&  Multi-agent& RNN-based DQN \\ \Xhline{4\arrayrulewidth}
 
 \parbox[t]{2mm}{\multirow{5}{*}{\rotatebox[origin=c]{90}{Satellite Nets}}}& Multi-beam satellite systems& Liu \textit{et al.}\cite{RN17}& Dynamic spectrum allocation& Single-agent& DQN \\ \cline{2-6}
 & Multi-beam satellite systems& Hu \textit{et al.}\cite{hu2018deep}& Dynamic spectrum allocation& Single-agent& DQN \\ \cline{2-6}
 & SIoT& Zhao \textit{et al.}\cite{zhao2020deep}& Dynamic spectrum allocation& Single-agent& DQN \\ \cline{2-6}
 & SIoT& Liu \textit{et al.}\cite{liu2020dynamic}& Dynamic spectrum allocation& Single-agent& DQN \\ \cline{2-6}
 \parbox[c][4mm]{2mm} {}& LEO satellites& Zheng \textit{et al.}\cite{zheng2020leo}& Joint channel pre-allocation \& dynamic channel scheduling&  Single-agent& $Q$-learning\\ [2.8pt] \Xhline{4\arrayrulewidth}
 
 \parbox[t]{2mm}{\multirow{5}{*}{\rotatebox[origin=c]{90}{Multi-RAT}}} & Small BSs cellular& Challita \textit{et al.}\cite{RN15}& Joint dynamic channel selection \& interference management& Multi-agent& DQN \\ \cline{2-6}
 & Small BSs cellular& Naparstek \textit{et al.}\cite{RN20}& Dynamic spectrum allocation& Single-agent& DQN \\ \cline{2-6}
 & LTE-WLAN HetNets& Wang \textit{et al.}\cite{wang2019deep}& Dynamic multi-channel access& Single-agent& prediction-DDPG \\ \cline{2-6}
 & MEC-based V2X& Peng \textit{et al.}\cite{peng2020deep}& Joint allocation of spectrum, computing, \& storing& Multi-agent& DDPG \\ \cline{2-6}
 & 4G LTE and 5G NR systems& Challita \textit{et al.}\cite{challita2021deep}& Dynamic spectrum sharing& Single-agent& DQN \\ \hline
\end{tabular}%
}  
\end{table*}
\subsection{DRL for Rate Control}
This refers to the adaptive rate control in the uplink and downlink of wireless networks. With the explosive increase in the number of user devices and the emergence of massive types of data-hungry applications \cite{RN234}, it becomes essential to keep high network KPIs in terms of data rates and users' QoE. Adaptive rate control schemes must ensure satisfactory QoS in highly dynamic and unpredictable wireless environments. In this context, DRL techniques can be efficiently deployed to solve adaptive rate control problems instead of conventional approaches that possess high complexity and heavily rely on accurate network information and instantaneous CSI. 

In the following paragraphs, we review works that employ DRL algorithms to address the rate control issue in cellular and HomNets.

Liu \textit{et al.} \cite{RN296} address the problem of network resource allocation, in terms of rate, for 5G network slices. The problem is decomposed into a master-slave, and a multi-agent DDPG-based DRL scheme is then proposed to solve it. The agents are located in every network slice, whose action space is continuous, defining the resource allocation to users in the network slice. The state space is continuous and has two main parts; the first one shows how much utility the user obtained compared to its minimum utility requirement, while the second part shows the auxiliary and dual variables from the master problem. The reward is a continuous function defined in terms of utility, utility requirements, and auxiliary and dual variables. Simulation results demonstrate that their proposed algorithm outperforms the baseline approaches and gives a near-optimal solution.

In \cite{RN293}, the authors propose a single-agent DQN-based DRL algorithm to address the problem of dynamic uplink/downlink radio resources allocation in terms of network capacity in high mobility 5G HetNets. Their proposed algorithm is based on Time Division Duplex (TDD )configuration in which the agent is the BS, whose action space is discrete, corresponding to the configurations of TDD sub-frame allocation at the BS. The sate space is discrete, comprised of different kinds of features of the BS, including uplink/downlink occupancy, buffer occupancy, and channel condition of all uplinks/downlinks to the BS. The reward is discrete, defined as a function of the uplink and downlink channel utility, which mainly depends on the channel occupancy with chosen TDD configuration. Using experimental results, the authors show that their proposed algorithm achieves performance improvement in terms of both network throughput and packet loss rate, compared to some conventional TDD resource allocation algorithms. 

\paragraph*{Synthesis and Reflections}
This section reviews the use of DRL techniques for adaptive rate control in modern wireless networks.  In general, there is limited research that is solely dedicated to addressing the rate radio resource. Most of the research is dedicated to video streaming applications, as summarized in \cite{RN1}. However, as we discussed in the previous sections, the data rate control issue is typically addressed via controlling other radio resources such as power and spectrum. In addition,  the adaptive rate control is typically addressed as a joint optimization with other radio resources, as we will elaborate in the next section, e.g., as in \cite{chkirbene2021deep, munaye2021deep}. 

We also observe that DRL approaches for cellular and HomNets receive more attention than other wireless networks, and there is a lack of research on adaptive rate control for IoT and satellite networks. This also deserves more in-depth investigation and analysis.   

\subsection{DRL for Joint RRAM} \
Due to the massive complexity and large-scale nature of modern wireless networks, it becomes necessary to design efficient schemes that account for the joint radio resources. In many scenarios, the design problem in wireless networks might end up with competing objectives. For example, in UDNs, increasing the power level is beneficial in combating path loss and enhancing the received signal quality. However, this might cause serious interference to the neighboring user devices and BSs. Hence, the joint design of power level control and interference management becomes mandatory. Conventional approaches for solving joint RRAM problems require complete and instantaneous knowledge about network statistics, such as traffic load, channel quality, and radio resources availability. However, obtaining such information is not possible in such large-scale networks. In this context, DRL techniques can be adopted to learn system dynamics and communication context to overcome the limited knowledge of wireless parameters. 

This section intensively reviews the works that implement DRL algorithms for the problem of joint RRAM in modern wireless networks. Particularly, we present related works that jointly optimize the radio resources shown in Fig. \ref{ClassificationOnResources}, such as power allocation, spectrum resources, user association, dynamic channel access, cell selection, etc. 

\subsubsection{In Cellular and HomNets} \
In the following paragraphs, we review works that employ DRL algorithms to address the joint RRAM issue in cellular and HomNets shown in Fig. \ref{ClassificationOnResources}.

The work in \cite{zhang2020energy} presents a single-agent DDPG-based DRL model to study the joint optimization of user mode selection, bandwidth allocation, power control, and channel selection with the aim for maximizing the EE in D2D-enabled cellular HetNets. The agent is a controller in the macrocell, whose state space is continuous, representing the user’s QoS satisfaction degree. The action space is continuous, which is to select power, channel, and bandwidth allocation factor, while the reward is a continuous function defined by the system EE. Experimental results demonstrate the efficiency of their proposed algorithm in terms of both the convergence speed and improving the EE in a D2D-enabled HetNets compared with other benchmark schemes.

In another interesting work in \cite{zhang2019deep}, the authors study the problem of joint optimization of transmission mode selection and resource allocation for CV2X. They propose single-agent settings in which DQN and federated learning (FL) models are integrated to improve the model's robustness. The agent in their model is each V2V pair. The action space is discrete, representing the resource block (RB) allocation, communication mode selection, and transmit power level of the V2V transmitter. The state space is a hybrid (continuous and discrete) consisting of five main parts; the received interference power at the V2V receiver and the BS on each RB at the previous subframe, the number of selected neighbors on each RB at the previous subframe, the large-scale channel gains from the V2V transmitter to its corresponding V2V receiver and the BS, current load, and remaining time to meet the latency threshold. The reward is a continuous function defined in terms of the sum-capacity of vehicular UEs as well as the QoS requirements of both vehicular UEs and V2V pairs. Using experimental results, the authors show that their proposed two-timescale federated DRL algorithm outperforms other decentralized baselines.

Jang \textit{et al.} \cite{jang2020deep} propose a multi-agent DQN-based algorithm to address the problem of sum-rate maximization via a joint optimization of resource allocation and power control in small cell wireless networks. The agents in their proposed model are the small cell BSs, whose action space is discrete, corresponding to selecting the resource allocation and power control of small BS on RB. The state space is continuous, including all the CSI that the small BS collects on RB, such as local CSI, local CSI at the transmitter, etc. The reward is a continuous function expressed by the average sum-rate of its own serving users and the other small BSs. Experimental results show that their proposed approach both outperforms the conventional algorithms under the same CSI assumptions and provides a flexible tradeoff between the amount of CSI and the achievable sum-rate.

In another interesting work in \cite{liao2019model}, the authors propose a model-driven multi-agent Double DQN-based framework for resource allocation in UDNs. In particular, They first develop a DNN-based optimization framework comprised of a series of ADMM iterative procedures that uses the CSI as the learned weights. Then, a channel information absent $Q$-learning resource allocation algorithm is presented to train the DNN-based optimization scheme without massive labeling data, where the EE, SE, and fairness are jointly optimized. The agents are each D2D transmitter, whose action space is discrete, corresponding to selecting a subcarrier and corresponding transmission power. The state space is a hybrid (continuous and discrete) consisting of two parts; user association information and interference power. The reward function is continuous, comprised of two components; the network EE and the fairness of service quality, which is expressed by the variance of throughput between authorized users. Using experimental results, it is demonstrated that their proposed algorithm has a rapid convergence speed, well characterizes the extent of optimization objective with partial CSI, and outperforms other existing resource allocation algorithms.

Liu \textit{et al.} \cite{liu2019deep} propose a single-agent Dueling DQN-based DRL model to address the problem of joint energy efficiency (EE) and spectral efficiency (SE) in 5G ultra-dense networks (UDNs). In their approach, the agent is a macro gNodeB (MgNB) whose state space is continuous, comprised of the number and throughput of all small cells as well as the allocation of all resource blocks (RBs) in the network. The agent's action space is to define which RBs are reused by the small cell in the network, while the reward is a function of the tradeoff between SE and EE. Experimental results demonstrate that their proposed algorithm performs better than both the conventional $Q$-learning and DQN.

The authors in \cite{zhang2020deepMulti} propose a multi-agent Double DQN-based scheme to address the problem of joint subcarrier assignment and power allocation in D2D underlying 5G cellular networks. The agents in their model are the D2D pairs, whose action space is discrete, corresponding to determining the transmit power allocation on the available subcarriers. The state space is a hybrid (continuous and discrete), comprised of four components: 1) local information (including the previous transmit power, previous SE achieved by transmitters, channel gain, and SINR), 2) the interference that each agent causes at the BS side, 3) the interference received from agent's interfering neighbors and the SE achieved by agent's neighbors, and 4) the interference that each agent causes to its neighbors. The reward is a continuous function comprised of three elements: 1) the SE achieved by each agent, 2) the SE degradation of the agent's interfered neighbors, and 3) the penalty due to the interference generated at the BS. Experimental results show that their proposed algorithm outperforms both the exhaustive and random subcarrier and even power (RSEP) assignment methods in terms of SE of D2D pairs.

In previous work in \cite{asheralieva2016autonomous}, the authors propose a multi-agent $Q$-learning-based scheme to address the problem of joint channel and power level selection in autonomous D2D-based heterogeneous cellular networks. Two spectrum usage scenarios are considered; when the D2D pairs transmit over the dedicated spectrum bands and when they shared cellular/D2D channels. The agents are the D2D pairs whose action space is discrete, defined as the set of their possible channel and power level decisions. The state space is also discrete, which takes the value of "1" if the SINR is greater than some predefined threshold; otherwise, the value of the state is "0". The agent's reward is a continuous function, defined by the current agent's state and its partially observable actions. Experimental results show that their proposed algorithm has relatively fast convergence and near-optimal performance after a few number of iterations.

Most recently in \cite{wang2021joint}, the authors propose a multi-agent DQN-based DRL scheme to address the problem of spectrum allocation and power control for Mission-critical communication (MCC) in 5G networks. In MCC, multiple D2D users reuse non-orthogonal wireless resources of cellular users without BS in order to enhance the network's reliability. The paper aims to help the D2D users autonomously select the channel and allocate power to maximize system capacity and SE while minimizing interference to cellular users. The agents are the D2D transmitters whose action space is discrete, corresponding to channel and power level selection. The state space is discrete, defined in a three-dimensional matrix, which includes information on the channel state of uses, the state of power level, and the number of the D2D pairs. The reward function is discrete, defined in terms of the total system capacity and constraints. Simulation results show that their proposed learning approach significantly improves spectrum allocation and power control compared to traditional methods.

The authors in \cite{RN297} propose a multi-agent DQN-based model to address the problem of joint user association and power control in OFDMA-based wireless HetNet. The agents are the UEs, whose action space is discrete, corresponding to jointly associate with the BS and determine the transmit power. The state space is discrete, which is defined by the situation of all UEs association with BS and power control. The reward function is continuous, which is defined in terms of the sum-EE of all UEs. Using simulation results, it is shown that their proposed method outperforms the $Q$-learning method in terms of convergence and EE.

The authors in \cite{yu2021resource} propose a decentralized multi-agent DQN-based scheme to address the problem of joint channel resource and transmit power in D2D underlay cellular networks. In their scheme, the agents are each D2D transmitter. The action space is discrete, corresponding to the selection of each D2D's channel resource and transmit power for data transmission. The state space is discrete consisting of six elements; the transmit power level of the cellular users, the maximum transmit power of each D2D transmitter, the distances between the cellular users and the D2D receiver devices, distances between D2D transmit devices and D2D receiver devices, selected channel of other agents, and selected transmit power level of other agents. The reward is a continuous function defined by the total effective throughput of D2D pairs in the network. Simulation results show that their proposed algorithm achieves better performance in terms of the total effective data rate than the random resource allocation method.

In \cite{zhang2019deepHetNet}, a single-agent DQN-based DRL model is proposed to address the problem of joint optimization of user association, resource allocation, and power allocation in HetNets. The agent is the BS, whose action is discrete, corresponding to power allocation to users. The state space is discrete, defined by the channel gain matrix and the set of users association. The reward function is continuous, defined by the utility function of users' achieved data rate. Using simulation results, the authors show that their proposed algorithm outperforms some of the existing methods in terms of SE and convergence speed.

Huang \textit{et al.} \cite{huang2020reinforcement} propose a single-agent DQN model to address the problem of delay minimization via joint spectrum and power resource allocation in mmWave mobile hybrid access network. The agent is located in the roadside BS, whose action space is discrete, corresponding to allocating spectrum and power resources for data. The state space is discrete, consisting of information about the current power and spectrum of the resource pool, required spectrum and power, and the number of spectrum and power levels. The reward signal is a continuous function defined in terms of queueing delay and the resource length required for each data. Using simulation results, it is shown that their proposed model guarantees the URLLC delay constraint when the load does not exceed 130\%, which outperforms other conventional methods such as random and greedy algorithms.

\subsubsection{In IoT and Other Emerging Wireless Networks} \ 
In the following paragraphs, we review works that employ DRL algorithms to address the joint RRAM issue in IoT and emerging wireless networks depicted in Fig. \ref{ClassificationOnResources}.

For the same system settings in \cite{nasir2019multi}, the authors in \cite{nasir2020deepjoint} extended their work and propose a multi-agent DDPG-based DRL framework to address the problem of the joint spectrum and power allocation in wireless networks. Two DRL-based algorithms are proposed, which are executed and trained simultaneously in two layers in order to jointly optimize the discrete subband selection and continuous power allocation. The agent in their approach is each transmitter. In the top layer, the action space of all agents is discrete, representing the discrete subband selection, while the bottom layer has a continuous action space corresponding to the transmit power allocation. The state space is a hybrid (continuous and discrete), containing information on achieved SE, transmit power, sub-band selection, rank, and downlink channel gain. The reward is shared by both layers, which is a continuous function defined in terms of the externality of agents to interference and the spectral efficiency. Using experimental results, the authors show that their proposed framework outperforms the conventional fractional programming algorithm.

Based on their initial work in \cite{tan2019deep}, the authors in \cite{tan2020deep} extended their work and propose a distributed multi-agent DQN-based DRL scheme to address the problem of joint channel selection and power control in D2D networks. The agents in their model are the D2D pairs, whose action space is discrete, corresponding to selecting a channel and a transmit power. The state space of each agent is a hybrid (continuous and discrete) which contains three sets of information; local information, non-local information from the agent's receiver-neighbor set, and non-local information from the agent's transmitter-neighbor set. The reward function of each agent is continuous, which is decomposed into the following elements; its own received signal power, its own total received SINR, its interference caused to transmitter-neighbors, the received signal power, and the total received SINR of transmitter-neighbors. Using simulation results, it is shown that the performance of their scheme closely approaches that of the FP-based algorithm even without knowing the instantaneous global CSI.

In \cite{RN14}, the authors extended their previous work in \cite{ ye2018deep, ye2018deepdistributed} and present a distributed multi-agent DQN-based model to address the problem of joint sub-band selection and power level control in V2V communication networks. Their proposed model is applicable to both unicast and broadcast scenarios. The agents are the V2V link or vehicles whose action space is discrete, corresponding to the selection of the frequency band and transmission power level that generate small interference to all V2V and V2I links while ensuring enough resources to meet latency constraints. The state space is continuous, containing the following information; the CSI of the V2I link, the received interference signal strength in the previous time slot, the channel indices selected by neighbors in the previous time slot, the remaining load for transmission, and the time left before violating the latency constraint. The reward function is continuous, consisting of three components; the capacity of V2I links, the capacity of V2V links, and the latency constraint. Using experimental results, it is shown that agents learn to satisfy the latency constraints on V2V links while minimizing the interference to V2I communications.

The authors in \cite{yuan2020deep} propose a single-agent Double DQN-based DRL to address the problem of joint channel selection and power allocation with network slicing in CRNs. The aim of their study is to provide high SE and QoS for cognitive users. The agent is the overall CRN, whose action space is discrete, corresponding to the channel selection and power allocation of SUs. The state space is continuous, defined by the SINR of the PU. The reward function is continuous, which is a function of the system SE, user QoS, interference temperature, and the interference temperature threshold. Experimental results show that their proposed algorithm improves the SE and QoS and provides faster convergence and more stable performance compared to the $Q$-learning and DQN algorithms.

\begin{table*}[htb!]
\centering
\caption{Summary of the Related Works that Address the Joint RRAM.}
\label{TableClass4_Version2}
\resizebox{\textwidth}{!}{%
\begin{tabular}{|l|l|c|c|c|c|}
\hline
\rowcolor[HTML]{C0C0C0} 
\multicolumn{2}{|l|}{\cellcolor[HTML]{C0C0C0}} & \cellcolor[HTML]{C0C0C0} & \cellcolor[HTML]{C0C0C0} & \multicolumn{2}{c|}{\cellcolor[HTML]{C0C0C0} \textbf{Learning Algorithm}} \\ \cline{5-6} 
\rowcolor[HTML]{C0C0C0} 
\multicolumn{2}{|l|}{\multirow{-2}{*}{\cellcolor[HTML]{C0C0C0} \textbf{Network Type}}} & \multirow{-2}{*}{\cellcolor[HTML]{C0C0C0} \textbf{Ref.}} & \multirow{-2}{*}{\cellcolor[HTML]{C0C0C0} \textbf{Types of Joint Radio Resources (or Issues Addressed)}} & \textbf{Mode} & \textbf{Algorithm} \\ \hline \hline
 \parbox[t]{2mm}{\multirow{12}{*}{\rotatebox[origin=c]{90}{Cellular \& HomNets}}} &   D2D-enabled cellular& Zhang \textit{et al.}\cite{zhang2020energy}& User mode selection, bandwidth allocation, power control, \& channel selection&  Single-agent& DDPG  \\ \cline{2-6}
 & CV2X& Zhang \textit{et al.}\cite{zhang2019deep}& Transmission mode selection \& resource allocation& Single-agent& DQN \\ \cline{2-6}
 & Small cell networks& Jang \textit{et al.}\cite{jang2020deep}& Resource allocation \& power control& Multi-agent& DQN \\ \cline{2-6}
 & UDNs& Liao \textit{et al.}\cite{liao2019model}& Subcarrier selection \& transmission power& Multi-agent& DDQN  \\ \cline{2-6}
 & 5G UDNs& Liu \textit{et al.}\cite{liu2019deep}& EE and SE& Single-agent& Dueling DQN \\ \cline{2-6}
 & D2D underlying 5G cellular& Zhang \textit{et al.}\cite{zhang2020deepMulti}& Subcarrier assignment \& power allocation& Multi-agent& DDQN \\ \cline{2-6}
 & D2D-based heterogeneous cellular& Asheralieva \textit{et al.}\cite{asheralieva2016autonomous}& Channel \& power level selection& Multi-agent& $Q$-learning \\ \cline{2-6}
 &Mission-critical in 5G & Wang \textit{et al.}\cite{wang2021joint}& Spectrum allocation \& power control&  Multi-agent & DQN \\ \cline{2-6}
 & OFDMA-based networks& Ding \textit{et al.}\cite{RN297} & User association \& power control& Multi-agent & DQN \\ \cline{2-6}
 & D2D underlay cellular& Yu \textit{et al.}\cite{yu2021resource}& Channel resource \& transmit power& Multi-agent & DQN \\ \cline{2-6}
 & HetNets Cellular & Zhang \textit{et al.}\cite{zhang2019deepHetNet}& User association, resource allocation, \& power allocation& Single-agent & DQN \\ \cline{2-6}
 & mmWave mobile hybrid access& Huang \textit{et al.}\cite{huang2020reinforcement} & Spectrum \& power resource allocation&  Single-agent & DQN  \\ \Xhline{4\arrayrulewidth}
 
 \parbox[t]{2mm}{\multirow{9}{*}{\rotatebox[origin=c]{90}{IoT \& Emerging Nets}}}&  Wireless networks& Nasir \textit{et al.}\cite{nasir2020deepjoint}&  Spectrum \& power allocation &  Multi-agent& DDPG   \\ \cline{2-6} 
 & D2D networks& Tan \textit{et al.}\cite{tan2020deep, tan2019deep}& Channel selection \& power control & Multi-agent & DQN  \\ \cline{2-6} 
 & V2V networks& Ye \textit{et al.}\cite{RN14, ye2018deep, ye2018deepdistributed}& Sub-band selection \& power level control& Multi-agent & DQN  \\ \cline{2-6} 
 & CRNs& Yuan \textit{et al.}\cite{yuan2020deep}& Channel selection \& power allocation& Single-agent & DDQN  \\ \cline{2-6} 
 & 5G-based NOMA systems& Zhang \textit{et al.}\cite{zhang2019energy}& Subcarrier assignment \& power allocation& Multi-agent & DQN \& DDPG  \\ \cline{2-6} 
 & Multi-carrier NOMA-based systems& He \textit{et al.}\cite{he2019joint}& Channel assignment \& power allocation& Single-agent & DQN  \\ \cline{2-6} 
 & Wireless networks& Jiang \textit{et al.}\cite{jiang2020partially}& Channel selection \& power allocation& Multi-agent & DQN  \\ \cline{2-6} 
 & NOMA-based V2X networks& Xu \textit{et al.}\cite{xu2020deep}& Spectrum \& power allocation& Multi-agent & DDPG  \\ \cline{2-6} 
 & UAV-assisted IoT networks& Munaye \textit{et al.}\cite{munaye2021deep}& Bandwidth, throughput, \& power& Multi-agent & DQN  \\ \Xhline{4\arrayrulewidth} 
 
 \parbox[c][4.5mm]{2mm}{\multirow{2}{*}{\rotatebox[origin=c]{90}{Multi-RAT}}} &  Hybrid RF/VLC systems& Shrivastava \textit{et al.}\cite{shrivastava2020deep}& Bandwidth, power, \& user association& Multi-agent & DQN  \\ \cline{2-6}
 \parbox[c][4.5mm]{2mm}{}& Multi-RAT HetNets& Alwarafy \textit{et al.}\cite{alwarafy2021DeepRAT}& RAT selection \& power control& Single\& multi-agent & DQN \& DDPG \\ \cline{2-6}
 \parbox[c][4.5mm]{2mm}{}& Heterogeneous health systems& Chkirbene \textit{et al.}\cite{chkirbene2021deep}& RAT selection, data split control, \& compression ratio control& Single-agent & DDPG \\ \hline
\end{tabular}%
}
\end{table*}

The problem of joint subcarrier assignment and power allocation in an uplink multi-user 5G-based NOMA systems is addressed in \cite{zhang2019energy}. A multi-agent two-step DRL algorithm is proposed; the first step employs the DQN algorithm to output the optimum subcarrier assignment policy, while the second step employs the DDPG algorithm to dynamically allocate the transmit power for the network's users. The agent is a controller located at the BS, whose action space is a hybrid (discrete and continuous), corresponding to the subcarrier assignment decisions and power allocation decisions. The state space is continuous, which is defined by the users' channel gains at each subcarrier. The reward function is defined as the sum EE of the NOMA system. Experimental results show that their proposed algorithm provides better EE than the fixed and DQN-based power allocation schemes. 

In \cite{he2019joint}, the authors propose a single-agent DQN-based algorithm to address the problem of joint channel assignment and power allocation in multi-carrier NOMA-based systems. The authors used an attention-based neural network (ANN) to perform the channel assignment process. The agent in their approach is the BS, whose action is discrete, representing the selection of a channel for a user. The state space is discrete, characterized by the channel information that is defined by user-channel assignment pairs. The reward function is continuous, defined by the user’s achieved data rate. Using simulation results, the authors show that their proposed algorithm achieves better system performance in terms of sum-rate, compared to the state-of-the-art exhaustive search approaches and the method reported in \cite{zhu2017optimal}.

In another work \cite{jiang2020partially}, the authors address the problem of dynamic optimization of channel selection and power allocation in wireless networks based on distributed multi-agent DQN algorithm. The authors adopted a partially distributed framework with the aim of minimizing the long-term average data backlog of the network's users. In their model, the agents are the users whose action space is discrete and finite, representing the users' transmit power allocation for data blocks in the data buffer. The state space is also discrete and finite comprised of two parts; the users' data size and data block remaining lifetime. The reward function is continuous, defined in terms of the size of the remaining data in the data block and the transmitted data taken from the data block. Experimental results demonstrate that their proposed solution outperforms the conventional exhaustive search and Block Coordinate Descent-based optimization algorithms in terms of EE and data backlog.

The authors in \cite{xu2020deep} propose a multi-agent DDPG-based model to address the problem of joint power and spectrum allocation in NOMA-based V2X networks. In particular, the authors are looking to maximize the sum-rate of V2I communications. The agents are the V2V communication links. The state space is discrete, defined by a set of actions performed by V2I and V2V communication links. The set includes the transmit power of both V2I and V2V links as well as the spectrum multiplexing factor of both V2V and V2V links. The state space is continuous, defined by five parts; the local channel gain information of each V2V link, interference channels from other V2V communication links, interference channel from each link's own transmitter to the BS, interference channel from all V2I transmitters, and the state of queue length in the buffer of each V2V transmitter. The reward function is continuous, defined by the achieved sum-rate of V2I communication links and the delivery probability of V2V communication links. Compared with both the DQN algorithm and random resource allocation scheme, simulation results show that their proposed algorithm outperforms both of them in terms of maximizing the sum-rate of V2I communication links while meeting the latency and reliability requirements of V2V communications. 

Munaye \textit{et al.} \cite{munaye2021deep} propose a multi-agent DQN-based DRL model to address the problem of joint radio resources of bandwidth, throughput, and power in UAV-assisted IoT networks. The agents are the UAVs, whose action space is discrete, corresponding to jointly selecting channel allocation of bandwidth, throughput, and power. The state space is discrete, comprising three components; the air-to-ground channel used by users, the rate of power consumption, and the interference. The reward is a discrete function, defined in terms of throughput, power allocation, bandwidth, and SINR levels. Simulation results show that their proposed algorithm outperforms other algorithms in terms of accuracy, convergence speed, and error.

\subsubsection{In Multi-RAT HetNets} \ 
In the following paragraphs, we review works that employ DRL algorithms to address the joint RRAM problem in multi-RAT HetNets. This includes the coexistence of various variants of the wireless networks as illustrated in Fig. \ref{ClassificationOnResources}. 

Most recently, the authors in \cite{shrivastava2020deep} present a multi-agent DQN-based algorithm to address the problem of joint optimization of bandwidth, power, and user association in hybrid RF/VLC systems. The APs are the agents whose action is discrete, representing the bandwidth, association function, and power level. The state space is discrete, which is a function of the problem constraints such as system bandwidth, association function, and power levels. The reward is discrete, which is a function of the rates delivered by the APs. Experimental results show that their algorithms improve the achievable sum-rate and number of iterations for convergence by 10\% and 54\% compared to that obtained using conventional optimization approaches.

Another interesting work is proposed by Alwarafy \textit{et al.} in \cite{alwarafy2021DeepRAT}. In that work, the authors propose a hierarchical multi-agent DQN and DDPG-based algorithm to address the problem of sum-rate maximization in multi-RAT multi-connectivity wireless HetNets. The authors addressed the problem of multi-RATs assignment and continuous power allocation that maximize the network sum rate. In their model, single and multi-agents are proposed. The edge server acts as a single agent employed by DQN, while RATs APs behave as multi-agents employed by DDPG. For the single-agent DQN model, the action space is discrete, corresponding to the RATs-EDs assigning process. The state space of the DQN is continuous, comprised of the link gains and the required data rates of users. The reward function of the DQN agent is continuous, defined by the difference between the achieved rate and the required rate by users. For the multi-agent DDPG models, the action space is continuous, representing the power allocation of each RAT AP agent. The state space is a hybrid (continuous and discrete) consisting of four elements: the RATs-EDs assignment process performed by the DQN agent, the minimum data rate of users, the gains of the links, and the achieved data rate. Experimental results show that their algorithm's performance is approximately 98.1\% and 95.6\% compared to the conventional CVXPY solver that assumes full knowledge of the wireless environment.

An interesting work is presented in \cite{chkirbene2021deep}, in which the authors address the problem of network selection with the aim of optimizing medical data delivery over heterogeneous health systems. In particular, an optimization problem is formulated in which the network selection problem is integrated with adaptive compression to minimize network energy consumption and latency while meeting applications’ QoS requirements. A single-agent DDPG-based DRL model is proposed to solve it. The agent is a centralized entity that can access all radio access networks (RANs) information and Patient Edge Node (PEN) data running in the core network. The action space is discrete, corresponding to the joint selection of data split over the existing RANs and the adequate compression ratio. The state space is a hybrid (continuous and discrete) defined by two elements: the fraction of time that the PENs should use over a particular RAN and the PEN investigated in the current timestamp. The reward is a continuous function, which is defined in terms of: the fraction of data of PEN that will be transferred through RAN, the energy consumed by PEN to transfer bits over RAN, distortion, expected latency of RANs, the monetary cost of PENs to use RANs, the resource share, the fraction of time that the PENs should use over a particular, and the data rate. Simulation results demonstrate that their proposed scheme outperforms the greedy techniques in terms of minimizing energy consumption and latency while satisfying different PENs requirements.

\paragraph*{Synthesis and Reflections}
This section reviews the use of DRL methods for joint radio resources. Table \ref{TableClass4_Version2} summarizes the reviewed papers in this section. We observe that DRL tools can be efficiently deployed to address different types of joint radio resources for diversified network scenarios. The results obtained using DRL models are better than the heuristic methods \cite{munaye2021deep, xu2020deep} and comparable to the state-of-the-art optimization approaches \cite{tan2020deep}.

We also observe that multi-agent DRL deployment based on value-based algorithms receives more attention than policy-based algorithms. The reason is that users tend to have more control over their channel selection, data control, and transmission mode selection, and hence we find a more popular implementation of DRL agents at local IoT devices. In addition, the integration of value-based and policy-based algorithms for joint RRAM is also an interesting concept as presented in \cite{zhang2019energy, alwarafy2021DeepRAT}, which requires more investigation, especially for multi-agent deployment scenarios.

We also observe that DRL methods for cellular and HomNets as well as IoT wireless networks gain more attention than multi-RAT networks, particularly for D2D communications. In addition, there is a lack of research on applications of DRL for emerging IoT applications, such as healthcare systems as investigated recently in \cite{chkirbene2021deep}, which is also a promising field that requires more attention. Furthermore, we observe a lack of research on DRL applications for joint RRAM in satellite networks, which also deserves more in-depth investigation. 

\section{Open Challenges and Future Research Directions} \label{Sec5:future}
Throughout the previous section, we have demonstrated the superiority of DRL algorithms over traditional methods in addressing complex RRAM problems for modern wireless networks. However, there are still several challenges and open issues that either not explored yet or need further exploration. This section provides highlights these open challenges and provides insights on future research directions in the context of DRL-based RRAM for next generation wireless networks.

\subsubsection{Open Challenges}
\paragraph{Centralized vs. Decentralized RRAM Techniques}
Future wireless networks are characterized mainly by their massive heterogeneity in wireless RANs, the number of user devices, and types of applications. Centralized DRL-based RRAM schemes are efficient in guaranteeing enhanced network QoS and fairness in allocating radio resources. They also ensure that RRAM optimization problems will not get stuck in local minima due to their holistic view of the system. However, formulating and solving RRAM optimization problems become tough tasks in such large-scale HetNets. Hence, centralized DRL-based RRAM solutions are typically unscalable. This motivates distributed multi-agent DRL-based algorithms that enable edge devices to make resource allocation decisions locally. Stochastic Game-based DRL algorithms are one promising research direction in this context \cite{Feriani2021Single}. However, the rapid increase in the number of edge devices (players) makes information exchange in such networks unmanageable. Also, the partial observability of agents might lead to suboptimal RRAM policies. Therefore, it is an open challenge to develop DRL-assisted algorithms that optimally balance between the centralization and distribution issue in RRAM.

\paragraph{Dimensionality of State Space in HetNets}
In modern wireless HetNets, service requirements and network conditions are rapidly changing. Hence, single-agent DRL algorithms must be designed to capture and respond to these fast network changes. To this end, it is required to reduce the state space and action space during the learning process, which inevitably degrades the quality of the learned policies. The existence of multi-agents and their interactions will also complicate the agents' environment and prohibitively increase the dimensionality of state space, which will slow down the learning algorithms.

\paragraph{Reliability of Training Dataset}
Although the DRL-based solutions for RRAM we reviewed previously demonstrate efficient performance results, almost all the models are developed based on simulated training and testing datasets. The simulated dataset is typically produced based on some stochastic models, which provide simplified versions of practical systems and greatly ignore hidden system patterns. This methodology greatly weakens the reliability of the developed policies as their performance on practical networks would be skeptical. Hence, it is imperative to develop more effective and reliable approaches that generate precise simulation datasets and capture practical system settings as much as possible. This ensures high reliability and confidence during the training and testing modes of the developed RRAM policies. Developing such approaches is still a challenge due to the large-scale nature and rapid variations of future wireless environments.

On the other hand, the DRL models are sensitive to any change in the input data. Any minor changes in the input data will cause considerable change in the models' output. This mainly deteriorates the reliability of DRL algorithms, especially when deployed for modern IoT applications that require ultra-reliability, such as remote surgery or any other mission-critical IoT applications. Hence, ensuring high reliability for DRL models is a challenging issue.

\paragraph{Engineering of DRL Models for RRAM}
Since DRL employs DNNs as function approximators for the reward functions, DRL models will inherit some of the challenges that exist in the DNN world. For example, it is still challenging to optimize the DNN hyperparameters, such as the type of DNNs used (e.g., convolutional, fully connected, or RNN), the number of hidden layers, the number of neurons per hidden layer, the learning rate, etc. This challenge is even exacerbated in multi-agent settings as all agents share the same radio resources and must converge simultaneously to some policies.

On the other hand, the engineering of DRL parameters such as state space and reward function is challenging for RRAM. The state space must be engineered to capture useful and representative information about the wireless environment, such as the available radio resources, users' QoS requirements, channel quality, etc. Such information is crucial and heavily defines the learning and convergence behaviors of DRL agents. Again, the presence of multi-agents will even make it more challenging, as discussed in \cite{Feriani2021Single}. Also, since DRL models are reward-driven learning algorithms, the design of the reward function is also essential to guide the agent during the policy-learning stage. Formulating reward functions that capture the network objective and account for the available radio resources is also challenging.

\paragraph{System Dependency of DRL models}
DRL models are system-dependent as they are trained and tested for specific wireless environments and networks. Therefore, they provide effective results when employed to solve specific types of problems for which they are trained. However, if there would be a significant change in the characteristics of the wireless environment or the nature of the RRAM problem, such as network topology and available radio resources, the DRL model must be retrained as the old model is no longer reflecting the new training experiences. In modern wireless HetNets, such cases are frequently encountered, especially with real-time applications or in highly dynamic environments. In such a case, it becomes quite challenging for DRL agents to update and retrain their DNNs with rapidly changing input information from the HetNet environment \cite{hussain2020machine}.

\paragraph{Continuous Training of DRL Models}
DRL algorithms require big datasets to train their models, which is typically associated with a high cost \cite{RN1}. The network system pays this cost during the information collection process due to, e.g., the high delays, extra overhead, and energy consumption. The emergence of a large number of real-time applications and services has even increased this training cost. In this context, DRL models require to be continuously retrained with fresh data collected from the wireless environment to be up-to-date and to ensure accurate and long-term control decisions. It is not practical to conduct manual retraining of the models in such large-scale HetNets settings. Therefore, continuous retraining is preferred, in which a dedicated autonomous system can be employed to continuously assess and retrain old DRL models. However, it would be quite challenging for the new autonomous system to promptly perform this operation due to the environment's rapid variations. Also, monitoring and updating DRL models in multi-agent scenarios becomes an expensive task.

\paragraph{Context of RRAM}
The implementation of DRL algorithms basically depends on the use-cases. The context and deployment scenarios in which RRAM is required must be considered during the development of DRL models. For example, RRAM in health-sector IoT applications is different from the environmental IoT applications counterparts. Due to the high sensitivity of data in the health-sector applications, extra data pre-processing must be performed, including data compression and encryption \cite{chkirbene2021deep}. This will directly affect the number of radio resources to be allocated for such applications. Hence, DRL models must be aware of the context aspect of applications, which is considered another challenge. 

\paragraph{Competing Objective Design of DRL Models}
Next generation wireless networks are expected to provide enhanced system QoS in terms of high data rate, high EE/SE, and reduced latency in order to support the emerging IoT vital applications. Depending on the deployment scenario, formulating multi-objective RRAM optimization problems usually ends with many competing objectives and/or constraints. For instance, in cellular UDNs, high resource utilization of, e.g., power allocation or channel may cause severe interference. Also, for IoT applications such as vehicular communications, we require to ensure ultra-reliable and low-latency communication links, which are usually competing objectives. Therefore, developing multi-objective DRL-based RRAM models that accommodate these competing requirements is still a persisting challenge.

\subsubsection{Future Research Directions}
\paragraph{DRL with Explainable AI (XDRL) for RRAM}
Explainable AI (XAI) has recently emerged as an efficient technology to improve the performance of DRL models. It is mainly envisioned to unlock the "black-box" nature of conventional ML approaches and provide interpretability for DRL models \cite{adadi2018peeking}. In particular, XAI explains the reasons behind certain predictions made by DRL models (or ML models in general) by fully understanding the precise working principle of these models. Hence, ensuring trust, reliability, and transparency in the DRL algorithms' policy devlopment and decision-making processes. The research on XAI technologies in wireless communication is still at its initial stages, and there are still some key issues for future research in the context of RRAM for next generation wireless networks. For example, DRL models can get stuck easily into local optimal solutions when utilized to solve complex RRAM problems. This issue can be significantly avoided with the help of XAI. Fortunately, the heterogeneity of information in modern wireless HetNets  will significantly help to achieve the interpretation for DRL algorithms. In this context, developing RRAM schemes for wireless HetNets, through entity recognition, entity-relationship extraction, and representation learning, makes the DRL models' interpretation more reliable, accurate, and intuitive, which is a promising research direction.

\paragraph{Integrating DRL and Blockchain Techniques}
Blockchain-based RRAM has emerged recently as one of the promising enabling technologies for future wireless HetNets \cite{RN234}. It has gained considerable momentum lately due to its ability to provide intelligent, secure, and highly efficient distributed resource sharing and management. The integration of DRL with Blockchain is also an interesting research direction, as in \cite{He2021Blockchain, Guo2020Adaptive, Hu2021Blockchain}. For example, DRL algorithms can be distributively deployed within participants or within the centralized spectrum-access systems to facilitate spectrum auctions and transactions \cite{Hu2021Blockchain}. Also, many of the auction's winner-determination problems in future wireless HetNets are expected to be extremely complex and intractable due to the massive increase in the number of participants, e.g., bidders and sellers. Hence, DRL algorithms are efficient tools that can be utilized to solve such types of problems.

\paragraph{Federated DRL (FDRL)-Based RRAM}
Federated learning (FL) framework is envisioned mainly to preserve data privacy in ML algorithms \cite{wahabfederated2021, tran2019federated}. In FL, ML algorithms are locally distributed at the wireless network edge, and the data is processed locally and not shared globally. The local ML models are then utilized for training a centralized global model.
In this context, the federated DRL learning (FDRL) scheme can be leveraged when many wireless user devices require making autonomous local decisions. In such a case, DRL agents do not exchange their local observations, and also, not necessarily all agents receive reward signals \cite{RN251}.

\begin{table*}[htb!]
\centering
\caption{Summary of Challenges and Future Research Directions in the Context of using DRL for RRAM in Future Wireless Networks.}
\label{Challenges_Future}
\begin{tabular} {|c|p{16.6cm}|} \hline 
\cellcolor[HTML]{C0C0C0}  &  Developing DRL-based algorithms that optimally balance the centralization and distribution issue of RRAM in future large-scale massive   HetNets. \\ \cline{2-2} 
\cellcolor[HTML]{C0C0C0}&  Reducing the dimensionality of state space in distributed MADRL algorithms during the learning process without slowing down or degrading the   quality of learned RRAM policies. \\ \cline{2-2} 
\cellcolor[HTML]{C0C0C0}  &  Developing more effective and reliable training approaches that generate accurate simulation datasets and capture practical system settings. \\ \cline{2-2} 
\cellcolor[HTML]{C0C0C0}  &  Optimizing DRL models' hyperparameters, especially in MADRL scenarios, and engineering the state space and reward functions to capture   representative information about system dynamics. \\ \cline{2-2} 
\cellcolor[HTML]{C0C0C0}&  Designing agile DRL algorithms that can quickly update and retrain   the DNNs in response to the rapid change of input information from the highly   dynamic HetNet environment. \\ \cline{2-2} 
\cellcolor[HTML]{C0C0C0}  &  Performing continuous retraining for the DRL models, especially MADRL, with fresh data in future large-scale and rapidly changing wireless environments. \\ \cline{2-2} 
\parbox[c]{2mm}{\multirow{-8}{*}{\rotatebox[origin=c]{90}{\textbf{ Open Challenges}}}} \cellcolor[HTML]{C0C0C0}&  Developing DRL models that are aware of the context aspect and use   cases of various emerging applications. \\ \cline{2-2} 
 \cellcolor[HTML]{C0C0C0}&  Developing DRL algorithms that accommodate competing multi-objectives relevant to emerging applications. \\ \hline \hline
  
 \cellcolor[HTML]{C0C0C0}& Developing efficient and reliable DRL algorithms for RRAM in next-generation HetNets based on the XAI concept through, e.g., entity recognition, entity-relationship extraction, and representation learning.    \\ \cline{2-2}
\cellcolor[HTML]{C0C0C0}&Developing DRL-based blockchain techniques to address the problem of distributed resource sharing and management for future large-scale HetNets, e.g., to facilitate spectrum auctions and transactions, solving the problem of auction's winner-determination, etc.    \\ \cline{2-2} 
\cellcolor[HTML]{C0C0C0}& Developing FDRL algorithms that ensure global solutions for complex RRAM optimization problems while guaranteeing data and models privacy during information sharing and models updating.\\ \cline{2-2} 
\cellcolor[HTML]{C0C0C0}& Developing DRL models to achieve intelligent load balancing in future self-sustaining (or self-organization) HetNets.    \\ \cline{2-2}  
  \cellcolor[HTML]{C0C0C0}& Developing light-weighted and agile networked MADRL algorithms that enable cooperation between agents with different heterogeneous reward functions and adapt to environments with rapid mobility.    \\ \cline{2-2} 
 \cellcolor[HTML]{C0C0C0}& Developing ultra-reliable RRAM schemes by integrating DRL algorithms and GANs techniques to support emerging IoT applications with high-reliability demands.    \\ \cline{2-2} 
 \cellcolor[HTML]{C0C0C0} \parbox[c]{2mm}{\multirow{-11}{*}{\rotatebox[origin=c]{90}{\textbf{Future Research Directions}}}} & Developing end-to-end DRL-based algorithms that jointly optimize the configuration of RIS systems, i.e., elements' phases and amplitudes, and radio resources of networks, e.g., downlink transmit power.    \\ \hline
\end{tabular}%
\end{table*}

Developing fine-grained policies in DRL becomes challenging when the state space is small and the training dataset is very limited \cite{wang2020federated}. In FDRL, the direct exchange of data between agents is not possible as this will preach the privacy promise of FL scheme. Instead, local DRL models can be developed and trained for agents with the help of other agents while preserving users' data privacy, as in \cite{RN144}. Hence, developing algorithms and schemes that guarantee data and models privacy during both information sharing and models updating is an interesting research direction.  

FDRL framework can also be exploited in the RRAM of modern wireless networks. For example, it can be deployed for solving complex wireless network optimization problems, such as power control in cellular UDNs. In this context, FDRL can ensure a global solution for complex network optimization problems without sharing information between BSs; each BS solves its optimization problem locally and shares the results with neighboring BSs.

\paragraph{DRL-Based Load Balancing for Self-Sustaining Networks}
Load balancing in modern wireless UDNs is another promising research direction. The objective is to balance the wireless networks by moving some users from the heavily congested BSs to uncongested ones, thus improving BSs utilization and providing enhanced QoS provisioning. Although the load balancing field has been heavily investigated in the literature using conventional resource management approaches, as in \cite{Das2013Load, Zhu2009Dynamic, Desai2016International}, there still a research gap in applying DRL for such a field. In this context, DRL can be adopted to realize the self-sustaining (or self-organization) vision of next generation wireless networks \cite{RN234}. Hence, developing single/multi-agent DRL models to achieve intelligent load balancing in future HetNets, is a possible research direction.

\paragraph{MADRL Algorithms in Support of Massive Heterogeneity and Mobility}
In modern wireless networks, massive heterogeneity is one of the main enabling technologies to provide ubiquitous coverage and enhance system reliability. However, some mathematical frameworks, such as the Markov game, are basically developed for homogeneous systems. To overcome this challenge, there are proposals on adopting networked MADRL algorithms that enable the cooperation between agents with different reward functions \cite{Feriani2021Single}, which is an interesting research direction. Also, RRAM for networks with high mobility, such as vehicular and railway communications, is a persistent challenge. In this context, developing light-weighted and agile MADRL algorithms that account for and adapt to rapid network mobility is another interesting research direction.

\paragraph{DRL-Based RRAM with Generative Adversarial Networks (GANs) for RRAM}
Ensuring the reliability of DRL algorithms is one of the major challenges and objectives in DRL-based RRAM methods. In many real-life scenarios, we may require to deploy DRL models to allocate resources in vital systems requiring ultra-reliability, such as IoT healthcare applications \cite{chkirbene2021deep}. In this context, there are proposals on Generative Adversarial Networks (GANs), which have emerged recently as an effective technique to enhance the reliability of DRL algorithms \cite{Kasgari2021Experienced}. 

In practice, the shortage of realistic training datasets that are required to train DRL models and learn optimal policies is a challenging issue. To overcome this, GANs are utilized, which generate large amounts of realistic datasets synthetically by expanding the available limited amounts of real-time datasets. From a DRL perspective, GANs-generated synthetic data is more effective and reliable than traditional augmentation methods \cite{naeem2020gentle}. This is because DRL agents will be exposed to various extreme challenging and practical situations by merging the realistic and synthetic data, enabling DRL models to be trained on unpredicted and rare events. Another advantage of GAN over traditional data augmentation methods is that it eliminates dataset biases in the synthetic data, which greatly enhances the quality of the generated data and leads to more reliability in DRL models' training and learning processes. 

In general, the research in the GANs-based DRL methods for RRAM is still in its early stages, and we believe that this research direction will take further pace in the future.

\paragraph{DRL for RRAM in RIS-Assisted Wireless Networks}
Reconfigurable Intelligent Surfaces (RIS) have emerged recently as an innovative technology to enhance the QoS of future wireless networks \cite{Alghamdi2020Intelligent, huang2020holographic}. RIS can be deployed in cellular networks as passive reflecting elements to provide near line-of-sight communication links to users, hence enhancing communication reliability and reducing latency \cite{Wu2020Towards, pan2020reconfigurable}. Deploying RIS to assist cellular communication, however, requires judicious RRAM schemes to optimize network performance. This research field is still nascent, and there is much to do for future research and investigation, especially in the context of DRL-based RRAM techniques. Towards this, it is required to develop end-to-end DRL-based algorithms that jointly optimize the configuration of the RIS system, i.e., elements' phases and amplitudes, and radio resources of BSs. For instance, designing DRL models that intelligently and optimally allocate the downlink BSs' transmit power from one side and the amplitude and phase shifts of the RIS elements on the other side is a promising research direction, as in \cite{lee2020deep}. We also believe that the currently ongoing research in RIS-assisted wireless networks, e.g., \cite{lee2020deep, Taha2020Deep, Huang2020Reconfigurable, Yang2020Deep} will be cornerstones.

Table \ref{Challenges_Future} summarizes the open challenges and future research directions provided in this section.

\section{Conclusion} \label{Conculsion}
This paper presented a comprehensive survey on the applications of DRL techniques in RRAM for next generation wireless HetNets. We thoroughly reviewed the conventional approaches for RRAM, including their types, advantages, and limitations. We have then illustrated how the emerging DRL approaches can overcome these shortcomings to enable DRL-based RRAM. After that, we illustrated how the RRAM optimization problems can be formulated as an MDP before solving them using DRL techniques. Furthermore, we conducted an extensive overview of the most efficient DRL algorithms that are widely leveraged in addressing RRAM problems, including the value- and policy-based algorithms. The advantages, limitations, and use-cases for each algorithm are provided. We then conducted a comprehensive and in-depth literature review and classified the existing related works based on both the radio resources they are addressing and the type of wireless networks they are considering. To this end, the types of DRL models developed in these related works and their main elements are carefully identified. Finally, we outlined important open challenges and provided insights into future research directions in the context of DRL-based RRAM.

\section*{Acknowledgment}
This publication was made possible by NPRP-Standard (NPRP-S) Thirteen ($13^\text{th}$) Cycle grant \# NPRP13S-0201-200219 from the Qatar National Research Fund (a member of Qatar Foundation). The findings herein reflect the work, and are solely the responsibility, of the authors.

\bibliographystyle{IEEEtran}
\bibliography{The_bibliography}

\end{document}